\documentclass[a4paper,10pt]{article}
 
 \usepackage{microtype}
\usepackage[utf8]{inputenc}
\usepackage[T1]{fontenc}
\usepackage{lmodern}
\usepackage{amsmath,amsthm,amsfonts,amssymb,bm}
\usepackage{dsfont}			
\usepackage{enumitem}
\usepackage{graphicx}
\usepackage{booktabs}
\usepackage[bottom]{footmisc} 

\usepackage{titlesec}

\titleformat{\paragraph}
{\normalfont\normalsize\bfseries}{\theparagraph}{1em}{}
\titlespacing*{\paragraph}
{0pt}{3.25ex plus 1ex minus .2ex}{1.5ex plus .2ex}

\usepackage{natbib}
\usepackage[right=2cm,left=2cm,top=2cm,bottom=3cm]{geometry}
\usepackage[font={small,it}]{caption}	
\usepackage[colorlinks,linkcolor=blue,citecolor=blue,urlcolor=blue]{hyperref}
\usepackage{multirow}
\usepackage{subfig}
\usepackage{quoting}
\quotingsetup{font=it}
\usepackage{url}
\usepackage{tikz}             
\usepackage{algorithm}
\usepackage[noend]{algpseudocode}

\makeatletter
\def\BState{\State\hskip-\ALG@thistlm}
\makeatother

\usetikzlibrary{bayesnet}     

\usepackage{DejaVuSansMono}
\usepackage[scaled=0.8]{beramono}

 \usepackage{array}
 \newcolumntype{L}[1]{>{\raggedright\let\newline\\\arraybackslash\hspace{0pt}}m{#1}}
\newcolumntype{C}[1]{>{\centering\let\newline\\\arraybackslash\hspace{0pt}}m{#1}}
 \newcolumntype{R}[1]{>{\raggedleft\let\newline\\\arraybackslash\hspace{0pt}}m{#1}}

\newcommand{\Y}{\mathbf{Y}}

\newcommand{\arch}{y_{ij}^{(k)} }
\newcommand{\ak}{\alpha^{(k)} }
\newcommand{\bk}{\beta^{(k)} }
\newcommand{\ti}{\theta_i }
\newcommand{\gj}{\gamma_j }
\title{ \textsc{Latent Space Modeling of Multidimensional Networks with Application to the Exchange of Votes in Eurovision Song Contest} }

\author{Silvia D'Angelo, Thomas Brendan Murphy, Marco Alfò}

\begin{document}
\maketitle

\begin{abstract}
The Eurovision Song Contest is a popular TV singing competition held annually among country members of the European Broadcasting Union. In this competition, each member can be both contestant and jury, as it can participate with a song and/or vote for other countries' tunes. Throughout the years, the voting system has repeatedly been accused of being biased by the presence of tactical voting, according to which votes would represent strategic interests rather than actual musical preferences of the voting countries. 
In this work, we develop a latent space model to investigate the presence of a latent structure underlying the exchange of votes.
Focusing on the period from 1998 to 2015, we represent the vote exchange as a multivariate network: each edition is a network, where countries are the nodes and two countries are linked by an edge if one voted for the other. The different networks are taken to be independent replicates of a common latent space capturing the overall relationships among the countries. Proximity denotes similarity, and countries close in the latent space are assumed to be more likely to exchange votes. Therefore, if the exchange of votes depends on the similarity between countries, the quality of the competing songs might not be a relevant factor in the determination of the voting preferences, and this would suggest the presence of bias.
A Bayesian hierarchical modelling approach is employed to model the probability of a connection between any two countries as a function of their distance in the latent space, and of network-specific parameters and edge-specific covariates. 
The inferred latent space is found to be relevant in the determination of edge probabilities, however, the positions of the countries in such space only partially correspond to their actual geographical positions.
\end{abstract}


\section{Introduction}
\label{intro}
The Eurovision Song Contest is a popular TV show, held since 1956, that takes place every year with participants the countries member of the European Broadcasting Union. 
The competition has undergone several various modifications through years and the number of participants has increased, together with the popularity of the show. Since its beginning, countries had to express their preferences for the competing songs through a voting system; representatives vote only for the songs that meet their tastes. Despite that, many issues of bias in the voting system have been raised during the years \citep{Yair}. In the press and the literature, it has been claimed that many votes are not the expression of preferences for the songs, but of preferences for the performing countries themselves. Therefore, it has been claimed that the exchange of votes is not random but rather it is determined by some kind of similarity measure: the more two countries are close according to the unknown proximity measure, the more they will tend to vote for each other. 

The exchange of votes in the Eurovision contest can be represented by means of network theory, where the countries are the nodes and the votes are recorded as edges. More specifically, within each edition, the data may be represented in the form of an adjacency matrix $\mathbf{Y}$, with generic element $y_{ij}=1$ if a representative of country $i$ votes for a song by a performer from the $j^{\text{th}}$ country and 0 otherwise, and $i,j = 1, \dots,n$ indexes countries. Network data can be represented by means of graph theory. More formally, a network is thought to be the realization of a graph $G(N, E)$, where $N$ denotes the set of nodes and $E$ the set of edges. The number of observed nodes and edges observed will be denoted, respectively, by $|N|=n$ and $ |E|=e$. 
Generally, the law generating the connections among units is unknown and several different models have been proposed to describe such networks. \cite{rg} and \cite{rg2} modelled arch formation in a network as arising from a random process: each dyad $(i,j)$ is independent and is associated to the same probability of forming a link. This first model was then generalized by other authors, both relaxing the assumption of constant edge probability over the network and the assumption of independence of the dyads. 
\cite{exp_rg} with models $p_1$ and $p_2$ kept the assumption of independence among the dyads but increased the number of parameters describing edge probabilities, to take into account the attractiveness of a node (the highest the value the highest the probability for this node to be connected with others) and the mutuality (the propensity of forming symmetric relations). The independence assumption on the dyads was then relaxed via the introduction of \emph{Markov graphs} by \cite{frank}, attempting to model triangular relations in a network. Later on $p^*$ models or ERGMs (Exponential Random Graph models) have extended the work done by \cite{frank} by the introduction of differnet summary statistics, see for example \cite{krivi} and \cite{robins}. A different approach is the so-called \emph{stochastic block model}, which attempts to decompose the nodes in different sub-groups, see \cite{sto_bloc}, \cite{mm_sto_bloc}. In its basic formulation, nodes  within a group have the same probability of forming edges, while this probability changes among groups. \cite{hoff} added an extra layer of dependency to model the complex structure of network data, via the assumption that the observed edge formation process could have been explained in terms of the nodes' coordinates in a (low-dimensional) latent space. Two different specifications are considered, the \emph{distance model}, where the latent space is euclidean, and the \emph{projection model}, where it is bilinear.
The model by \cite{hoff} has been extended to perform clustering on the latent coordinates of the nodes by \cite{mbcsn}. Another approach that makes use of latent variables has been proposed by \cite{snj}. This model formulates the stochastic block model by \cite{sto_bloc} introducing latent variables in the determination of the nodes' group memberships. A more exhaustive review of models for statistical network analysis can be found in \cite{surv_net}, \cite{nreview} and \cite{brendanHB}.

The works presented above refer to models for a single network, that is, in the present context, to the modelling of one single edition of the Eurovision Song Contest or a summary of several editions.
If a group of subsequent editions of the Contest is considered, many replications of the adjacency matrices, representing the preferences expressed by countries towards others, are available. Therefore, the data can be represented in the form of a multidimensional network (or multiplex), where $\mathbf{Y} = (\mathbf{Y}^{(1)}, \dots, \mathbf{Y}^{(K)})$ is the realization of a collection of graphs $G = (G^{(1)}, \dots, G^{(K)})$, $k=1, \dots , K$ indexing editions. The generic graph $G^{(k)} = (N, E^{(k)})$ has the same set of nodes $N$ as the others $(K-1)$ graphs in the collection (the participants to the group of editions of interest), but potentially different set of edges $ E^{(k)}$ (the different preferences expressed in each edition). Hence, a multidimensional networks describes different (independent) realizations of a relation among the same group of nodes. Different models have been developed to deal with this kind of data.
\cite{fienberg} adapted a log linear model to the context of multiplex data. \cite{greene} proposed to summarize the information coming from all the different networks (views) aggregating them into a single one. \cite{gollini} extended the model by \cite{hoff} to multiplex data, assuming that the edge probabilities are explained by a single latent variable. To estimate the joint latent space, they use a variational Bayesian algorithm and decompose the posterior distribution, fitting a different latent space to each network. Then, the separate estimates are employed to recover the joint latent space. \cite{salter} proposed a method to jointly model the structure within a network and the correlation among networks via a Multivariate Bernoulli model. Another approach developed to describe the (marginal) correlation among different networks in a multiplex has been proposed by \cite{butts}. \cite{hoff_multi} proposed to model multiplex data as multiway arrays and applied low-rank factorization to infer the underlying structure.  
\cite{durante} proposed a Bayesian non-parametric approach to latent space modelling, where clustering is performed on the latent space dimensions in order to discriminate the most relevant ones for each view.

The present work aims at recovering the similarities among countries, modelling the exchange of votes during several editions of the Eurovision Song Contest. We adopt a framework similar to that of \cite{gollini} and we consider the projection of the countries into a common latent space. Similarities among countries are then expressed in terms of distances in this latent space. We introduce network-specific coefficient parameters to weight the relevance of the latent space in the determination of edge probabilities in each network. We consider the editions that took place after the introduction of the televoting system and focus on the period 1998 to 2015. Further, we consider geographical and cultural covariates in the analysis.

The paper is organised as follows. Section \ref{euro_sec} summarizes the history of Eurovision Song Contest together with the principal related works on the subject (section \ref{euro_sec1}) and presents the analysed data (section \ref{euro_sec_2}). Latent space models for network data are introduce in section \ref{into_lsmmn} and the proposed model is outlined in section \ref{modello}. Model estimation is discussed in section \ref{model_estimation}. Further issues are discussed in section \ref{furt_issues}, such as model identifiability (section \ref{identif}), missing data (section \ref{missing}) and the introduction of edge-specific covariates (section  \ref{covariates}). The application is presented in section \ref{application} and the results are displayed in section \ref{results}. The simulation study is outlined in section \ref{sim_study}, where also the main findings are reported. Section \ref{lsjm} presents the results of a comparison between the proposed model and the \emph{lsjm} \citep{gollini}. We conclude discussing the model and its application on the Eurovision data in section \ref{disc}.

\section{Eurovision Song Contest}
\label{euro_sec}

\subsection{History of the contest and previous works on the subject}
\label{euro_sec1}
The Eurovision Song Contest, held since 1956, is a TV singing competition where the participants are countries members of the EBU (European Broadcasting Union). Despite its name, the European Broadcasting Union includes both European and non European countries. Indeed, Eurovision's fame has spread all over the world during the last years and it has been broadcast from South America to Australia. It is the non-sportive TV program with the highest number of viewers in the world and one of the oldest ones \citep{guwr}. 

From its first edition, where only seven countries competed, there have been many changes in the number of participants, the voting system and the structure of the competition. Due to the increasing popularity of the program,  many countries have been included in the contest. The current structure of the contest consists of two preliminary stages used to select the finalists; then, the selected countries compete in the final stage for the title. The voting system has been modified several times, in the voting procedure and the grading scheme. In the early years of the competition, a jury elected the winning song. Later, the system has been supported by televoting, introduced in 1998 in all the competing countries. As for the grading scheme, it is positional since 1962, but the method used to rank the countries has been modified across the different editions. From 1975 to 2015, each country had to express its top ten preferences ranking them from the most to the least favourite using the following scores: 12, 10, 8, 7, 6, 5, 4, 3, 2, 1. Each country had to vote exactly ten others, could not vote for itself and each grade could be used only once. At the end, the country receiving the highest overall score would have won the competition.  
A restriction has been imposed on the lyrics in the past, as the participants were required to perform a song written in their national language. However, this rule was definitively abolished after 1998. 

Every year the singer and the song representing a country both change, making each edition of the Eurovision independent from the previous one. Indeed, the structure of the competition is built in such a way that the past results will not influence the future performances. Countries should vote only according to their tastes and, as musical evaluation has no objective criteria, the voting results should not depend on the countries themselves, but only on the songs. 
However, this claim was often doubted, especially after the introduction of televoting. Several issues have been raised on the voting system, which was said to be biased.
The first paper investigating the presence of bias in the voting system is \cite{Yair}. This work considers voting relations among 22 of the 24 countries competing in the period 1975 to 1992 and claims that they can be clustered in three regional blocks, according to their voting preferences: Mediterranean, Western and Northern. Countries tend to vote for the others in the same block, hence following a non-objective (\emph{non-democratic}) behaviour. Nevertheless, the paper does not provide an in-depth statistical evaluation of the results.  
The author supports the theory that the geographic location of a country influences its voting behaviour; this assumption has been further investigated by \cite{Fenna}. In this work, the dynamic evolution of votes exchanged in the competition 1992 to 2003 has been analysed, with the aim at looking for sub-groups of countries; however, the sub-groups found are not fully explained by the geographical positions of the countries.  
\cite{clerides} developed an econometric framework to analyse the data and arrive to similar conclusions, in the sense that the authors do not find any \textquotedbl strategic\textquotedbl\  vote exchange in the period 1981 to 2005. \cite{Saavedraa} investigated the structural properties of the dynamic network related to the period 1984-2003 via q-analysis and found that clustering arises mainly between countries closed to each other in a geographical sense. 
\cite{spierdj} applied multilevel models to look for the influence of geographic and cultural factors in the vote exchange from 1975 to 2003 and found that these do not explain the behaviour of all the competing countries.
\cite{ginsib} claim that having a similar culture may influence the votes expressed by a country. 
Cultural proximity, as well as geographic proximity and migration flows in the period 1998-2012 have been investigated as sources of bias in the paper by \cite{blangiardo}. The authors discovered the presence of a mild positive bias among few couples of countries but no evidence of a negative bias overall.
\cite{mantza} analysed the editions 1975 to 2005 searching for couples of countries exhibiting preferential voting. They investigate the hypothesis of random allocation in the votes. The authors found evidence that geographic proximity is influential up to some extent, however, many countries do not tend to vote according to this rule.

The aforementioned works show that there has been a growing interest in the structure underlying the exchanging of votes in the Eurovision Song Contest in the past twenty years. The authors investigate the influence of social, geographical, cultural and political factors on the mechanism forming preferences and agree that, at least to some extent, these components are relevant. However, none of the factors above is able to explain completely the votes exchanged in the competition in the last years.  

\subsection{Data}
\label{euro_sec_2}
The assumption that the exchange of votes is driven by similarities among countries may be a reasonable hypothesis. However, similarities among countries might not coincide with social, geographical, cultural and political factors. In fact, there could be some unobservable (latent) factors influencing such process. The aim of this work is at recovering the underlying latent similarities among countries.
We focus on years subsequent to the introduction of the televoting system, 1998 to 2015. In doing so, we assume that the televoting preferences reflect the preferences of the whole population, and that they are more representative when compared to the opinions of the jury alone.
We do not consider years 2016 and 2017 because the voting system was modified in that period.
In fact, in the period from 1998 to 2015, votes given by the jury and by televoting were jointly considered: the final top ten for each country was determined looking at the intersection of the most voted songs by the two sources. Instead, from 2016 onwards, the final preferences expressed by each country are given by the union of the ten favourites of the jury and the ten favourites of the televoting. Hence, in the last two years of the competition, each country could elicit more than ten preferences.
In the period 1998-2015, the countries were allowed to sing in any language and most of the songs were in English. The period we consider is homogeneous both with respect to the voting system and (for the large part) to the language used in the performance.

In each edition, the votes exchanged among the countries can be described by a network, with nodes being the countries.
As we are interested in the exchange of votes and not in the ranking, an edge going from country $i$ to country $j$ will denote that $i$ has $j$ in its top ten. Vice versa, the edge will be from $j$ to $i$ if the latter has been voted by the $j$. 
Indeed, recall that altough the grading scheme is positional, each country can express a limited amount of preferences $r$, with $r=10$ in the analysed period. \cite{rosen} proved that when sampling $r$ units from a population of size $n$\footnote{In the present context, $n=48$ is the number of countries that country $i$ can potentially vote.}, if $r << n$, the first $r$ units are independent with respect to the extraction order. That is, the ranking might not be relevant.

The resulting network is then directed, acyclic (as countries can not vote for themselves) and un-weighted. If we consider a group of editions for the contest, we will have a collection of networks, defined on the same group of countries (see paragraph \ref{missing}), and this object is indeed a multidimensional network.
The following section introduces a more general framework to analyse such data.

\section{Latent space model for multidimensional networks}
\label{into_lsmmn}

Latent space models have been introduced by \cite{hoff} with the aim at reducing the complexity typical to the dependence structure observed in network data. This purpose is achieved via geometric projection of the nodes into a low dimensional space. The probability of observing an edge between two nodes is assumed to be dependent on some function of the unknown nodes' coordinates in the latent space. Conditionally on the set of latent positions, the observed binary indicators are assumed to be  independent. \cite{hoff} distinguish between distance and projection latent space models, depending on the choice of the function summarizing the latent coordinates. The distance model assumes that the function employed to describe the dependence of the edge probabilities on the latent coordinates is indeed a distance function.
\cite{gollini} extended the distance model described by \cite{hoff} to the case of multiplex data, with the introduction of the Latent Space Joint Model (\emph{lsjm} in the following). The authors assumed that each network depends on a specific latent space and a network specific intercept. Moreover, the latent spaces (one for each network) are thought to be generated from a common latent space, which captures the average latent coordinates of the nodes and is behind all the networks. The variational approach used to carry out the estimation of the model is fast but suffers from computational issues when the dimension of the multiplex is relatively big, either in the number of nodes $n$ or in the number of networks $K$.

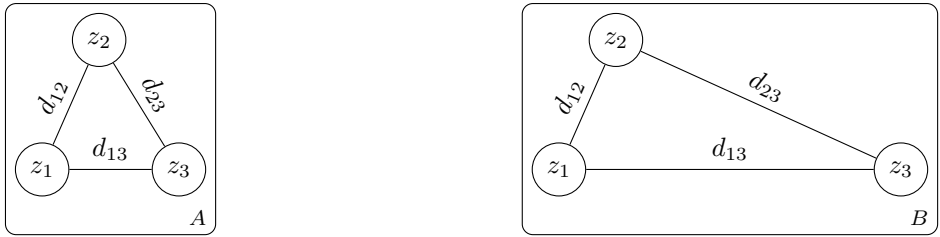
\begin{figure}[b]
  \begin{center}
    \begin{tabular}{cc}
%
%
%
%


\begin{tikzpicture}

  \node[latent, xshift=-3.8cm]    (z1) {$z_1$};
  \node[latent, above=of z1, xshift=.75cm] (z2) {$z_2$};
  \node[latent, xshift=-2cm]  (z3) {$z_3$};
  
  \node[latent, xshift=3cm]    (z11) {$z_1$};
  \node[latent, above=of z11, xshift=.75cm] (z21) {$z_2$};
  \node[latent, xshift=7.5cm]  (z31) {$z_3$};
 
  \path[every node/.style={sloped,anchor=south,auto=false}]
        (z1)   edge node {$d_{12}$} (z2);
  \path[every node/.style={sloped,anchor=south,auto=false}]
        (z1)   edge node {$d_{13}$} (z3);
 \path[every node/.style={sloped,anchor=south,auto=false}]
        (z2)   edge node {$d_{23}$} (z3);
        
 \path[every node/.style={sloped,anchor=south,auto=false}]
        (z11)   edge node {$d_{12}$} (z21);
  \path[every node/.style={sloped,anchor=south,auto=false}]
        (z11)   edge node {$d_{13}$} (z31);
 \path[every node/.style={sloped,anchor=south,auto=false}]
        (z21)   edge node {$d_{23}$} (z31);
   
 \plate {z1z2z3} {(z1)(z2)(z3)} {$A$} ;
 \plate {z11z21z31} {(z11)(z21)(z31)} {$B$} ;
 
\end{tikzpicture}

    \end{tabular}
  \end{center}
  \caption{Latent space representations of a network with 3 nodes.}
  \label{fig:ls}
\end{figure}

The present work builds on the model of \cite{gollini} with the aim at recovering the similarities among the countries participating in the Eurovision Song Contest. 
Notice that subsequent editions of the Eurovision are assumed not to depend on one another, as the singer and the song performed change every year without any pre-determined criteria. This assumption  allows us to consider the exchange of votes in different editions of the contest as replications of the same phenomenon, which is the expression of musical appreciation between couples of countries. These different replications of preferences among countries will then be used to recover the similarities. Indeed, the basic assumption is that the more two countries are similar, the more they tend to vote for one another through the editions.
As our interest lies in recovering such similarities, the choice of a distance based on a latent space model to reconstruct the network is quite appropriated. In fact, distances correspond to symmetric relations, which is a characteristic for similarity measure. 
Figure \ref{fig:ls} shows an example for the latent space representation of a $3$ nodes network. Node $z_3$  in space $A$ has been moved in space $B$, so that $d_{13}^{(A)} < d_{13}^{(B)} $ and  $d_{23}^{(A)} < d_{23}^{(B)} $. In our model, this correspond to a higher probability to observe a link between node $1$ and $3$ (or node $1$ and $2$) in space $A$ when compared to space $B$.

The distances in the latent space are scaled by a network specific coefficient, in order to weight the influence of the latent space in the determination of the votes expressed in a given edition of the Contest. The lowest the value of this coefficient, the more the structure of edge probabilities resembles a random graph This could lead to reject the claim that the pattern observed in votes for the Eurovision contest is biased by some pre-existing preferences among countries.   

The latent space is taken to have dimension $p=2$, to allow graphical visualization and to compare the estimates of latent coordinates and the geographical positions (latitude and longitude) of the analysed countries. In this way we might be able to tell whether the latent configuration obtained resembles the geographical one and, if so, we may conclude that the position of a country on the map is indeed of some relevance in the competition. However, the choice of $p$ is still an open problem in the literature and for other applications there might not be a unique criterion for choice.  
As in \cite{gollini}, the distance function is taken to be the squared Euclidean distance function,  to penalize more the probability of an edge linking nodes that are far apart in the latent space when compared to edge linking closer nodes.

The model is formulated in section \ref{modello} and also allows for the introduction of edge-specific covariates. In the present context, we will make use of cultural covariates, such as presence of a common language among a couple of participants, to see whether any cultural factors contribute to the formation of preferences.
The multidimensional network is defined on all the countries that took part at least once in the Eurovision for the period 1998-2015 (see section \ref{missing} for a more detailed explanation).

The set or collection of adjacency matrices will be denoted by $\mathbf{Y} = \bigl\{ \mathbf{Y}^{(1)}, \dots, \mathbf{Y}^{(K)}\bigr\}$. The generic element of each matrix in the collection is binary, with $\arch=1$ if there is an edge from node $i$ to node $j$ in the $k_{th}$ network, $\arch=0$ else. Indexes $i,j = 1, \dots,n$ are used to denote nodes in the network (countries) and index $k = 1, \dots, K $ refers to the $K$ different networks in the multiplex.

\subsection{The proposed model}
\label{modello}
Given the assumptions made in the section \ref{into_lsmmn} and following \cite{hoff} and \cite{gollini}, the probability of observing an edge between node $i$ and node $j$ in the $k^{\text{th}}$ network of the multiplex is given by:
\begin{equation}
\label{eq:edge_prob}
P\Bigl( \arch = 1 \mid \mathbf{\Omega}^{(k)},d( z_i, z_j )\Bigr)=
P\Bigl( \arch = 1 \mid \alpha^{(k)},\beta^{(k)} , d( z_i, z_j )\Bigr)=
\frac{\exp\bigl\{ \ak  -\bk  d( z_i, z_j ) \bigr\} }{1 + \exp\bigl\{ \ak  -\bk  d( z_i, z_j ) \bigr\} }
\end{equation}
where $\mathbf{\Omega} = \bigl(\mathbf{\Omega}^{(1)},\dots, \mathbf{\Omega}^{(K)} ) = \bigl( \mathbf{\alpha}, \mathbf{\beta}\bigr)$ is the set of model parameters, with $\mathbf{\alpha} = \bigl(\alpha^{(1)}, \dots,\alpha^{(K)} \bigr)$, $\mathbf{\beta} =\bigl( \beta^{(1)}, \dots,\beta^{(K)} \bigr)$.
For ease of notation, we will denote $P\Bigl( \arch = 1 \mid \mathbf{\Omega}^{(k)},d( z_i, z_j )\Bigr)=p_{ij}^{(k)}$.

Following \cite{gollini}, the function $d(\cdot, \cdot )$ is taken to be the squared Euclidean distance, hence $ d( z_i, z_j )= \sum_{l = 1}^p \bigl( z_{il} -z_{jl})^2  = d_{ij}$. The matrix of the distances among the nodes will be denoted by $\mathbf{D}$. This choice of the distance function allows to penalize more heavily the probability of an edge linking two nodes that are far apart in the latent space when compared to the one linking two closer nodes. Therefore, the latent space part of the model will push towards a non-random structure for the matrix of edge probabilities. 

The parameters $\beta^{(k)}$ and $\alpha^{(k)}$, with $k = 1, \dots, K$, are network-specific.  
$\beta^{(k)}$ is a coefficient scaling the weight/impact of the latent space in the determination of the edge probabilities in the $k^{\text{th}}$ network. According to the assumption that the probability of observing an edge decreases with growing distances, the constraint $\beta^{(k)} \geq 0$ must be imposed.
As the coefficient is bounded and can not take negative values, it follows that the lowest the value of $\beta^{(k)}$, the closer to a random graph the structure of the network will be. In fact, if $\beta^{(k)} = 0$, then:
\[
P\Bigl( \arch = 1 \mid \alpha^{(k)},\beta^{(k)} , d( z_i, z_j )\Bigr) = \frac{\exp\bigl\{ \ak  \} }{1 + \exp\bigl\{ \ak  \bigr\} } =p_{RG},
\]
and the model for the $k^{\text{th}}$ network reduces to a random graph \citep{rg} with edge probability  $p_{RG}$.
Thus, the coefficient $\beta^{(k)}$ induces sparsity in the graph and, when it is non-zero, edge probabilities are bounded by $p_{RG}$.
To counterbalance the effect of the coefficient, the intercept parameter $\alpha^{(k)}$ is defined so that the graph corresponding to $p_{RG}$ is not disconnected. Indeed, if $p_{RG} > \frac{(1-\epsilon) \log(n)}{n}$, then the graph will almost surely be connected, from the properties of random graphs \citep{rg2}. Taking $\epsilon =0$, as $n \rightarrow \infty$, this property can be expressed in terms of $\alpha^{(k)}$ as:
\[
\ak > \log \Biggl( \frac{\log(n)}{n -\log(n)}\Biggr) = LB\bigl(\ak\bigr) = LB(\alpha).
\]
The last equality comes from the node set being constant across the multidimensional network. 
Defining a lower bound prevents from assigning large negative values to the intercept parameters. Indeed, if $\ak$ is too low, the effect of the latent distances would be dominated by the intercept parameter, even if this effect is relevant ($\bk > 0$). That is, large negative values of $\ak$ in the argument of the exponential in equation \ref{eq:edge_prob} would correspond to edge probabilities tending to $0$ and numerically undistinguishable. In such case, two dstinct matrices of edge probabilities having elements $p_{ij}^{(k)}\approx 0$ would lead numerically to the same likelihood. 

\section{Parameter estimation}
\label{model_estimation}
\subsection{Likelihood and posterior}
\label{lsmmn_like}
Given the edge probability function defined in equation \ref{eq:edge_prob}, the likelihood function for the model is a product of $K n (n-1)$ terms:
\begin{equation}
\label{eq:likel}
L \bigl(\mathbf{\Omega}, \mathbf{D} \mid \Y \bigr) = \prod_{k = 1}^K \prod_{i = 1}^n  \prod_{j \neq i} L_{ij}^{(k)} = \prod_{k = 1}^K \prod_{i = 1}^n  \prod_{j \neq i}  \frac{\exp\bigl\{ \ak  -\bk  d( z_i, z_j ) \bigr\} }{1 + \exp\bigl\{ \ak  -\bk  d( z_i, z_j ) \bigr\} },
\end{equation}
the corresponding log likelihood is: 
\begin{equation}
\label{eq:log_likel}
\begin{split}
\ell\bigl(\Omega, D \mid \Y \bigr) & = \sum_{k = 1}^K \sum_{i = 1}^n  \sum_{j \neq i} \ell_{ij}^{(k)}\\
&   =\sum_{k = 1}^K  \sum_{i = 1}^n  \sum_{j \neq i} \arch \log \Biggl( \frac{\exp\bigl\{ \ak  -\bk  d_{ij} \bigr\} }{1 + \exp\bigl\{ \ak  -\bk  d_{ij} \bigr\} } \Biggr)+ 
   ( 1 - \arch) \log  \Biggl( \frac{1}{1 + \exp\bigl\{ \ak  -\bk d_{ij} \bigr\} }  \Biggr)\\
 & =  \sum_{k = 1}^K \sum_{i = 1}^n  \sum_{j \neq i} \arch \bigl(\ak  -\bk  d_{ij} \bigr) - \log \Bigl(
  1 + \exp\bigl\{ \ak  -\bk d_{ij} \bigr\} \Bigr) 
  \end{split}
\end{equation}
As the matrices of edge probabilities are symmetric in all the analysed networks, one could equivalently consider only their upper or lower triangular part, and the number of terms to  be considered in the product for the likelihood reduces to $K$ $n \choose 2$.

Similarly to \cite{gollini} and \cite{mbcsn}, we adopt a Bayesian approach to estimate the model. 
The latent coordinates are assumed to be independent random variables distributed according $p$-variate Gaussian distribution, as in \cite{gollini}:
 \[
 \mathbf{z}_i \thicksim MVN_{p} \bigl( \mathbf{0}, \mathbf{I}\bigr).
 \]
In the present context, the dimension of the multivariate Gaussian is fixed to $p=2$. Indeed, the aim is to recover latent coordinates that, for each country, could be compared with the actual geographical ones. However, in other applications the selection of the dimension $p$ of the latent space $p$ could be a relevant issue. The determination of $p$ is a model selection problem and it is quite a debated topic in the literature.

\begin{figure}[b]
  \begin{center}
    \begin{tabular}{cc}
%
%
%
%


\begin{tikzpicture}

  \node[obs]                               (y) {$y$};
  \node[latent, above=of y, xshift=-1.5cm] (a) {$\alpha$};
  \node[latent, above=of y, xshift=1.5cm]  (b) {$\beta$};
  \node[latent, right=2cm of y]            (z) {$\mathbf{z}$};
  
  \node[latent, above=of a, xshift=-.9cm]  (ma) {$\mu_{\alpha}$};
  \node[latent, above=of a, xshift=.9cm]  (sa) {$\sigma_{\alpha}^2$}; 
  \node[latent, above=of b, xshift=-.9cm]  (mb) {$\mu_{\beta}$};
  \node[latent, above=of b, xshift=.9cm]  (sb) {$\sigma_{\beta}^2$}; 

  \edge {b,a,z} {y} ; %
  \edge{ma,sa} {a};
  \edge{sa}{ma};
  \edge{mb,sb}{b};
  \edge{sb}{mb};


\end{tikzpicture}

    \end{tabular}
  \end{center}
  \caption{Hierarchy structure of the model.}
  \label{fig:model_structure}
\end{figure}
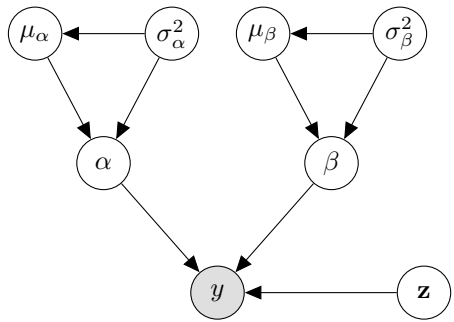

The parameter space for both the intercepts and the coefficients is bounded, as described in paragraph \ref{modello}. For this reason, the prior distributions for these parameters are described by truncated Gaussian distributions. As no a priori information is available on their relationship, they are assumed to be independent, both inside and across the networks:
\[
\ak \thicksim N_{\bigl[ LB(\alpha), \infty \bigl]} \bigl( \mu_{\alpha}, \sigma_{\alpha}^2\bigr) \qquad \bk \thicksim N_{\bigl[ 0, \infty \bigl]}  \bigl( \mu_{\beta}, \sigma_{\beta}^2\bigr)
\]
The unknown $\mu_{\alpha}, \sigma_{\alpha}^2,\mu_{\beta}, \sigma_{\beta}^2$ are nuisance parameters. Indeed, their value is of no interest but their specification is relevant, as they determine the solutions for the parameters of interest, $\ak$ and $\bk$. Given their relevant role in the model, these parameters will be estimated, to avoid subjective specifications of their values. To estimate such parameters, an extra layer of dependence is introduced in the model, as described in figure \ref{fig:model_structure}; that is, the model has a  hierarchical structure. The prior distributions specified for the nuisance parameters are:

$$
\mu_{\alpha} | \sigma_{\alpha}^2 \thicksim N_{\bigl[ LB(\alpha), \infty \bigl]} \bigl(m_{\alpha}, \tau_{\alpha}\sigma_{\alpha}^2  \bigr)  \quad 
\sigma_{\alpha}^2 \thicksim \text{\small{Inv}} \chi_{\nu_{\alpha}}^2 \qquad
\mu_{\beta} | \sigma_{\beta}^2 \thicksim N_{\bigl[ 0, \infty \bigl]} \bigl(m_{\beta}, \tau_{\beta}\sigma_{\beta}^2  \bigr) 
\quad \sigma_{\beta}^2 \thicksim \text{\small{Inv}} \chi_{\nu_{\beta}}^2 
$$
The distributions for the nuisance parameters depend on a set of hyperparameters $\eta = \bigl(\nu_{\alpha}, \nu_{\beta}, \tau_{\alpha}, \tau_{\beta} \bigr)$, that have to be specified. However, their choice is not as influential as the one of the nuisance parameters for the estimation of $\ak$ and $\bk$; in the following, we will present some criteria for the determination of $\eta$ that were found to work well in practice.   

The posterior distribution is therefore defined as:
\begin{equation}
\begin{split}
P \bigl( \alpha, \beta, \mathbf{z},\mu_{\alpha}, \mu_{\beta},\sigma_{\alpha}^2 ,\sigma_{\beta}^2 |\mathbf{Y} \bigr)=&  L \bigl( \alpha, \beta, \mathbf{z} |\mathbf{Y} \bigr) \pi \bigl( \mathbf{z}\bigr)  \pi \bigl( \alpha | \mu_{\alpha} , \sigma_{\alpha}^2 \bigr) \pi \bigl(\mu_{\alpha} | \sigma_{\alpha}^2, \tau_{\alpha}\bigr) \pi \bigl(\sigma_{\alpha}^2 | \nu_{\alpha} \bigr) \\
&
\pi \bigl( \beta | \mu_{\beta} , \sigma_{\beta}^2 \bigr) \pi \bigl(\mu_{\beta} | \sigma_{\beta}^2, \tau_{\beta}\bigr) \pi \bigl(\sigma_{\beta}^2 | \nu_{\beta} \bigr)
\end{split}
\label{eq:post}
\end{equation}
The posterior distributions for the parameters $\ak, \bk$ and for the latent coordinates are not available in closed form. To estimate these parameters, proposal distributions have been developed and are presented in appendix \ref{app1}, together with the distributions of the nuisance parameters.

\subsection{Algorithm for model estimation}
\label{lsmmn_est}
Estimation of model parameters estimation is carried out using Markov Chain Monte Carlo. A detailed specification of the full conditional and the proposal distributions implemented can be found in the appendix \ref{app1}. 
Within each iteration of the chain, the nuisance parameters are updated from the corresponding full conditional; updated estimates for the intercepts, the coefficients and the latent coordinates are then proposed. The updates for the intercept and the coefficient in each network is jointly carried out, as it was empirically found that they may be correlated. The joint update helps improving the speed of convergence for these parameters, while the latent coordinates are updated separately and sequentially. Indeed, it could be the case that the current estimates for a subset of the latent positions have already converged, while the remaining $\mathbf{z}_i$'s are still far from the "true" values. Updating the whole $\mathbf{z}$ block jointly would not respond to the need to adjust just the \emph{mislocated} coordinates; hence, updating them separately has been found to be a better strategy.
After the set of latent coordinates has been updated at a given iteration of the algorithm, it has to be compared with the set of estimates obtained at the previous iteration. That is, since the likelihood in equation \ref{eq:likel} considers the distances between the latent coordinates, it is invariant to rotation or translation of the latent positions $\mathbf{z}_i$. Then, it has to be ensured that the current set is not a "rigid" transformation of the previous ones, to prevent from non-optimal stationary solutions for the latent coordinates. To achieve such aim, a Procrustes \citep{proc} test is performed and the current configuration is discharged if the correlation with the previous one is above a certain threshold, fixed to be $0.85$. The choice of this value, that ranges in $[0,1]$ is arbitrary and should reflect the presence of high correlation. Values of the threshold above $0.80$ have been found to work well in practice. 
Appendix \ref{app5} reports the pseudo-code describing the estimation procedure.   

As for the initialization, the set of hyperparameters $\eta$ needs to be defined before starting the algorithm. 
The degrees of freedom of the Inverse Chi-squared prior distribution for the variance parameters $\sigma_{\alpha}^2$ and $\sigma_{\beta}^2$ are fixed to be $\nu_{\alpha}=\nu_{\beta}=3$, as values in the range $[2,6]$ have been tried and it has been found that the different specifications do not have substantial impact on the parameter estimates. 
The variance-scale hyperparameters are set to be $\tau_{\alpha}=\tau_{\beta}=\frac{K-1}{K}$, so that the means of the proposal distributions for $\mu_{\alpha}$ and $\mu_{\beta}$ reduce to the corrected sample means (see appendix \ref{app1}). 

Starting values for the distances are taken to be the geodesic distances between the nodes in a randomly chosen network of the multiplex. From these distances, given a fixed $p$, starting values for the latent coordinates are computed via multidimensional scaling as in \cite{hoff} and the starting values for the distances are obtained taking the squared Euclidean distances between the starting values for the latent coordinates.
These are then used to model, via logistic regression, the adjacency matrices. The estimated intercepts and coefficients are taken to be the starting values for $\alpha$ and $\beta$. If these starting values fall outside the bounds specified for the parameters, they are replaced with these bounds.
The nuisance parameters $\mu_{\alpha}$ and $\mu_{\beta}$ are initialized as the sample means of the initial estimates for $\alpha$ and $\beta$, respectively. In a similar fashion, $\sigma_{\alpha}^2$ and $\sigma_{\beta}^2$ are initialised as the sample variances of $\alpha$ and $\beta$.

\section{Further issues}
\label{furt_issues}
\subsection{Identifiability}
\label{identif}
As it can be easily noticed, the likelihood in equation (\ref{eq:likel}) is invariant to linear transformations of the coefficient parameters. Indeed, for some constant $c$, 
\[
\ak -\bk d_{ij} = \ak -\frac{\bk}{c} (d_{ij} c) = \ak - \beta^{(k) \star} d_{ij}^{\star}, \quad k = 1, \dots,K.
\]
For this reason, the coefficient for the reference network is fixed to $\beta^{(r)} = 1$ (index $r$ denotes the reference network). The role of the $\beta$ parameters is to rescale the distances and, therefore, the corresponding values are meaningful only when compared with each other and the reference. Thus, there is no loss of information in fixing $\beta^{(r)} = 1$; however a further identifiability issue is:
\[
\alpha^{(r)} - d_{ij} =   (\alpha^{(r)}+c) - (d_{ij} +c) = \alpha^{(r) \star}  -d_{ij}^{\star}.
\]
To overcome this second issue, the intercept for the same network $\alpha^{(r)}$ needs to be fixed. As the intercept term defines an upper bound for the edge probabilities in the corresponding network, the value chosen for $\alpha^{(r)}$ should not underestimate this bound. We propose to fix it accordingly to the observed density in the network. More specifically, let us consider the expected value for the edge probability in network $r$:
\[
\bar{p}^{(r)} = E \Bigl[ \sum_{i =1}^n \sum_{j \neq i } P\Bigl( y_{ij}^{(r)} = 1 \mid \alpha^{(r)},\beta^{(r)} , \mathbf{D }\Bigr) \Bigr] = 
E \Biggl[  \sum_{i =1}^n \sum_{j \neq i }
\frac{\exp\bigl\{ \alpha^{(r)} -d_{ij} \bigr\} }{1 + \exp\bigl\{ \alpha^{(r)}  -  d_{ij}  \bigr\} }   \Biggr] .
\]
A naive empirical approximation to this value, which is not available in closed form, is given by
\[
 \bar{p}^{(r)} \simeq E \Biggl[  \sum_{i =1}^n \sum_{j \neq i }
\frac{\exp\bigl\{ \alpha^{(r)} -2 \bigr\} }{1 + \exp\bigl\{ \alpha^{(r)}  -2 \bigr\} }   \Biggr] = \frac{\exp\bigl\{ \alpha^{(r)} -2 \bigr\} }{1 + \exp\bigl\{ \alpha^{(r)}  -2 \bigr\} },
\]
where the distances have been replaced by the constant $2$ as this value is the mean empirical distance among coordinates simulated form a Gaussian distribution. 
This leads to: 
\[
\hat{\alpha}^{(r)} = \log \Bigl( \frac{\hat{\bar{p}}^{(r)}}{1-\hat{\bar{p}}^{(r)}}  \Bigr) +2,
\]
where $ \bar{p}^{(r)}$ with $\hat{\bar{p}}^{(r)} = \sum_{i =1}^n \sum_{j \neq i }  y_{ij}^{(r)} / \bigl(n(n-1)\bigr)$. 
Thus, $\hat{\alpha}^{(r)}$, or any number greater than $\hat{\alpha}^{(r)}$, can be taken as the value for the intercept in the reference network.

A network that is of particular interest can be taken as the reference network or, alternatively, if there is no reason to prefer a network, a randomly chosen one can be selected.
In the present work, for ease of interpretation of the results, the first network of the multiplex has been set as the reference network. 

\subsection{The issue of absent countries}
\label{missing}
Due to the increasing popularity that Eurovision gained over the years, many countries have requested to participate in the contest and have been accepted. With the increasing number of participants, preliminary stages had to be introduced to select a smaller sub-group of countries accessing to the final, where they could compete for the title. The winner of the previous edition enters the final straightforwardly, while the remaining countries have to compete in the qualifications round. Therefore, the selection of the finalists in the $k^{\text{th}}$ edition does not depend on the results in the previous edition, apart from the specific case mentioned above. 
With the introduction of semi-finals, countries that do not make it to the last stage are allowed to vote for their favourite ten participants in the final; therefore, their participation is passive, as they can vote but can not receive votes.
Last, several countries have abandoned the competition for years, for a series of particular reasons. 
The pre-selection process, together with the presence of passive countries and the drop outs imply that the set of participants in two consecutive editions may not be exactly the same; however, the phenomena above are structural and for that reason we decided not to treat them as a missing values problem.

To model absent countries, we will define the set of nodes $N$ in a more general way, as the set of countries that have voted at least once in the considered period. We can then rewrite the log-likelihood introduced in section \ref{lsmmn_like} as:

\begin{equation}
\ell\bigl(\mathbf{\Omega}, \mathbf{D} \mid \Y \bigr) = \sum_{k = 1}^K \sum_{i = 1}^n  \sum_{j \neq i} h_{ij}^{(k)} \ell_{ij}^{(k)},
\end{equation}
where $h_{ij}^{(k)}$ is an indicator variable, with $h_{ij}^{(k)}=1$ if the $i^{\text{th}}$ node was present in the $k^{\text{th}}$ edition and could have voted for node $j$; while $h_{ij}^{(k)}=0$ implies that the $i^{\text{th}}$ node was not allowed to vote for node $j$ in the $k^{\text{th}}$ edition. 
Let us denote as $\mathbf{H}^{(k)}$ the binary matrix indicating whether or not a country was present in the $k^{\text{th}}$ edition. The rows of the $\mathbf{H}^{(k)}$ matrix denote whether the corresponding countries may vote at the $k^{\text{th}}$ occasion; more formally, if $\sum_{j = 1}^n h_{ij} = 0$, then the $i^{\text{th}}$ node was absent from that edition of the contest. 
Instead, the columns of $\mathbf{H}^{(k)}$ refer to the possibility of being voted for the corresponding countries. That is, if $h_{ji}^{(k)}=1$ the $i^{\text{th}}$ node has been voted in the $k^{\text{th}}$ edition. 

\subsection{Covariates}
\label{covariates}
Edge-specific covariates can be considered in the application. All the covariates used do not depend on the specific network of the multiplex, as they are constant over the editions; therefore, the effect on edge probabilities is assumed to be constant over the networks. 
Each covariate is stored in a $n \times n$ matrix, that will be denoted by $\mathbf{X}_f$, where $f=1, \dots, F$ is the index for the set of $F$ covariates.
To maintain the characterization of the intercept term, the effect of the covariates is taken to be inversely related to edge probabilities (see section \ref{application}). 
Therefore, the scaling coefficient applied to each matrix of covariates will be characterized in similar fashion as the coefficients $\bk$, $k = 1, \dots, K$ (see appendix \ref{app1}).
The edge probability in equation (\ref{eq:edge_prob}) can be modified in presence of covariates to the following:
\begin{equation}
\label{eq:edge_prob_cov}
P\Bigl( \arch = 1 \mid \alpha^{(k)},\beta^{(k)} , d_{ij}, \lambda , \mathbf{x}_{ij}\Bigr)=
\frac{\exp\bigl\{ \ak  -\bk  d_{ij} -\sum_{l =1 }^F \lambda_l x_{ijl} \bigr\} }{1 + \exp\bigl\{ \ak  -\bk   d_{ij} -\sum_{l=1}^F \lambda_l x_{ijl} \bigr\} }
\end{equation}

The proposal distribution used to update the $\lambda_l$ parameters is derived in appendix \ref{app1}, where we briefly explain how to modify the proposal distributions for the other parameters when considering covariates in the model.
 
\section{The Eurovision song contest data}
\label{application}
The period we considered for this analysis covers 18 different editions of the Eurovison Song Contest. The editions took place after the introduction of televoting, from 1998 to 2015. The last two years of the show have been discarded from the analysis, as the voting system changed in 2016 (see section \ref{euro_sec}). The mechanism underlying the generation of the votes has changed, which means that the measure used to quantify the appreciation of a country for another has been modified. Nowadays countries can in fact express from ten to twenty preferences, depending on the degree of overlapping among the tastes of the jury and those of the public voting at home.

In the interval we considered two major changes have occurred in the structure of the program, due to the growing number of countries willing to participate in the show. First, in 2004, a semi-final stage was added to select participants. In 2008, after the 50th anniversary of the competition, the event was rebuilt and two semi final stages were introduced. 
countries that participate to the semi-finals are entitled to vote in the final, even if they have not passed the selection stage. Of course, the songs that do not go to the final can not compete for the title and can not receive any vote in the final. The voting structure in the final induced by the introduction of qualifying stages is modelled by the auxiliary variables $h_{ij}^{(k)}$, defined in section \ref{missing}.
During the period 1998-2015, a total of $n=49$ countries took part to the competition and each year. After 2004, on average $14$ countries were completely absent(not voting nor competing). A list of the $49$ countries and their ISO3 codes is given in appendix \ref{app2}.

Figures \ref{fig:descrittive_Y} describe some features of the 18 networks. In particular, in \ref{fig:descrittive_Y}(\emph{a}) we give an overview of the countries' participation per year, distinguishing the role that each country had in a given edition: absent (A), present but can not be voted (Pa) or fully present (Pp). It is easy to see from the plot that some countries, such as UK or France, have been constantly present to the competition, as the  while others had only made some sporadic appearances. Monaco for example competed from to 2004 to 2006, but never made it to the final.
Figure \ref{fig:descrittive_Y}(\emph{b}) reports the values for the association\footnote{ The association for the generic couple of editions $(k,l)$ is defined as:
\[
A_{(k,l)} = \frac{\sum_{i,j}^n \mathbf{I} \bigl( h_{ij}^{(k)} y_{ij}^{(k)} = h_{ij}^{(l)} y_{ij}^{(l)}\bigr) }{\sum_{i,j}^n \mathbf{I} \bigl( h_{ij}^{(k)} y_{ij}^{(k)} = h_{ij}^{(l)} y_{ij}^{(l)}\bigr) +\sum_{i,j}^n \mathbf{I} \bigl( h_{ij}^{(k)} y_{ij}^{(k)} \neq h_{ij}^{(l)} y_{ij}^{(l)}\bigr)}
\]
} in the exchanging of votes two editions.
The index above is limited between $0$ and $1$ and the values observed for the data range from $0.4$ to $0.8$. However, there seems to be evidence  that countries tend to repeat their patterns of votes through the analysed editions. 
The plots in \ref{fig:descrittive_Y}(\emph{c}) and \ref{fig:descrittive_Y}(\emph{d}) represent the number of joint participations for each couple of countries in the period 1998-2015 and the average number of votes they have exchanged while competing together. The matrix in \ref{fig:descrittive_Y}(\emph{d}) is not symmetric and the $i^{\text{th}}$ row shows the average number of votes that country $i$ gave to others. Instead, the $j^{\text{th}}$ column reports the average number of votes that country $j$ has received from the other participants. The last plot shows that many couples consistently voted or avoided to vote for the same group of countries, regardless of the edition.  

\begin{figure}[b]%
    \centering
    \subfloat[Association values for the couples adjacency matrices-covariates.]{{\includegraphics[width=8.2cm]{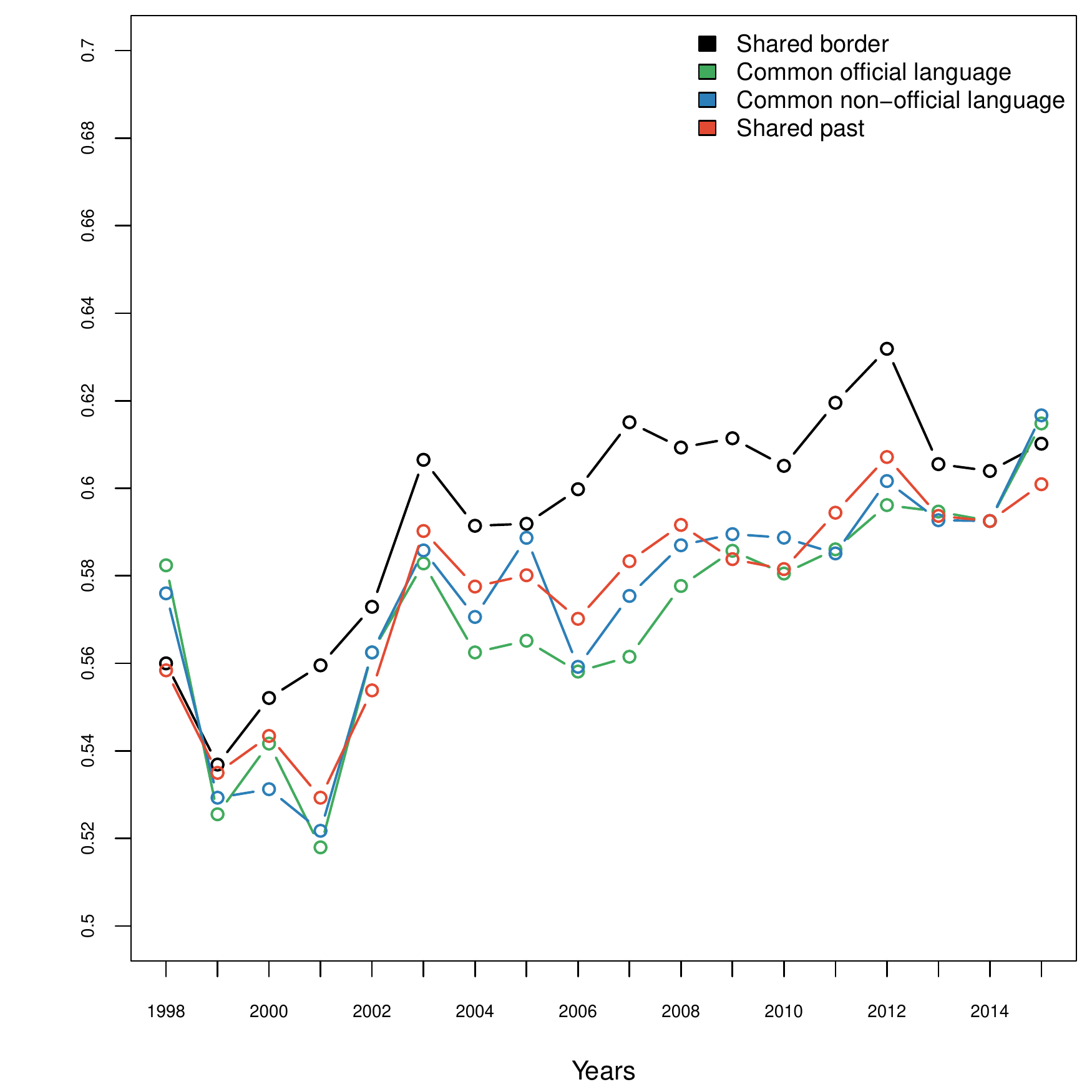} }}%
    \quad 
    \subfloat[Log-gegraphic distances vs observed arches, by year.]{{\includegraphics[width=8.2cm]{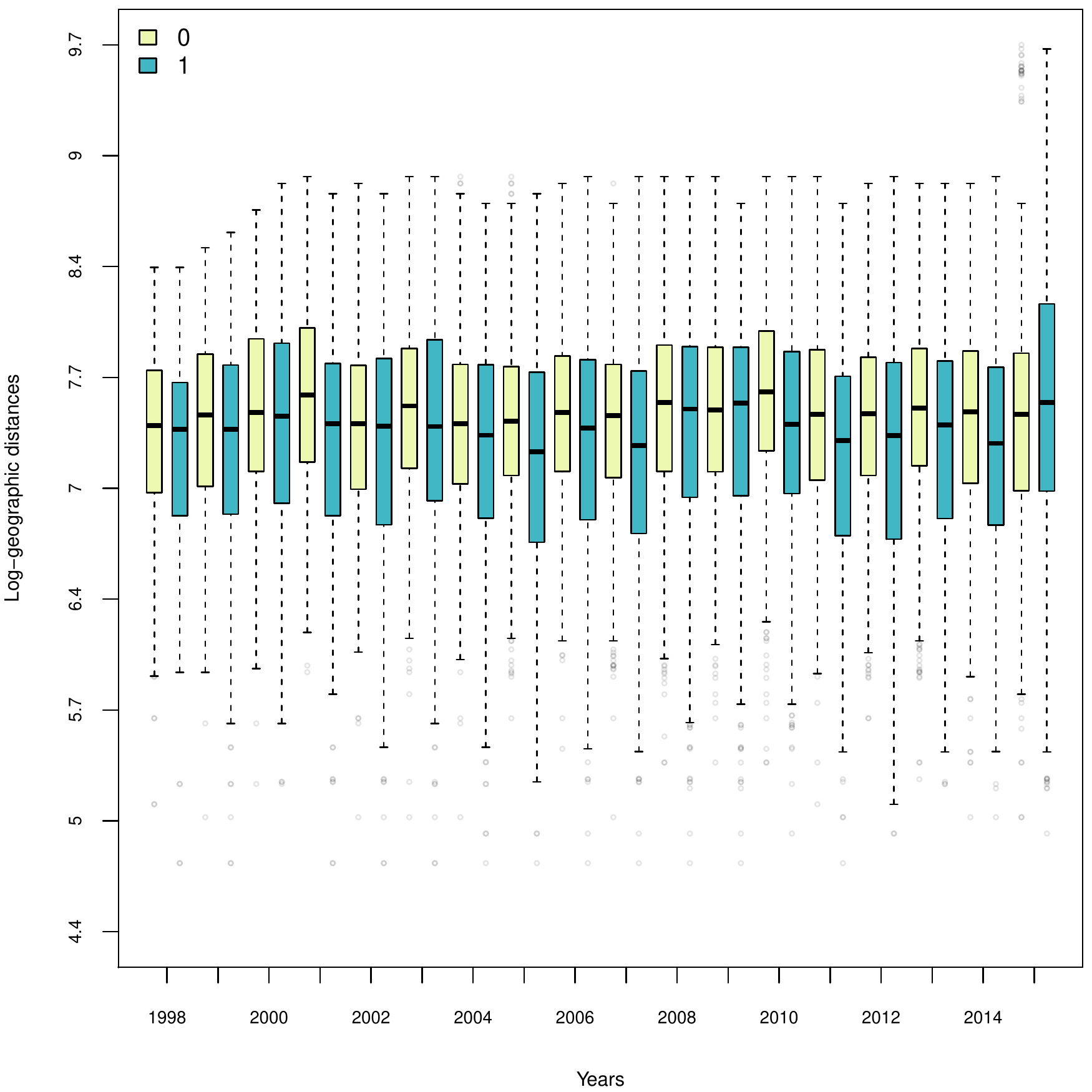} }}%
    \caption{Covariates.}%
    \label{fig:descrittive_covariate}%
\end{figure}

At a second stage, covariates have been included in the analysis, similarly to what has been done, among others, by \cite{blangiardo}, \cite{spierdj}, in order to see whether the exchanges of votes in the period could be partly explained by \textquotedbl cultural\textquotedbl\ factors.
The covariates we considered are listed below:
\begin{enumerate}
\item the log geographic distance between two countries. These distances were computed using the coordinates of the centroids of each country, obtained from \url{https://developers.google.com/public-data/docs/canonical/countries_csv}. The centroids have been estimated considering latitude and longitude of the main cities of each country, and the log geographic distances are stored in an $n \times n$ matrix denoted by $\mathbf{X}_1$;
\item the presence of a border common to a couple of countries. To maintain the characterization of the intercept, this information is coded as a binary variable that takes value $0$ if there is a common border and $1$ otherwise, so that there is a negative relation between covariate and edge probabilities. This same reason is behind the definition of the other following covariates. The common border covariate is stored in an $n \times n$ matrix denoted by $\mathbf{X}_2$;
\item the fact that two countries share an official language. This information is coded as a binary variable that takes value $0$ if they share the official language and $1$ otherwise, and it is stored in an $n \times n$ matrix denoted by $\mathbf{X}_3$;
\item the fact that two countries share a major language, defined as a language spoken at least by $9\%$ of the population. This information is coded as a binary variable that takes value $0$ if two countries share a major language and $1$ if not, and it is stored in an $n \times n$ matrix denoted by $\mathbf{X}_4$;
\item the presence of a common past \textquotedbl history\textquotedbl\ shared by two countries (they were colonized by the same country, they were the same country, etc.).This information is coded as a binary variable that takes value $0$ if two countries share a common past and $1$ otherwise, and it is stored in an $n \times n$ matrix denoted by $\mathbf{X}_5$.
\end{enumerate}
The figures in \ref{fig:descrittive_covariate} describe the association between the covariates and the adjacency matrices. The plot in \ref{fig:descrittive_covariate}(\emph{a}) displays the association\footnote{The associations are computed as 
\[
A_{(\mathbf{Y}^{(k)}, \mathbf{X}_l)} = \frac{\sum_{i,j}^n \mathbf{I} \bigl( h_{ij}^{(k)} y_{ij}^{(k)} = 1 - x_{l, i,j}\bigr) }{\sum_{i,j}^n \mathbf{I} \bigl( h_{ij}^{(k)} y_{ij}^{(k)} = 1 - x_{l, i,j}\bigr) + \sum_{i,j}^n \mathbf{I} \bigl( h_{ij}^{(k)} y_{ij}^{(k)} \neq 1 - x_{l, i,j}\bigr)}
\]} between the set of binary covariates $\mathbf{X}_2$ to  $\mathbf{X}_5$. 
The set of covariates which seems to be most relevant, when compared to the others, is the one indicating the presence of a shared border ($\mathbf{X}_2$). The values for the different associations are quite constant over time, slightly increasing with the introduction of the semi-final stage. This leads to assume that the influence of the covariates on the edge probabilities is quite constant with the edition.
Figure \ref{fig:descrittive_covariate}(\emph{b}) reports the boxplots for the couple $\bigr($adjacency matrix;log geographic distances ($\mathbf{X}_1$) $\bigl)$. There seems to be no clear relation between the distances and the presence of an arch in the adjacency matrix (that is, a vote). However, for each year, if we look at the median geographic distance for the block where an arch is present, we observe that is almost always lower then the one of the complementary block. Regressing the adjacency matrices on the log-geographic distances gives a negative estimate for every edition. That supports the claim that the geographic distances are indeed negatively correlated with the propensity to vote for a country.

Last, two sub-periods will be analysed separately, to check for large changes in the latent space position of a country according to the analysed years. That is, the analysis of the two sub-periods would give an idea on the stability of the average coordinates in the latent space recovered for the fill interval 1998-2015.

The set of covariates  $\mathbf{X}_2-\mathbf{X}_5$ have been collected from the CEPII database, \url{http://www.cepii.fr/CEPII/en/welcome.asp}, while the data analysed are available at \url{http://eschome.net/}. 
\begin{figure}[b]%
    \centering
    \subfloat[Countries participation by year.]{{\includegraphics[width=8cm]{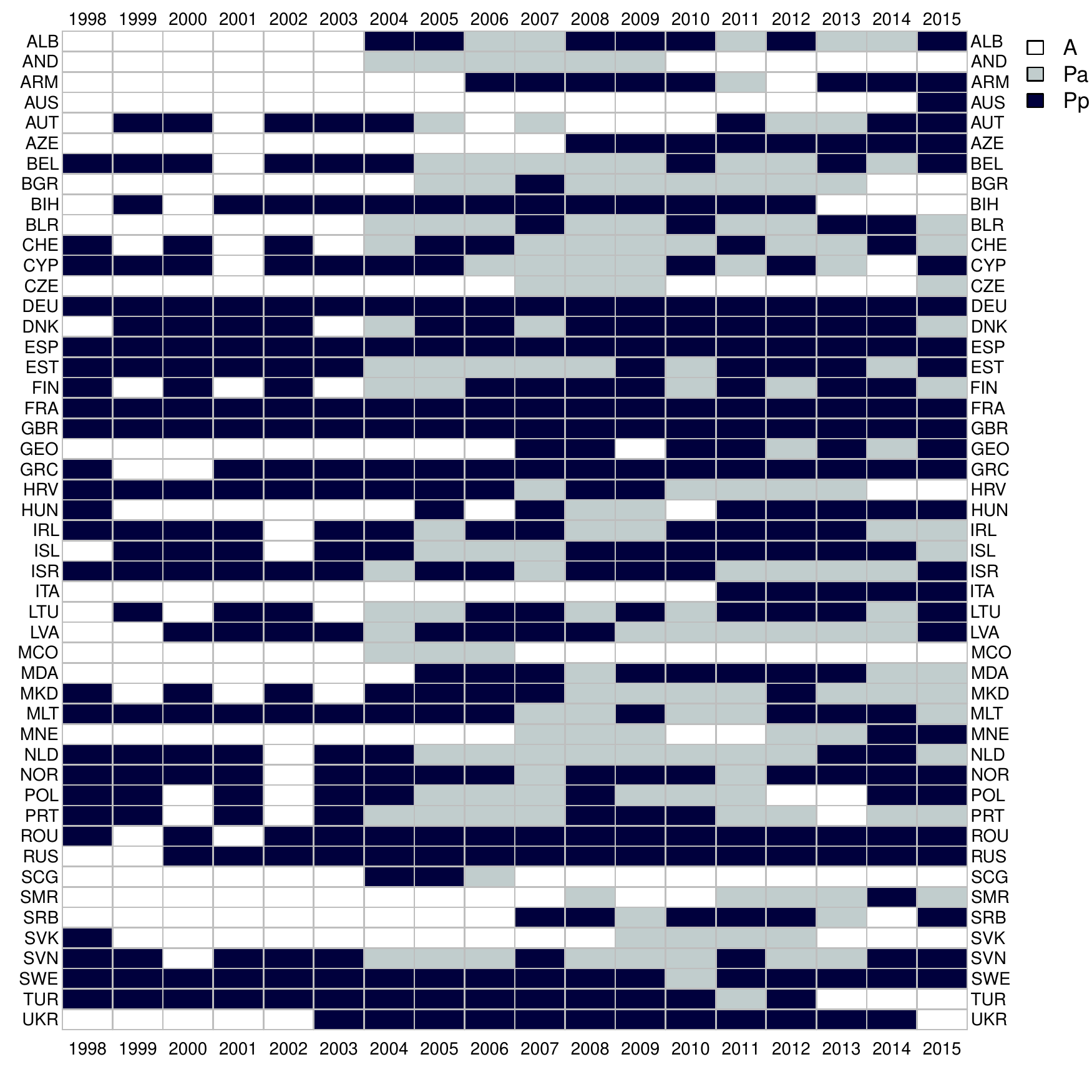} }}%
    \quad 
    \subfloat[Association between adjacency matrices for couple of years.]{{\includegraphics[scale = .45]{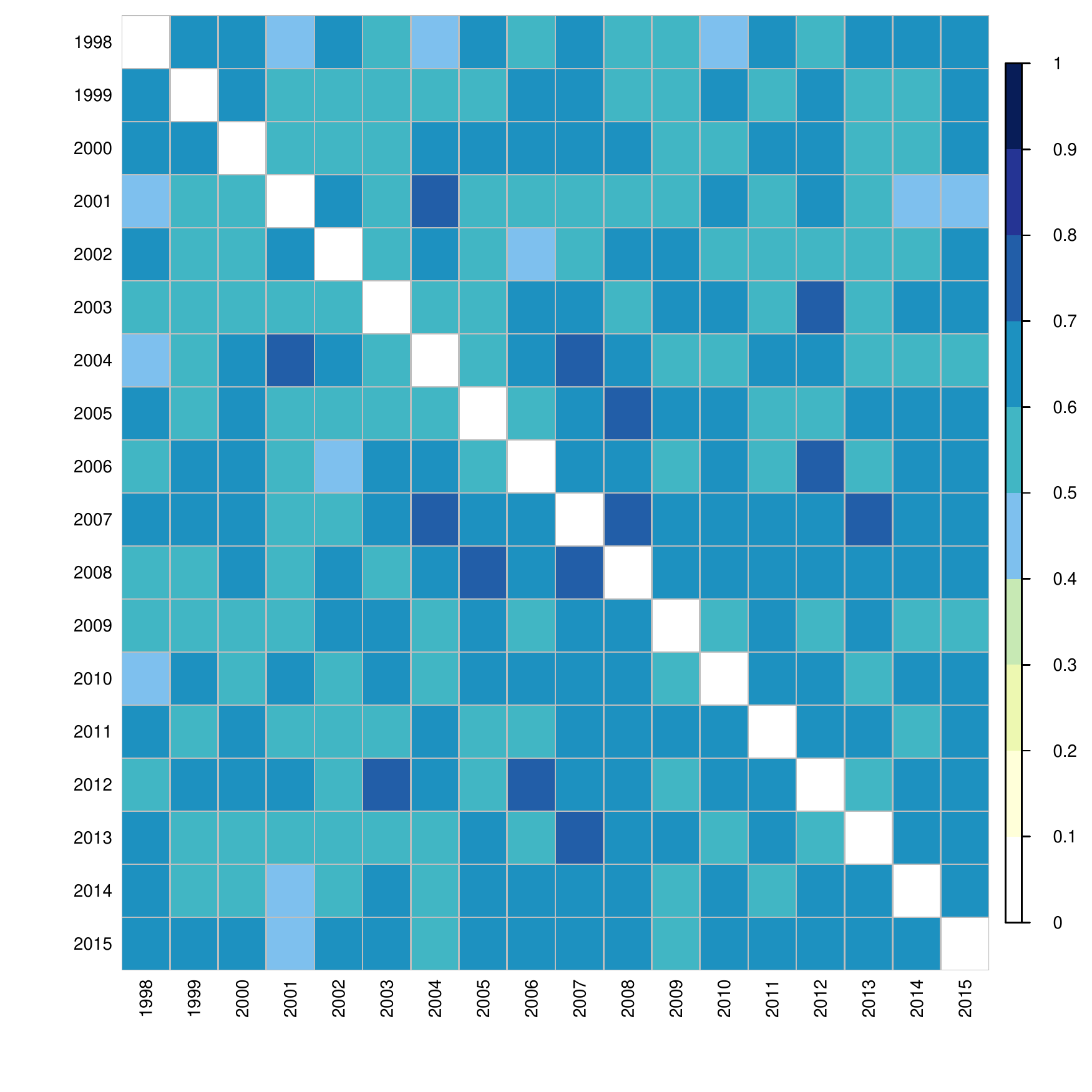} }}%
    \qquad
      \subfloat[Number of times two countries have participated jointly to the competition. The value "N" corresponds to the null elements in the diagonal.]{{\includegraphics[width=8cm]{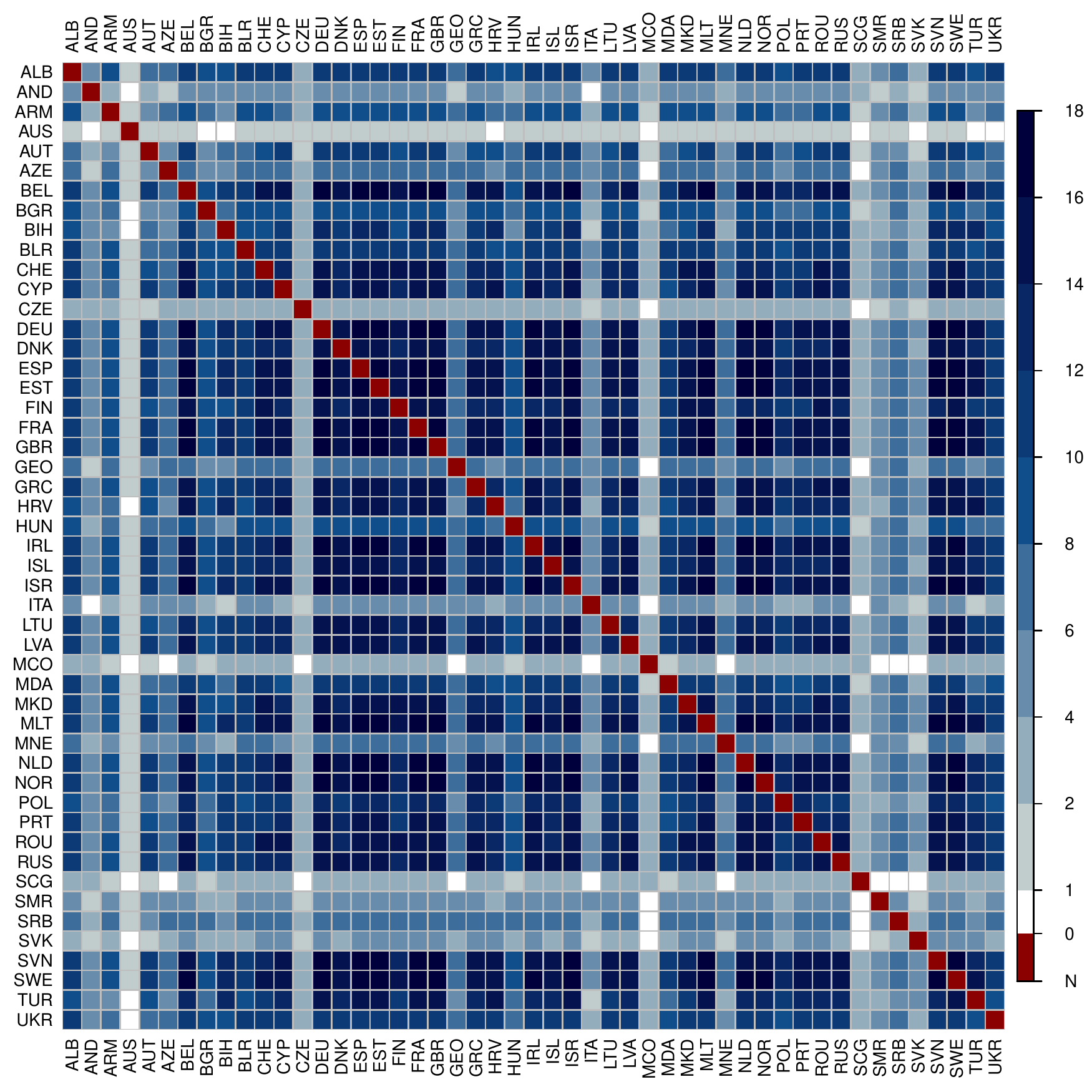} }}%
    \quad 
    \subfloat[Number of votes exchanged between couples of countries between 1998 and 2015, weighted by the numbers of times they have participated jointly to the competition. The value "N" corresponds to the case where two countries have never attended the contest together.]{{\includegraphics[width=8cm]{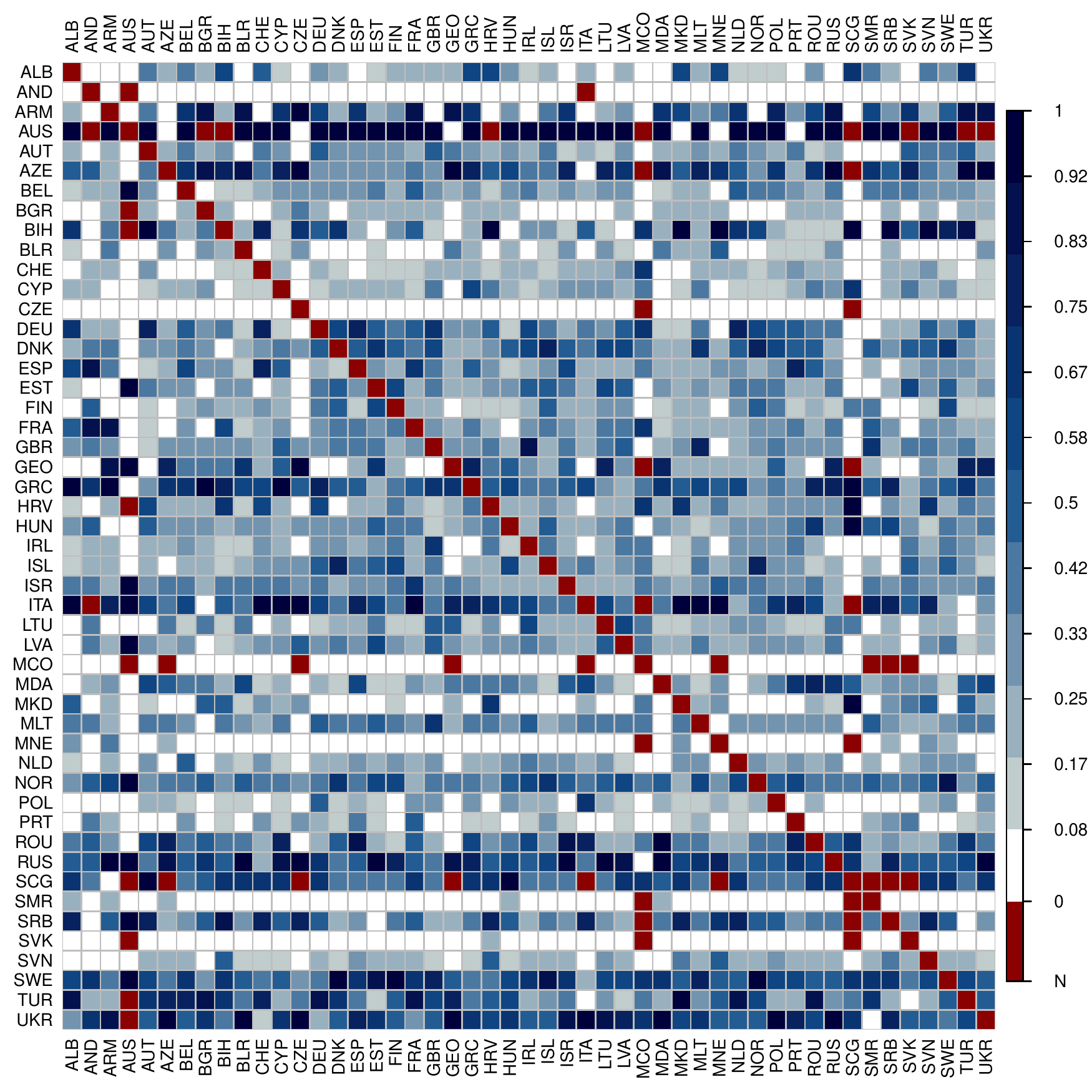} }}%
    \caption{}%
    \label{fig:descrittive_Y}%
\end{figure}

\section{Results}
\label{results}
The following models have been considered in the analysis
\begin{enumerate}
\item \emph{Model 1}: covariates not included;
\item \emph{Model 2}: covariates $\mathbf{X}_1-\mathbf{X}_5$ included;
\item \emph{Model 3}: log-geographic distance included ($\mathbf{X}_1$);
\item \emph{Model 4}: information on shared borders included ($\mathbf{X}_2$);
\item \emph{Model 5}: no latent space ($\bk = 0$ $\forall k$) and covariate $\mathbf{X}_2$ included;
\item \emph{Model 6}: random graph model (no covariates and $\bk = 0$ $\forall k$).
\end{enumerate}
Models \emph{5} and \emph{6} have been estimated to test for the need to consider an additional layer in the modelling structure (the latent space) for these data. Each model was estimated running the MCMC algorithm for 50000 iterations and with a burn of 5000 iterations. The intercept parameter in the first network was fixed to $\alpha^{(1)} = 0$, as $\hat{p}^{(1)}\approx 0.11$ \footnote{The intercept in the reference network is fixed as\[
\hat{\alpha}^{(1)} = \log \Bigl(\frac{0.11}{1-0.11} \Bigr) +2 \simeq -0.09 \approx 0\].} and $\beta^{(1)}=1$. 
The estimated models have been compared using the Deviance Information Criterion (DIC), see \cite{dic}:
\[
DIC = D(\hat{\Theta}) +2(\bar{D}(\Theta) - D(\hat{\Theta}))
\]
where $D(\Theta) = -2 \log L(\Theta)$ and $\Theta = (\mathbf{\Omega}, \mathbf{D}, \lambda)$. The deviance term $D(\hat{\Theta})$ is computed with the posterior estimates, while $\bar{D}(\Theta)$ is the mean deviance of the posterior distribution.
The best model is determined to be the one associated to the lowest value of the DIC; the values are reported in table \ref{tab:dic_values}.
\begin{table}[b]
\centering
\small
\begin{tabular}{@{}lllllllllllll@{}}
\toprule
 & \emph{Model 1} & \emph{Model 2} & \emph{Model 3} & \emph{Model 4} & \emph{Model 5} & \emph{Model 6} \\
\midrule
DIC & $19584.12$ & $19494.74$ & $19461.52$ & $ \mathbf{19412.12}$ & $23340.84$ & $ 23899.62$ \\
\bottomrule
\end{tabular}
\caption{DIC values for models $M_1-M_6$.}
\label{tab:dic_values}
\end{table}

\emph{Model 4} is found to be the best one, including both the latent space and the information on a shared border. The hypothesis of a random mechanism determining the exchange of votes can then be discarded in favour of a more complex solution, where similarities among countries, described by distances in a latent space, play a role in the formation process for observed preferences. Indeed, under \emph{Model 4}, the coefficient estimates for the latent space distances are quite high in each edition, see table \ref{tab:par-98-15}.
\begin{table}[b]
\centering
\small
\begin{tabular}{@{}lllllllllll@{}}
\toprule
Year  & $\hat{\alpha}$ & $sd(\alpha)$ & $\hat{\beta}$ & $sd(\beta)$ &
& Year & $\hat{\alpha}$ & $sd(\alpha)$ & $\hat{\beta}$ & $sd(\beta)$  \\
\midrule
1998 & $0$    &   -    & $1$    & -      & & 2007 & $1.15$ & $0.15$ & $0.92$ & $0.14$ \\
1999 & $0.72$ & $0.18$ & $0.36$ & $0.14$ & & 2008 & $1.01$ & $0.16$ & $0.92$ & $0.15$ \\
2000 & $0.89$ & $0.18$ & $0.67$ & $0.15$ & & 2009 & $0.66$ & $0.17$ & $0.49$ & $0.14$ \\
2001 & $0.58$ & $0.17$ & $0.22$ & $0.12$ & & 2010 & $0.69$ & $0.15$ & $0.54$ & $0.12$ \\
2002 & $0.77$ & $0.19$ & $0.47$ & $0.15$ & & 2011 & $0.84$ & $0.17$ & $0.70$ & $0.15$ \\
2003 & $0.81$ & $0.16$ & $0.80$ & $0.16$ & & 2012 & $0.79$ & $0.15$ & $0.75$ & $0.14$ \\
2004 & $0.82$ & $0.18$ & $0.59$ & $0.16$ & & 2013 & $0.76$ & $0.16$ & $0.73$ & $0.13$ \\
2005 & $0.86$ & $0.16$ & $0.65$ & $0.13$ & & 2014 & $0.57$ & $0.16$ & $0.48$ & $0.13$ \\
2006 & $0.76$ & $0.17$ & $0.50$ & $0.16$ & & 2015 & $0.91$ & $0.16$ & $1.01$ & $0.17$ \\
\bottomrule
\end{tabular}
\caption{Estimated averages and standard deviations for the network parameters in the multiplex 1998-2015.}
\label{tab:par-98-15}
\end{table}
The covariate $\mathbf{X}_2$ seems to be the only relevant one in the analysis. Indeed, the estimated coefficients associated with the other sets of edge covariates in \emph{model 2} where all close to $0$, which supports the model choice obtained with the deviance information criterion.
The procrustes correlations among the latent space estimated in \emph{model 1} and the one estimated in \emph{model 2} is quite high, namely $0.975$. That is because the matrix of covariates $\mathbf{X}_2$ is quite dense and acts like a fixed effect on the edge probabilities, with a decreasing effect each time there is no common border between two countries. Therefore, the introduction of the set of covariates $\mathbf{X}_2$ seems not to have direct effect on the distances between the nodes. The mean of the posterior estimate of the effect $\lambda$ associated with the border covariates is $0.60$ with a standard deviation of $0.10$. 

\begin{figure}[b]%
    \centering
    {{\includegraphics[width=10cm]{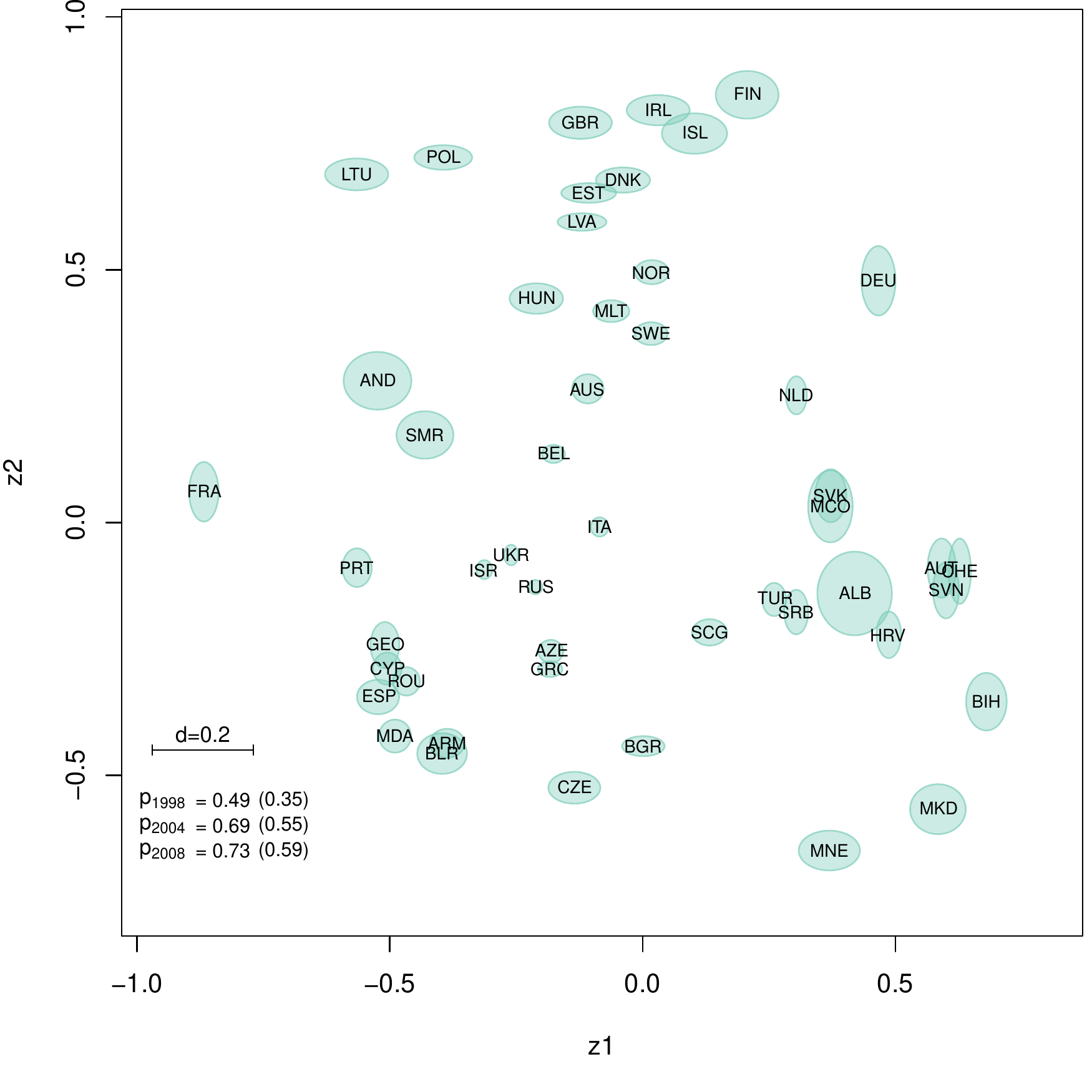} }}%
\caption{Estimated latent positions 1998-2015. The legend reports the probabilities corresponding to a distance in the latent space of $0.2$ for the years 1998, 2004 and 2008. The values refer to the case of $x_{2,ij} = 0$, within the brackets are reported the values for $x_{2,ij} = 1$.}
\label{fig:risult_contig2-1}%
\end{figure}

\begin{figure}[b]%
    \centering       
    {{\includegraphics[width=10cm]{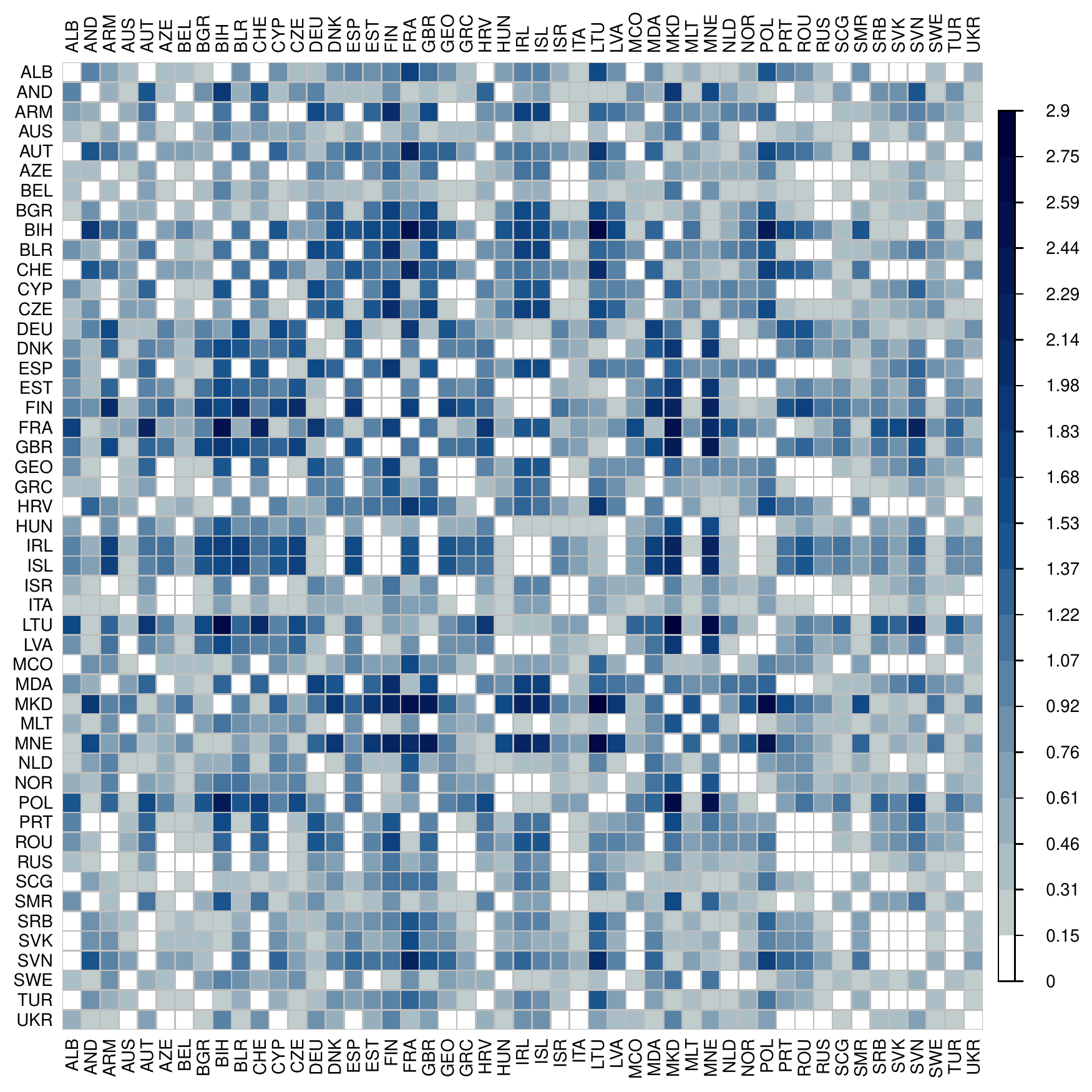} }}%
\caption{Estimated distances between couple of countries for the period 1998-2015.}
   \label{fig:risult_contig2-2}%
\end{figure}

\begin{figure}[b]%
    \centering    
      {{\includegraphics[scale =.45]{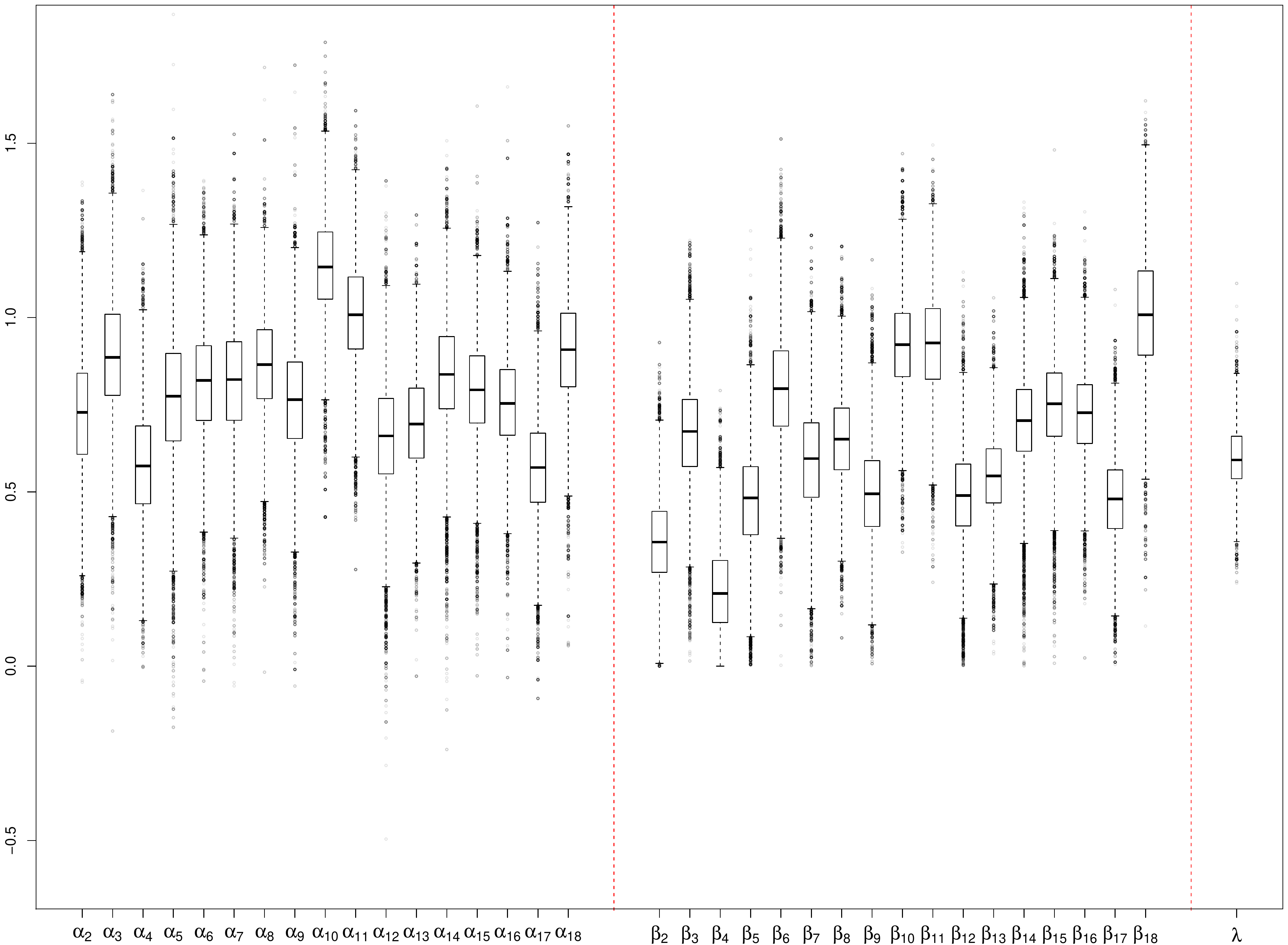} }}%
    \caption{Boxplots for the estimates of the logistic parameters and the coefficient for the set of covariates $\mathbf{X}_2$ in the period 1998-2015.}%
    \label{fig:risult_contig2-3}%
\end{figure}

\begin{figure}[b]%
    \centering
    \subfloat[$neigh = 1$]{{\includegraphics[width=7.2cm]{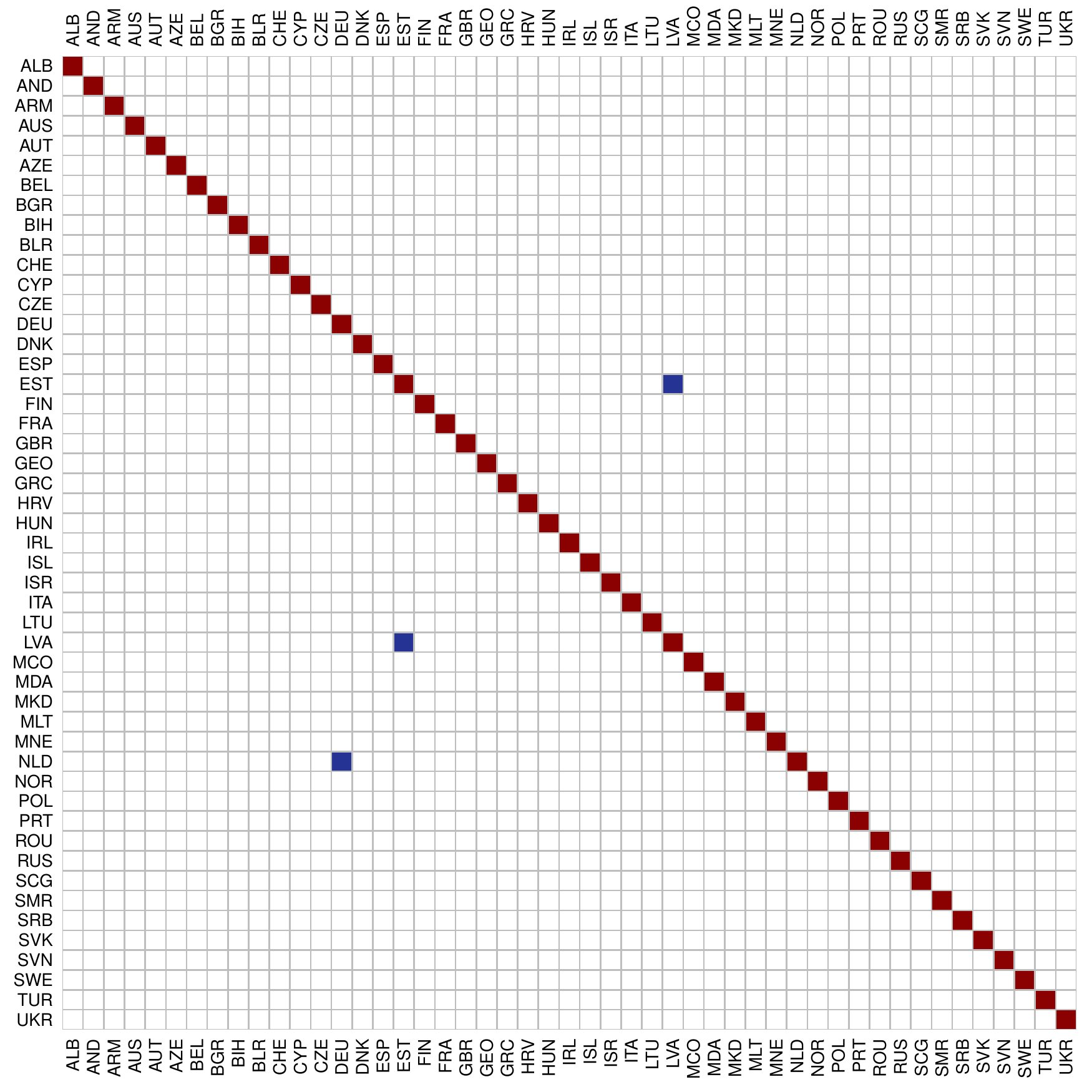} }}%
    \quad 
    \subfloat[$neigh = 2$]{{\includegraphics[width=7.2cm]{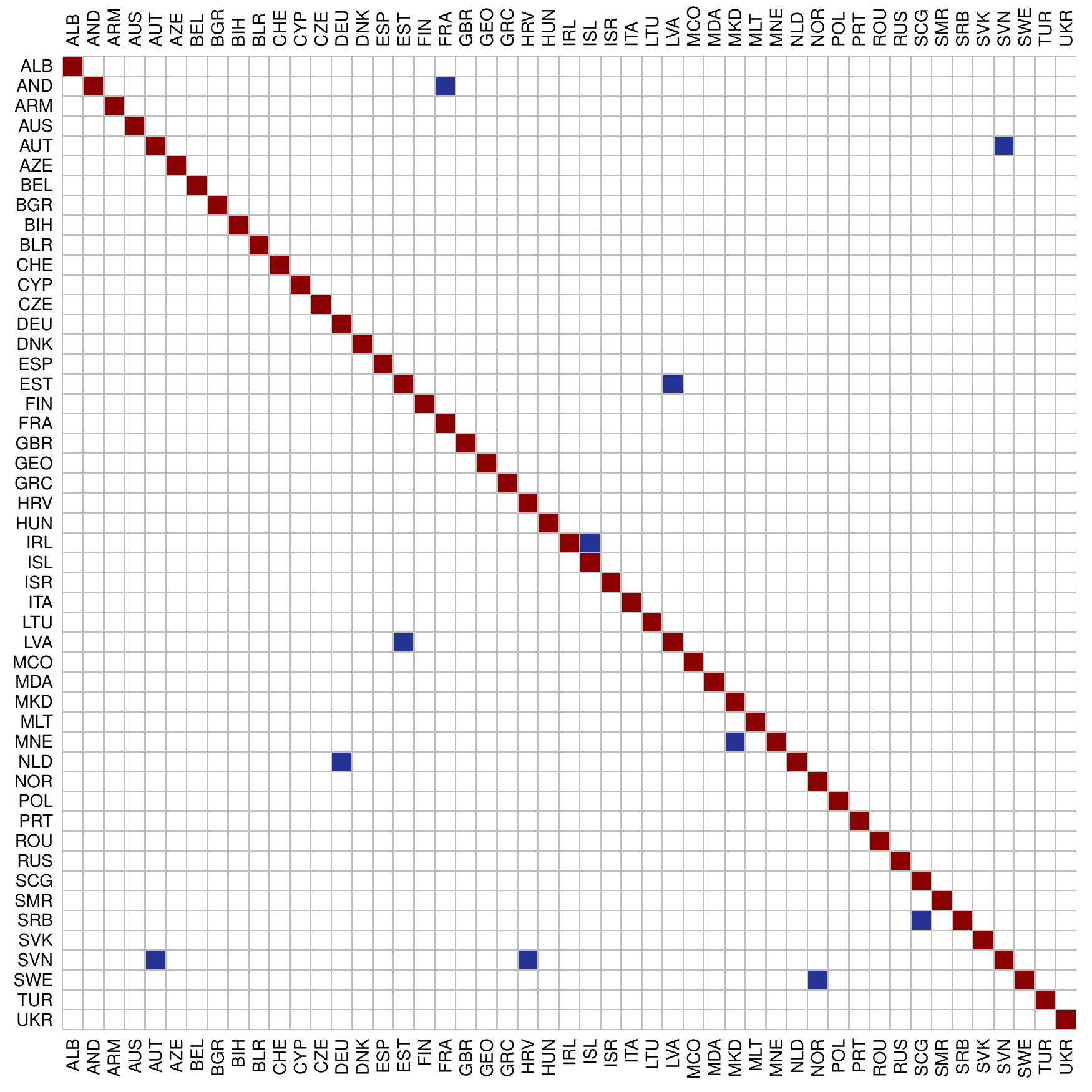} }}%
    \qquad
    \subfloat[$neigh = 3$]{{\includegraphics[width=7.2cm]{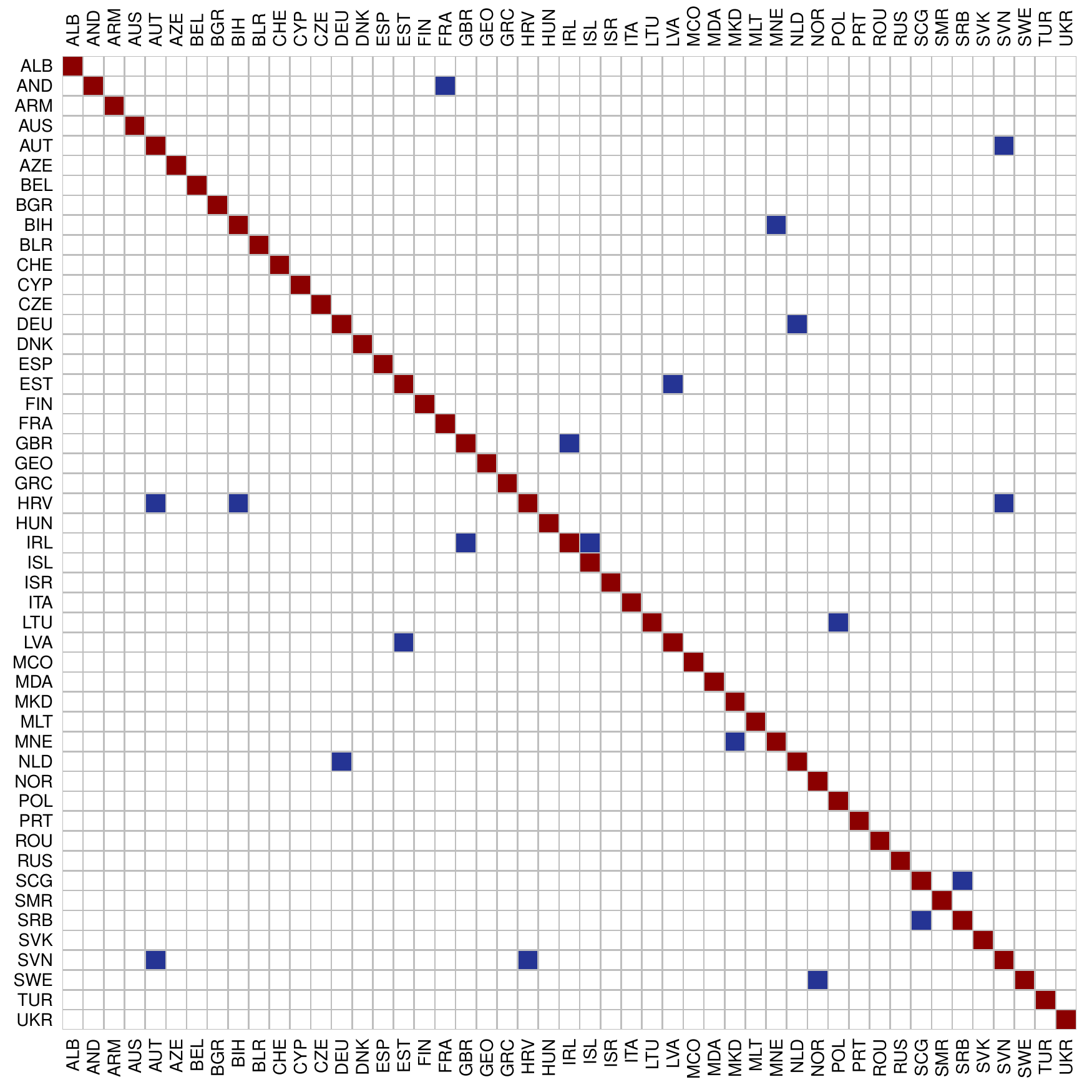} }}%
    \quad 
    \subfloat[$neigh = 5$]{{\includegraphics[width=7.2cm]{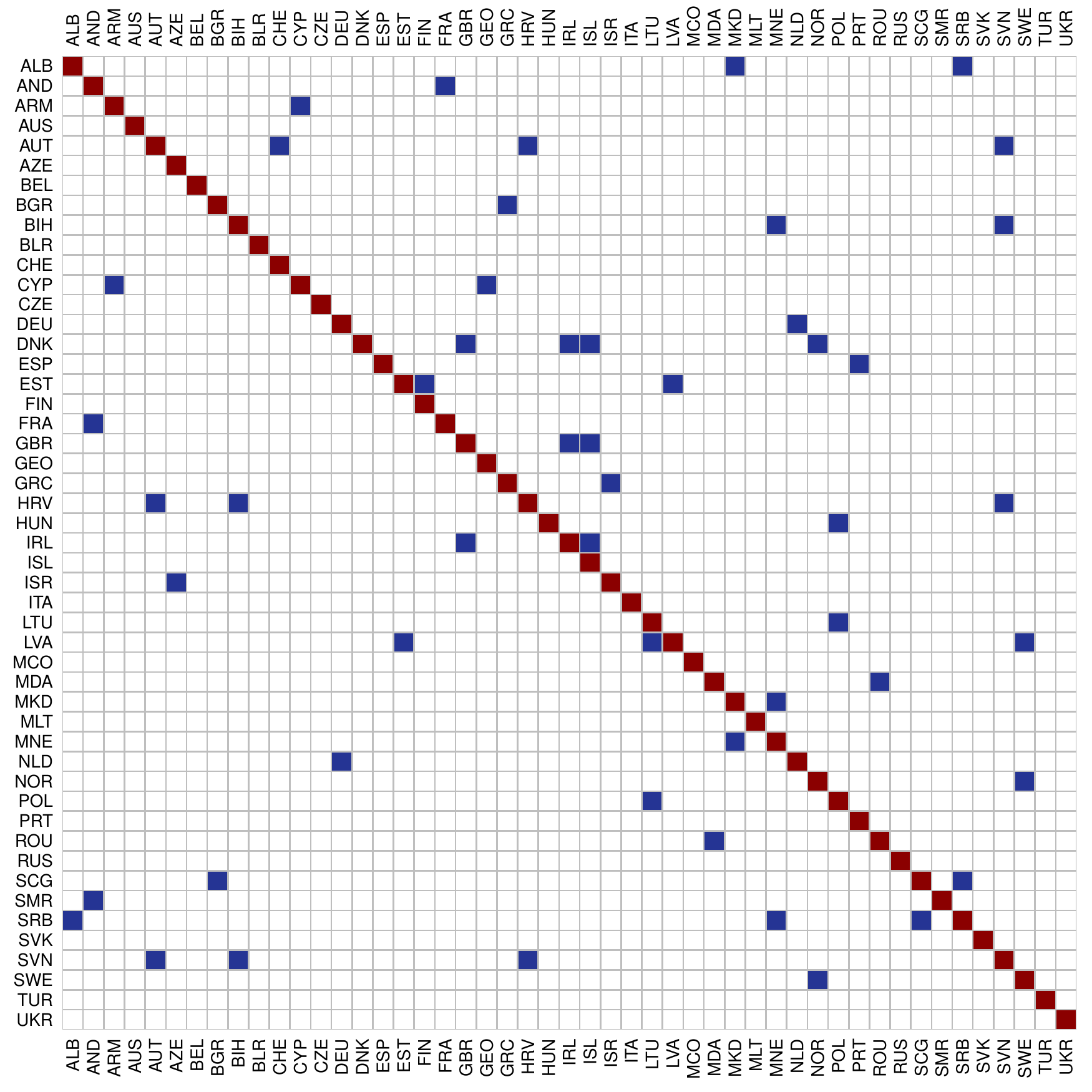} }}%
    \qquad
    \subfloat[$neigh = 10$]{{\includegraphics[width=7.2cm]{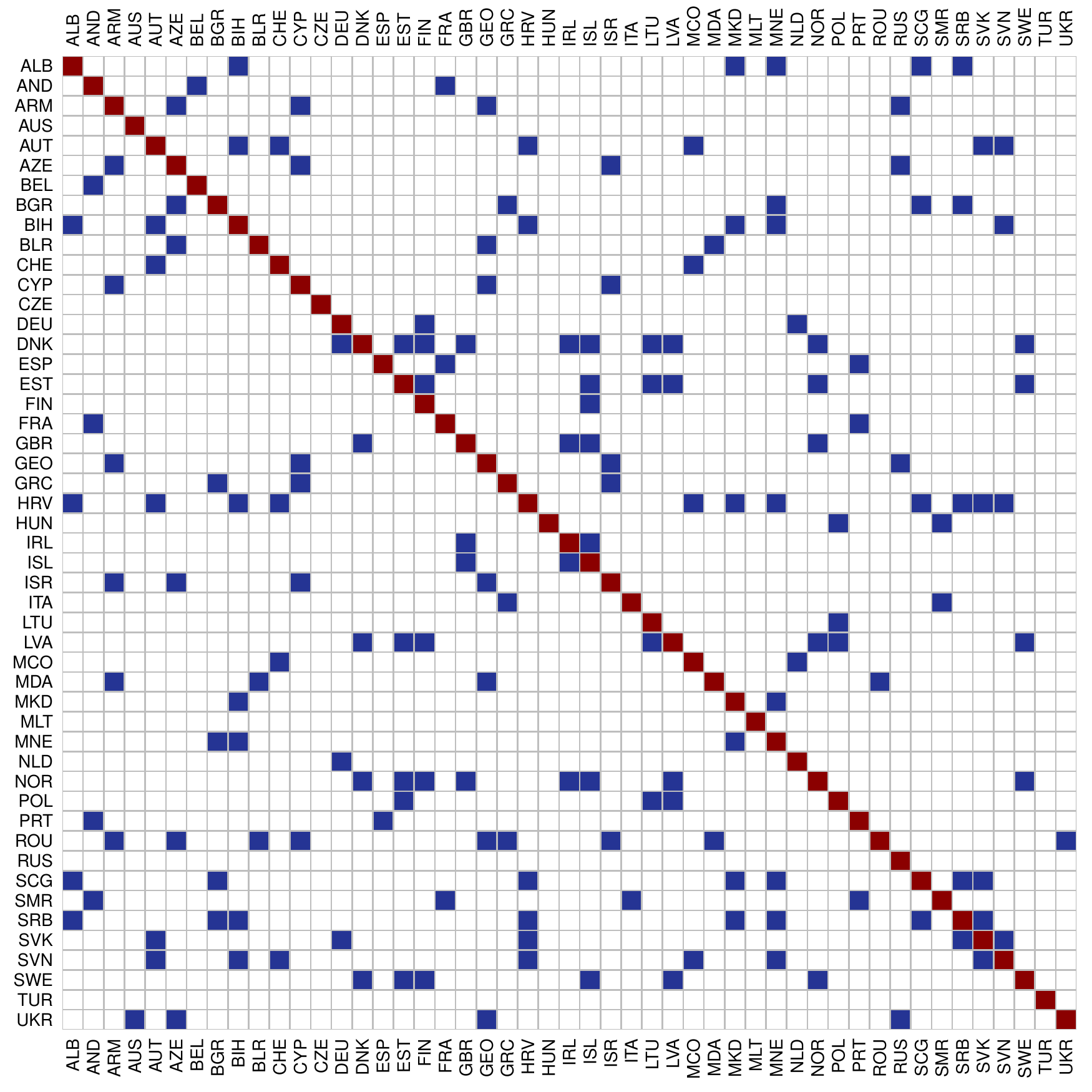} }}%
    \quad 
    \subfloat[$neigh = 15$]{{\includegraphics[width=7.2cm]{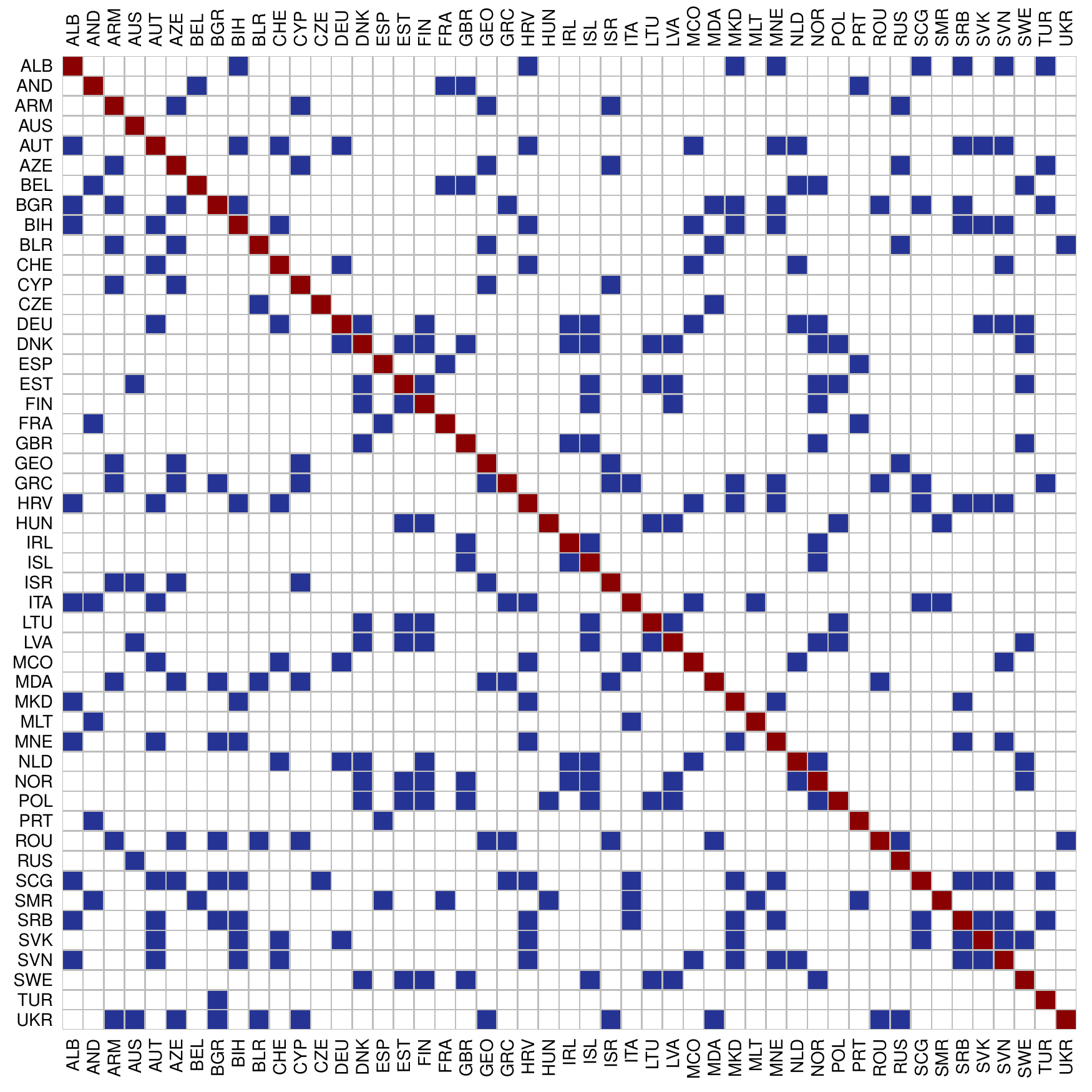} }}%
    \caption{Intersections of the set of neighbours $LN_{i,r}$ and $GN_{i,r}$, for the period 1998-2015}
    \label{fig:eurof7}%
\end{figure}

\begin{figure}[b]%
    \centering
{{\includegraphics[width=12cm]{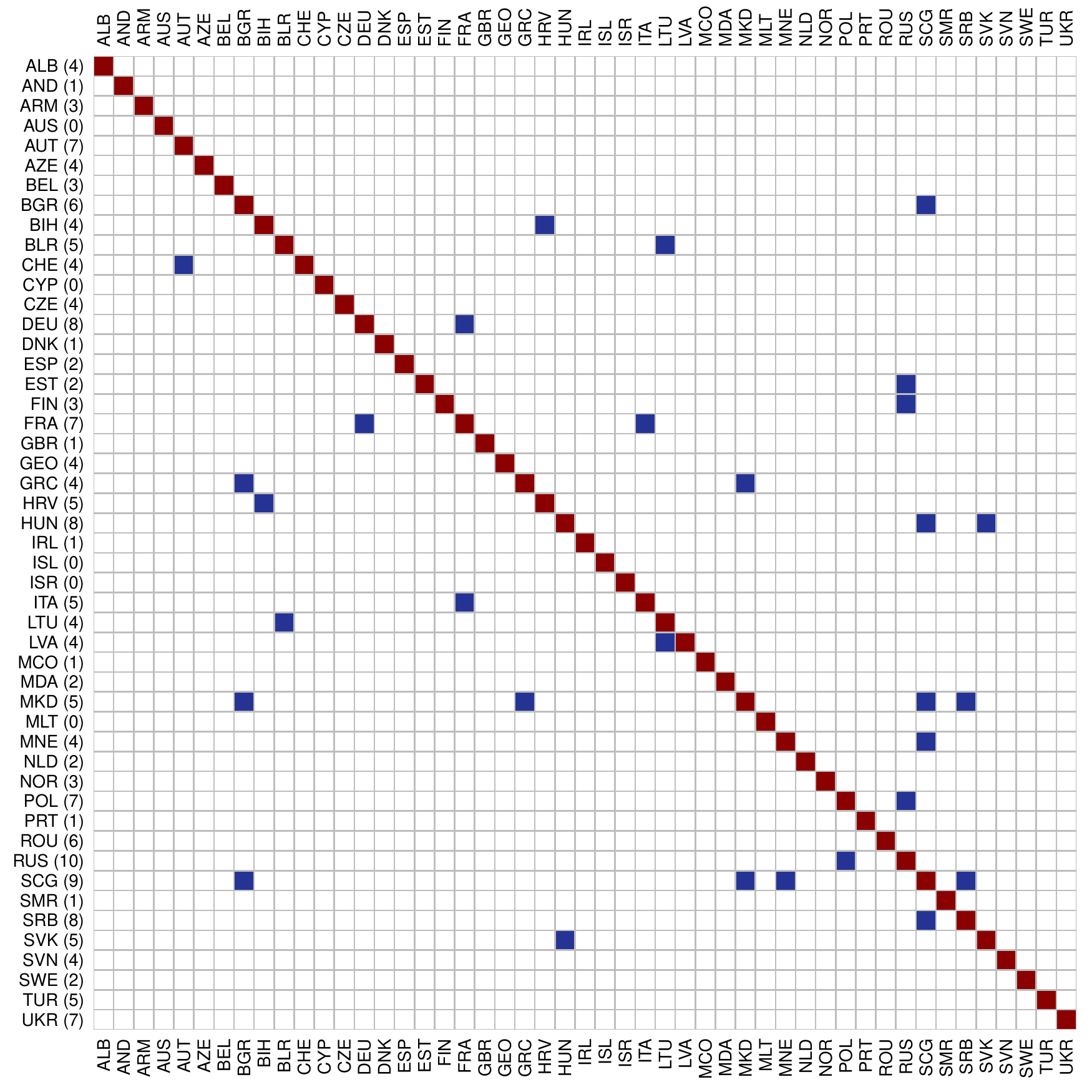} }}%
    \caption{Intersection of the set of neighbours $LN_{i,r_i^*}$ and $CN_{i,r_i^*}$, for the period 1998-2015. On the left column, in brackets, are reported the values of $r_i^*$.}%
    \label{fig:risult_contig_b}%
\end{figure}

Figures \ref{fig:risult_contig2-1}, \ref{fig:risult_contig2-2} and \ref{fig:risult_contig2-3} show the estimates obtained in \emph{model 2} for the latent positions, the distances and the posterior distributions for the parameters of interest. Figure \emph{(a)} reports the posterior means of the estimates for the country latent coordinates (reported with their ISO3 codes, see appendix \ref{app2}) together with the corresponding standard deviations. 
Note that the model does not necessarily place in the center of the latent space those countries that have been most successful throughout the editions. Indeed, for example, Sweden and Denmark won the largest number of editions in the period 1998-2015, respectively $3$ and $2$ titles. However, Denmark is not in the middle of the space, but it is rather placed close to a group of countries from northern Europe. If we look at a specific country, its neighbours on the latent space are the countries that were estimated to be more similar in \emph{tastes}, expressed in terms of voting patterns. 
The latent space presents a number of denser sub-groups of locations, that partly resemble northern Europe, eastern Europe and North Eastern Europe. However, these subgroups are not completely faithful to the geographic locations of countries, as, for example, Spain is closer to Romania than to Portugal or France. Nevertheless, the population of Romanians in Spain defines one of the major immigrant group in the country as well as Latvians is one of the largest immigrant group in Ireland. Therefore, some of the \emph{geographical misplacements} within the dense sub-groups in the latent space may be explained in terms of migration flows. What is certain is that geographical locations are not able to fully explain the observed patterns of votes.
Indeed, figures \ref{fig:risult_contig2-2} and \ref{fig:eurof7} show the presence of large differences between the estimated latent distances and the geographical ones. In particular, for a given country $i$, the rows in the matrices of figure \ref{fig:eurof7} represent the intersection between its $r$ nearest neighbours in the latent space (we denote this set as $LN_{i,r}$) and its $r$ closer neighbours in terms of geographical distances ($GN_{i,r}$); we consider the values $r = 1,2,3,5,10,15$. Given the number of neighbours $r$, the average\footnote{the average number of common neighbours is given by: \[\frac{\sum_{i=1}^n|LN_{i,r} \cap GN_{i,r}|}{ r}\]} and the maximum number of common neighbour countries is reported in table \ref{tab:neigh}. The table confirms what was already visible from figure \ref{fig:eurof7}: there is poor association between the coordinates in the latent space and the geographical ones.
A similar comparison can be made with the information on the shared border. Given a country $i$. let us define $r_i^*$ the number of bordering countries, $LN_{i,r_i^*}$ the $r^*$ nearest neighbours in the latent space and $CN_{i,r_i^*}$ the set of bordering countries of node $i$. The average number of geographical bordering countries that are also neighbouring in the latent space\footnote{this average is given by: \[\frac{\sum_{i=1}^n|LN_{i,r_i^*} \cap CN_{i,r_i^*}|}{ r_i^*}\]} is $0.11$, where $\hat{r}^* \approx 4$. This low correspondence between bordering countries and closest countries in the latent space confirms that $\mathbf{X}_2$ is only partially relevant in the description of the votes' exchange in the contest. Figure \ref{fig:risult_contig_b} reports the matrix of the intersections between the sets of neighbours $LN_{i,r_i^*}$ and $CN_{i,r_i^*}$, for the period 1998-2015. Bulgaria, Lithuania and Serbia and Montenegro are the countries that more tend to vote to their bordering countries. Indeed, their closest countries in the latent space are often countries with which they share a border ( $ |LN_{i,r_i^*} \cap CN_{i,r_i^*}|/ r_i^* \geq 0.5$). In general, there is no strong association between the presence of a border between a couple of countries and their closeness in the latent space.

\begin{table}[b]
\centering
\small
\begin{tabular}{@{}lllllllllllll@{}}
\toprule
 & $r=1$ & $r=2$  & $r=3$ & $r=5$  & $r=10$  & $r=15$  \\
\midrule
average number & $0.06$ & $0.11$ & $0.14$ & $0.21$ & $0.37$ & $0.47$\\
maximum number & $1$    & $1$ & $2$ & $3$ & $8$ & $12$\\
\bottomrule
\end{tabular}
\caption{Average and maximum number of neighbours in the intersection of the set of the closest latent positions and the set of the closest geographic positions for the countries.}
\label{tab:neigh}
\end{table}

The estimated values for the network intercept parameters are quite similar for the different networks corresponding to the editions in the period 1998-2015 (figure \ref{fig:risult_contig2-3}). Indeed, the voting rule (in the Eurovision song contest) for that period required the participating countries to vote for exactly 10 others, which implied a fixed outdegree for each node in the corresponding networks. The observed densities are then quite similar and this is reflected in similar estimates for the $\ak$ parameters, which define the upper bound for the edge probabilities in a given network.


Appendix \ref{app3} reports the results for the analysis of the two sub-periods 1998-2007 and 2008-2015. The model considered for the sub-periods does not include any covariate, as the interest lays primarily in recovering the latent coordinates. The findings confirmed a weak correspondence between the estimated latent position of a country and its actual latitude and longitude coordinates on the globe. A direct comparison of the latent space in the period 1998-2015 and the ones for the periods 1998-2007 and 2008-2015 is not available, as the number of countries is different. Indeed, not all the participants in the period 1998-2015 were also participant in both of the sub-periods. Just to give an example, Italy rejoined the competition in 2011, after being absent in the period 1998-2007. Although some of the distances vary with respect to those in the longer period 1998-2015, the sub-groups observed in figure \ref{fig:risult_contig2-1} are still present. Indeed, for example, Northern Europe countries tend always to be closer to each other, as well as Eastern Europe countries do.
 

\section{A simulation study}
\label{sim_study}
Different simulations scenarios have been considered to test the proposed model. In particular, simulations have been exploited to assess the large-sample behaviour of the parameters estimates, when the dimension $K$ of the multiplex is large, and to verify the robustness in the latent coordinates estimates with respect to different underlying distributions. 
In all the different scenarios, the reference network was taken to be the first of the multiplex and the reference parameters have been fixed to $\beta^{(1)}=1$ and $\alpha^{(1)}=0$ (the value of the intercept in the reference network in the application). 
The intercept and the coefficient parameters have been simulated from their prior distributions (see section \ref{lsmmn_like}), with $\sigma_{\alpha}^2 = \sigma_{\beta}^2=1$, $\mu_{\alpha}=\mu_{\beta}=0$.
Four simulation scenarios have been defined and divided in two blocks:
\begin{itemize}
\item \textbf{Block I}
This block has been built to verify the large-sample behaviour and the robustness of the estimates when the distribution of the latent coordinates is modified. Indeed, not all the observed data might be well described by latent Gaussian coordinates.
Within each scenario in the block, we have considered $4$ types of multidimensional networks, with relatively small dimension of $K$ but increasing number of nodes:
\begin{enumerate}
\item $n = 25$ and $K =3$,
\item $n = 50$ and $K =3$,
\item $n = 50$ and $K =5$,
\item $n = 100$ and $K =3$.
\end{enumerate} 
The values for the coefficients and the intercepts are constants in scenarios I-III, conditionally on the type of multiplex considered.

\begin{itemize}
\item \textbf{Scenario I}: the latent coordinates have been simulated from a bivariate normal distribution (the prior distribution specified in the model). 
\item \textbf{Scenario II}:  the latent coordinates have been simulated from a mixture of bivariate normal distributions, where the number of components was set is $G \approx n /7$. The mean vector for each group has been simulated from a standard bivariate normal distribution and the covariance matrices are diagonal with elements randomly sampled in the interval $(0.1,1)$.
This scenario corresponds to the case of data representing different kind of relations among separated groups of nodes or communities. Indeed, the probability for node $i$ in group $c$ to link with node $j$ will be higher if $j \in c$ then if $ j \notin c $;
\item \textbf{Scenario III}: the latent coordinates have been simulated from a standard bivariate Hotelling's T-squared distribution with $4$ degree of freedom. This scenario allows for some nodes to be located far from the center in the latent space. Thus, this case reflects the presence of inactive or semi-inactive nodes in the network, that tend to interact poorly with the rest not forming many edges;
\end{itemize}
\item \textbf{Block II}
This block has been built to test the large-sample properties of the estimates when the number of networks $K$ is large. That is, the case of the application considered in the present work.
\begin{itemize}
\item \textbf{Scenario IV}: the latent coordinates are simulated according to the prior distribution specified in section \ref{lsmmn_like}. Three different multidimensional networks have been simulated:
\begin{enumerate}
	\item $K = 10$ and $n = 50$,
	\item $K = 20$ and $n = 50$,
	\item $K = 30$ and $n = 50$.
\end{enumerate}
\end{itemize}
\end{itemize}
In all the considered scenarios, we have treated both the case where all the nodes are present in each network (denoted by $P$) and the case where some of the nodes are absent in some networks (denoted by $A$) (see section \ref{missing}). 
In the second case, the missing data process resembles the one observed in the Eurovision data with respect to the average number of absent nodes per network and the number of absences for each node. The reference network was taken to be the first of the multiplex and the corresponding parameters have been fixed to $\beta^{(1)}=1$ and $\alpha^{(1)}=0$ (the value in the application). 
The latent coordinates, the intercept and the coefficient parameters have been simulated from their prior distributions (see section \ref{lsmmn_like}), with $\sigma_{\alpha}^2 = \sigma_{\beta}^2=1$, $\mu_{\alpha}=\mu_{\beta}=0$.

To estimate the model on the simulated data (\textbf{Block I} and \textbf{Block II}), we fixed $\nu_{\alpha}=\nu_{\beta}=3$, $\tau_{\alpha} = \tau_{\beta} = (K-1)/K$, $\alpha^{(1)} =0$ and $\beta^{(1)}=1$ (see section \ref{lsmmn_est} for details).
Each model was estimated $10$ times, performing $40000$ mcmc iterations and discarding the first $5000$. The parameter estimates were consistent with the simulated values in all different scenarios and the estimates of the coordinates have been found to be robust to the different specifications of the distribution for the latent positions. 
In appendix \ref{app4} we present in details the results for the different scenarios. Boxplots and tables with mean and standard deviations for the parameter estimates are presented, as well as mean and standard deviation of the procrustes correlation between the estimated and the simulated latent space coordinates.
Overall, the proposed method returns reliable estimates for the true parameter values. The simulated values fall within the $95\%$ credible interval built on the posterior distributions, with a couple of exceptions that occur when the number of networks increases. However, in these cases, the true magnitude of the value is always recovered, as the estimates are still quite close to the actual simulated values.
The simulated latent spaces are always recovered with high correlation, even in the stressed scenarios I and II. There is no substantial difference in estimates for cases $P$ and $A$. Thus, the absence of some of the nodes in some networks of the multiplex does not impact the estimation of the latent positions.

\section{Comparison with the \emph{lsjm} model}
\label{lsjm}
The model proposed in the present work is similar to the latent space joint model (\emph{lsjm}) proposed by \cite{gollini}, which considers a similar specification for the edge probability (\ref{eq:edge_prob}), with the difference that no coefficient term is associated to the distances, and these are network-specific. The model by \cite{gollini} also considers a mean latent space, which originated the network-specific latent coordinates. 
To compare the two models, we have simulated different types of multiplex, all constructed fixing $\bk = 1$ for $k = 1, \dots, K$, that is, according to the \emph{lsjm}. The simulated multidimensional networks come from the following scenarios:
\begin{enumerate}
\item A sparse multiplex with $n=50$, $K=3$ and $\alpha =(-0.66, -0.70, -0.54)$,
\item A multiplex with $n=70$, $K=2$ and $\alpha =(-0.73, -1.12)$,
\item A small but denser multiplex with $n=25$, $K=3$ and $\alpha =(0.00, 1.02, 0.28)$.
\end{enumerate}
Simulating from the model presented in this work (which will be referred to as \emph{lsmmn}) when all the coefficients are fixed to $1$ corresponds to simulating from the latent space joint model when all the network-latent spaces are the same. Therefore, in the present setting, the two models can be compared. 
Both the \emph{lsjm} and the \emph{lsmmn} models have been estimated 10 times for each of the simulated multidimensional networks.
In appendix \ref{app4} we report the results of such a comparison. Overall, the model we propose outperforms the \emph{lsjm} both in the quality of parameter estimates and in recovering the latent coordinates. 

\section{Discussion}
\label{disc}
In the present work, we have introduced a general and flexible model for the analysis of multidimensional networks. In particular, the model is defined to recover similarities among the nodes when the structure of the observed networks is complex and there is not a clear information on which external information can be used to explain the observed patterns. 
The model extends the latent space model by \cite{hoff} and the latent space joint model by \cite{gollini} with the introduction of network-specific coefficient parameters to weight the relevance of the latent space in the edge-probabilities determination. When these coefficients are null, the model reduces to a random graph model for multidimensional networks. Moreover, missing data (not at random) and edge-specific covariates are considered. A hierarchical Bayesian approach is employed to define the model and its estimation is carried out via mcmc. We have defined hyperprior distributions for the hyperparameters of the model, to avoid their (subjective) specification. The latent coordinates allow for an efficient visualization of the network, a well desired feature for large multidimensional networks.   

The model is applied to the votes exchanged among countries in a popular TV show, the Eurovision Song Contest, from 1998 to 2015. 
Cultural and geographical covariates have been included in the analysis and only the presence of a shared boarder between two countries was found to be relevant to explain the observed voting pattern. The recovered similarities among the participants in the period 1998-2015 have been found to only partially resemble the actual geographical locations of the countries. Indeed, a group structure was  found in the latent space which does not completely agree to geographical criteria. These findings sustain the claim of bias in the voting structure observed in the Eurovision, which, however, can not be attributed to geographical reasons alone. 

In the simulation study we have applied the proposed model on a large varieties of multidimensional networks and successfully recovered the latent coordinates and the network-specific parameters. That has proved the ability of the model to recover the (latent) association structure among the nodes in a multiplex, also when the number of networks is large.   

The latent space model for multivariate networks is implemented in the \textit{R} package \textit{spaceNet} and it is available on \textit{CRAN}. 

\appendix


\bibliographystyle{agsm}
\bibliography{bibliography}

\clearpage
\section{Posterior distributions for the parameters}
\label{app1}
The posterior distribution for the model has been presented in section \ref{lsmmn_like} in equation \ref{eq:post}. 
The present appendix will present the full conditional distributions and the proposal distributions that have been derived from equation \ref{eq:post} and used to estimate the model, as described in section \ref{lsmmn_est}.
For ease of calculation, the log-posterior distribution is considered:
\[
\begin{split}
\log\Bigl( P \bigl( \alpha, \beta, \mathbf{z},\mu_{\alpha}, \mu_{\beta},\sigma_{\alpha}^2 ,\sigma_{\beta}^2 |\mathbf{Y} \bigr) \Bigr) \propto & 
\sum_{k = 1}^K \sum_{i = 1}^n  \sum_{j \neq i} h_{ij}^{(k)} \Bigl[ \arch \bigl(\ak  -\bk  d_{ij} \bigr) - \log \Bigl( 1 + \exp\bigl\{ \ak  -\bk d_{ij} \bigr\} \Bigr) \Bigl] \\
& -\frac{1}{2} \Biggl\{ \sum_{i = 1}^n z_i^2 + \frac{\sum_{k=1}^K (\ak -\mu_{\alpha})^2}{\sigma_{\alpha}^2} + \frac{\sum_{k=1}^K (\bk -\mu_{\beta})^2}{\sigma_{\beta}^2} +K \log(\sigma_{\alpha}^2 ) \\
&  +K \log(\sigma_{\beta}^2 ) + \log(\tau_{\alpha}\sigma_{\alpha}^2 ) +\log(\tau_{\beta}\sigma_{\beta}^2 )+\frac{\mu_{\alpha}^2}{\tau_{\alpha}\sigma_{\alpha}^2} +\frac{\mu_{\beta}^2}{\tau_{\beta}\sigma_{\beta}^2} +\frac{1}{\sigma_{\alpha}^2} + \frac{1}{\sigma_{\beta}^2} \Biggr\}\\
& + \Bigl( -\frac{\nu_{\alpha} }{2} -1\Bigr)\log(\sigma_{\alpha}^2) + \Bigl( -\frac{\nu_{\beta} }{2} -1\Bigr)\log(\sigma_{\beta}^2)
\end{split}.
\]
Without any loss of information, the latent coordinates are assumed to be univariate. A generalization to the multivariate case is straightforward and will be considered when the proposal distribution for the latent positions is introduced.

\subsection{Nuisance parameters}
All the posterior distributions for the nuisance parameters have closed form.
In particular, the variances follow Inverse Gamma distributions
\[
\sigma_{\alpha}^2 | \alpha, \mu_{\alpha}, \tau_{\alpha}, \nu_{\alpha}, K \thicksim \text{\small{Inv}} \Gamma\bigl( r_{\alpha}, R_{\alpha} \bigr); \qquad
\sigma_{\beta}^2 | \beta, \mu_{\beta}, \tau_{\beta}, \nu_{\beta} , K\thicksim \text{\small{Inv}} \Gamma\bigl( r_{\beta}, R_{\beta} \bigr),
\]
with parameters:
\[
r_x = \frac{\nu_x +K +1}{2}, \qquad R_x = \frac{\tau_x + \tau_x \sum_{k =1}^K (x^{(k)} -\mu_x)^2 +\mu_x^2}{2 \tau_x}.
\]
The mean parameters $\mu_{\alpha}, \mu_{\beta}$ are distributed as truncated normal distributions:
\[
\mu_{\alpha} | \alpha , \sigma_{\alpha}^2, \tau_{\alpha}, m_{\alpha}, K \thicksim N_{\bigl[ LB(\alpha), \infty \bigl]} \Biggl( \frac{\tau_{\alpha} \sum_{k =1}^K \ak + m_{\alpha}}{1 + K\tau_{\alpha} }, \frac{\tau_{\alpha}\sigma_{\alpha}^2}{1 + K\tau_{\alpha}}\Biggr),
\]
\[
\mu_{\beta} | \beta , \sigma_{\beta}^2, \tau_{\beta},m_{\beta},K \thicksim N_{\bigr[0, \infty \bigl]} \Biggl( \frac{\tau_{\beta} \sum_{k =1}^K \bk +m_{\beta}}{1 + K\tau_{\beta} }, \frac{\tau_{\beta}\sigma_{\beta}^2}{1 + K\tau_{\beta}}\Biggr) .
\]
  
\subsection{Latent positions}
\label{lat_pos_prop}
In order to derive a proposal distribution for the latent coordinates, only the terms containing the variables $z_i$ in the log-posterior distribution are considered:
\[
\log \Bigl( P(z_i \mid \mathbf{Y}, \mathbf{D}, \mathbf{H}, \alpha, \beta ) \Bigr) = 
-\frac{1}{2} z_i^2 -  \sum_{k = 1}^K \sum_{j \neq i}  h_{ij}^{(k)} \Biggl[ \arch \bk d_{ij} +\log \Bigl( 1 + \exp\bigl\{ \ak  -\bk d_{ij}\bigr\}\Bigr)\Biggr].
\]
To approximate the logarithmic term in the above equation, let us recall the definition of the LSE (Log Sum Exponential) function,
\[
\log \Bigl( \sum_{g= 1}^G \exp(x_g)\Bigr),
\]
that is known to be bounded between:
\[
\max \bigl\{ x_1, \dots , x_G \bigr\} \leq  \log \Bigl( \sum_{g= 1}^G \exp(x_g)\Bigr) \leq \max \bigl\{ x_1, \dots , x_G \bigr\} +\log(G).
\]
The logarithmic term in the log-posterior part of interest for the latent coordinates has indeed LSE form, thus it can be bounded between:
\[
\max \bigl\{ 0, \ak  -\bk d_{ij} \bigr\} \leq  \log \Bigl( \exp\bigl\{ 0 \bigr\}  \exp\bigl\{ \ak  -\bk d_{ij}\bigr\}\Bigr) \leq \max \bigl\{ 0, \ak  -\bk d_{ij} \bigr\}+\log(2).
\]
The lower bound is met when only one of the term in the summation is non-zero, which is the case.
Defining the binary variable $w_{ij}^{(k)}$ as,
\[
w_{ij}^{(k)}= 
\begin{cases}
1 \qquad \text{if} \quad \ak  -\bk d_{ij} >0
  \\
 0 \qquad \text{if} \quad \ak  -\bk d_{ij} \leq 0
   \end{cases},
\]
then the logarithmic term of interest can be approximated using its lower bound.
The log-posterior terms in $z_i$ can be approximated as:
\[
\begin{split}
& -\frac{1}{2} z_i^2 -  \sum_{k = 1}^K \sum_{j \neq i} h_{ij}^{(k)} \Biggl[ \arch \bk d_{ij} +\max \bigl\{ 0, \ak  -\bk d_{ij} \bigr\} \Biggr] \\
 \propto & -\frac{1}{2} z_i^2 -  \sum_{k = 1}^K \sum_{j \neq i} h_{ij}^{(k)} \Biggl[ \arch \bk( z_i^2 +z_j^2 -2 z_iz_j ) 
 +  w_{ij}^{(k)} \Bigl( \ak \ti \gj -\bk (z_i^2 +z_j^2 -2 z_iz_j) \Bigr) \Biggr] \\
  \propto & -\frac{1}{2} z_i^2 -  \sum_{k = 1}^K \sum_{j \neq i} h_{ij}^{(k)} \Biggl[ \arch \bk( z_i^2 -2 z_iz_j ) 
 +  w_{ij}^{(k)} \Bigl(  -\bk (z_i^2  -2 z_iz_j) \Bigr) \Biggr] \\
 = & -\frac{1}{2} z_i^2 \Biggl[ 1 + 2 \sum_{k = 1}^K  \bk \sum_{j \neq i} h_{ij}^{(k)} (\arch -w_{ij}^{(k)}) \Biggr]+ z_i \Biggl[ 2 \sum_{k = 1}^K  \bk \sum_{j \neq i} h_{ij}^{(k)} (\arch -w_{ij}^{(k)}) z_j \Biggr].
\end{split}
\]
The last equality returns a quadratic form in $z_i$ and could be used to specify a proposal distribution for the latent coordinates. However, it is not guaranteed that the quantity multiplying $z_i^2$, which would be the precision of the proposal distribution for latent position $i$, is strictly positive.
Indeed, defining $n_i^{(k)} = \sum_{j \neq i} \arch$,
\[
\begin{split}
1 + 2 \sum_{k = 1}^K  \bk \sum_{j \neq i} h_{ij}^{(k)} (\arch -w_{ij}^{(k)}) & \geq \min \Biggl\{
1 + 2 \sum_{k = 1}^K  \bk \sum_{j \neq i} h_{ij}^{(k)} (\arch -w_{ij}^{(k)}) \Biggr\} \\
& \propto  1 + 2 \Bigl[\sum_{k = 1}^K  \bk n_i^{(k)}  - \max \bigl\{ \sum_{k = 1}^K  \bk \sum_{j \neq i} w_{ij}^{(k)} \bigr\} \Bigr]\\ 
 & = 1 + 2 \Bigl[\sum_{k = 1}^K  \bk \bigl( n_i^{(k)} - (n -1) \bigr) \Bigr]
 \end{split}.
\]
In the second row node indicator variables are fixed $h_{ij}^{(k)}=1$.
Notice that $ n_i^{(k)} \leq n -1 \implies \bigl( n_i^{(k)} - (n -1) \bigr) \leq 0$. 
Hencefor the precision to be strictly positive it should be that: 
\[
\begin{split}
 \min \Biggl\{
& 1 + 2 \sum_{k = 1}^K  \bk \sum_{j \neq i}  (\arch -w_{ij}^{(k)}) \Biggr\} > 0 \iff  2 \Bigl[\sum_{k = 1}^K  \bk \bigl( n_i^{(k)} - (n -1) \bigr) \Bigr] > -1\\
\iff  & \sum_{k = 1}^K  \bk \bigl( n_i^{(k)} - (n -1) \bigr)  > -\frac{1}{2}.
\end{split}
\]
One of the coefficient has to be fixed for identifiability issue (\ref{identif}). Taking $\beta^{(1)} = 1$, the above inequality may be rewritten as:
\[
\begin{split}
\sum_{k = 1}^K  \bk \bigl( n_i^{(k)} - (n -1) \bigr) &\geq \sum_{k = 1}^K  \bk ( 1 -n ) \\
&= (1-n) + \sum_{k = 2}^K  \bk ( 1 -n ) \geq -\frac{1}{2}.
\end{split}
\]
The last inequality holds only when $n =1$. This implies that the quantity 
\[
1 + 2 \sum_{k = 1}^K  \bk \sum_{j \neq i} h_{ij}^{(k)} (\arch -w_{ij}^{(k)}) 
\]
may not always be positive. Therefore, an alternative specification for the proposal precision is needed.
The binary variables $w_{ij}^{(k)}$ can be thought as potential-link indicator, as they depend on the argument of the numerator in the edge probability of equation (\ref{eq:edge_prob}). In that sense, the difference term $(\arch -w_{ij}^{(k)})$ can be considered as a measure of the goodness of link classification.
However, the sum of these terms, $\sum_{j \neq i} (\arch -w_{ij}^{(k)})$, 
could be null either in presence of perfect classification or in the presence of perfect misclassification. In fact, $\sum_{j \neq i} \arch = \sum_{j \neq i} w_{ij}^{(k)} \implies \sum_{j \neq i} (\arch -w_{ij}^{(k)}) = 0$. 
To overcome this problem, the absolute value of the difference is considered.
The proposal distribution specified for latent coordinate $i$ is then:
We can then define the proposal distribution for the $i^{\text{th}}$ latent coordinate as:
\[
\tilde{z}_i \mid \mathbf{Y},\alpha, \beta, \mathbf{D}, K \thicksim N \Bigl(\mu_{\tilde{z}_i}, \sigma_{\tilde{z}_i}^2 \Bigr),
\]
where 
\[
\mu_{\tilde{z}_i} = \sigma_{\tilde{z}_i}^2 \Bigl(2 \sum_{k = 1}^K \bk \sum_{j \neq i} h_{ij}^{(k)} \bigl(\arch -w_{ij}^{(k)} \bigr) z_j \Bigr), \quad 
 \sigma_{\tilde{z}_i}^2  = \Biggl( 1 + 2 \sum_{k = 1}^K  \bk \sum_{j \neq i} h_{ij}^{(k)} |\arch -w_{ij}^{(k)}| \Biggr)^{-1}.
\]
In the case of multivariate latent coordinates, with number of dimensions $p$, $\mu_{\tilde{z}_i}$ would be a $p-$dimensional vector and the covariance matrix would be diagonal, with generic element $ \sigma_{\tilde{z}_i}^2$.

\subsection{Intercept parameters}
The part of interest of the log-posterior distribution in \ref{eq:post} to derive a proposal distribution for the intercept terms is: 
\[
\begin{split}
\log\Bigl( P(\ak \mid \mathbf{Y}, \mathbf{D}, \mathbf{H} , \bk, \mu_{\alpha}, \sigma_{\alpha}^2 ,n \Bigr) \propto &
\sum_{i = 1}^n  \sum_{j \neq i} h_{ij}^{(k)} \Bigl[ \arch \bigl(\ak  -\bk  d_{ij} \bigr) - \log \Bigl( 1 + \exp\bigl\{ \ak  -\bk d_{ij} \bigr\} \Bigr) \Bigl]  \\
&-\frac{1}{2} \frac{(\ak -\mu_{\alpha})^2}{\sigma_{\alpha}^2}.
\end{split}
\]
The logarithmic term is approximate with its second order Taylor expansion around $\mu_{\alpha}$: 
\[
\begin{split}
\log \Bigl( 1 + \exp\bigl\{ \ak  -\bk d_{ij} \bigr\} \Bigr) \Bigr\lvert_{\ak = \mu_{\alpha}} \approx & \log \Bigl( 1 + \exp\bigl\{ \mu_{\alpha} -\bk d_{ij} \bigr\} \Bigr)+ (\ak -\mu_{\alpha})\frac{\exp\bigl\{ \mu_{\alpha} -\bk d_{ij} \bigr\} }{1 + \exp\bigl\{ \mu_{\alpha} -\bk d_{ij} \bigr\}}\\
& + \frac{1}{2}(\ak -\mu_{\alpha})^2 \frac{\exp\bigl\{ \mu_{\alpha} -\bk d_{ij} \bigr\} }{\bigl( 1 + \exp\bigl\{ \mu_{\alpha} -\bk d_{ij} \bigr\}\bigr)^2}.
\end{split}
\]
Substituting the logarithmic term with its approximation in the log posterior yields to:
\[
\begin{split}
\log\Bigl( P(\ak \mid \mathbf{Y}, \mathbf{D}, \mathbf{H} , \bk, \mu_{\alpha}, \sigma_{\alpha}^2) \Bigr) \propto &
-\frac{1}{2} \frac{(\ak -\mu_{\alpha})^2}{\sigma_{\alpha}^2}
+ \sum_{i = 1}^n  \sum_{j \neq i} h_{ij}^{(k)} \Biggl[ \arch \ak - \ak \frac{\exp\bigl\{ \mu_{\alpha} -\bk d_{ij} \bigr\} }{1 + \exp\bigl\{ \mu_{\alpha} -\bk d_{ij} \bigr\}} \\
&-  \frac{1}{2}(\ak -\mu_{\alpha})^2 \frac{\exp\bigl\{ \mu_{\alpha} -\bk d_{ij} \bigr\} }{\bigl( 1 + \exp\bigl\{ \mu_{\alpha} -\bk d_{ij} \bigr\}\bigr)^2}\Biggl] 
\end{split}
\]
that is a quadratic form in $\ak$. 
Defining $E^{(k)} = \sum_{i = 1}^n  \sum_{j \neq i} h_{ij}^{(k)} \arch $,
the proposal distribution for intercept $\alpha^{(k)}$ is taken to be:
\[
\tilde{\alpha}^{(k)} \mid \mathbf{Y}, \mathbf{D}, \mathbf{H}, \bk, \mu_{\alpha}, \sigma_{\alpha}^2, n \thicksim N \Bigl(\mu_{\tilde{\alpha}^{(k)}}, \sigma_{\tilde{\alpha}^{(k)}}^2 \Bigr),
\]
with
\[
\mu_{\tilde{\alpha}^{(k)}} = \sigma_{\tilde{\alpha}^{(k)}}^2 \Biggl\{ E^{(k)} -  \sum_{i = 1}^n  \sum_{j \neq i}  \frac{h_{ij}^{(k)} \exp\bigl\{ \mu_{\alpha} -\bk d_{ij} \bigr\} }{1 + \exp\bigl\{ \mu_{\alpha} -\bk d_{ij} \bigr\}} \Biggr\} +\mu_{\alpha}, \quad
\sigma_{\tilde{\alpha}^{(k)}}^2 = \Biggl\{ \sum_{i = 1}^n  \sum_{j \neq i} \frac{ h_{ij}^{(k)} \exp\bigl\{ \mu_{\alpha} -\bk d_{ij} \bigr\} }{\bigl( 1 + \exp\bigl\{ \mu_{\alpha} -\bk d_{ij} \bigr\}\bigr)^2}  +\frac{1}{\sigma_{\alpha}^2}\Biggr\}^{-1}.
\]

\subsection{Coefficient parameters (distances)}
The part of interest for the coefficient parameters in equation \ref{eq:post} is:
\[
\begin{split}
\log\Bigl( P(\bk \mid \mathbf{Y}, \mathbf{D}, \mathbf{H} ,\ak, \mu_{\beta}, \sigma_{\beta}^2,n \Bigr) \propto &
\sum_{i = 1}^n  \sum_{j \neq i} h_{ij}^{(k)} \Bigl[ \arch \bigl(\ak  -\bk  d_{ij} \bigr) - \log \Bigl( 1 + \exp\bigl\{ \ak  -\bk d_{ij} \bigr\} \Bigr) \Bigl]  \\
&-\frac{1}{2} \frac{(\bk -\mu_{\beta})^2}{\sigma_{\beta}^2}.
\end{split}
\]
The logarithmic term is approximated with its second order Taylor expansion in $\mu_{\beta}$: 
\[
\begin{split}
\log \Bigl( 1 + \exp\bigl\{ \ak  -\bk d_{ij} \bigr\} \Bigr) \Bigr\lvert_{\bk = \mu_{\beta}} \approx & \log \Bigl( 1 + \exp\bigl\{ \ak -\mu_{\beta} d_{ij} \bigr\} \Bigr)- (\bk -\mu_{\beta})\frac{ d_{ij}  \exp\bigl\{ \ak -\mu_{\beta} d_{ij} \bigr\} }{1 + \exp\bigl\{ \ak -\mu_{\beta}d_{ij} \bigr\}}\\
& + \frac{1}{2}(\bk -\mu_{\beta})^2  \frac{ d_{ij}^2 \exp\bigl\{ \ak -\mu_{\beta} d_{ij} \bigr\} }{\bigl( 1 + \exp\bigl\{ \ak -\mu_{\beta} d_{ij} \bigr\}\bigr)^2}.
\end{split}
\]
The logarithmic term is replaced with its approximation, leading to: 
\[
\begin{split}
\log\Bigl( P(\bk \mid \mathbf{Y}, \mathbf{D}, \mathbf{H} ,\ak, \mu_{\beta}, \sigma_{\beta}^2) \Bigr) \propto &
-\frac{1}{2} \frac{(\bk -\mu_{\beta})^2}{\sigma_{beta}^2} 
+ \sum_{i = 1}^n  \sum_{j \neq i} h_{ij}^{(k)} \Biggl[ -\arch \bk  d_{ij} + 
\bk \frac{ d_{ij}  \exp\bigl\{ \ak -\mu_{\beta} d_{ij} \bigr\} }{1 + \exp\bigl\{ \ak -\mu_{\beta}d_{ij} \bigr\}}\\
& - \frac{1}{2}(\bk -\mu_{\beta})^2  \frac{ d_{ij}^2 \exp\bigl\{ \ak -\mu_{\beta} d_{ij} \bigr\} }{\bigl( 1 + \exp\bigl\{ \ak -\mu_{\beta} d_{ij} \bigr\}\bigr)^2}.
\end{split}
\]
The above expression is quadratic in $\bk$. 
Then, the proposal distribution specified for intercept coefficient $k$ is:
\[
\tilde{\beta}^{(k)} \mid \mathbf{Y}, \mathbf{D}, \mathbf{H} ,\ak, \mu_{\beta}, \sigma_{\beta}^2 \thicksim N \Bigl(\mu_{\tilde{\beta}^{(k)}}, \sigma_{\tilde{\beta}^{(k)}}^2,n \Bigr) ,
\]
where
\[
 \mu_{\tilde{\beta}^{(k)}} = \sigma_{\tilde{\beta}^{(k)}}^2 \Biggl\{  \sum_{i = 1}^n  \sum_{j \neq i} h_{ij}^{(k)} d_{ij} \Biggl( \frac{\exp\bigl\{ \ak -\mu_{\beta} d_{ij} \bigr\} }{1 + \exp\bigl\{ \ak -\mu_{\beta} d_{ij} \bigr\}}  -\arch  \Biggr) \Biggr\} +\mu_{\beta},
\]
\[
 \sigma_{\tilde{\beta}^{(k)}}^2 = \Biggl\{ \sum_{i = 1}^n  \sum_{j \neq i} \frac{ h_{ij}^{(k)}  d_{ij}^2 \exp\bigl\{ \ak  -\mu_{\beta} d_{ij} \bigr\} }{\bigl( 1 + \exp\bigl\{ \ak -\mu_{\beta} d_{ij} \bigr\}\bigr)^2}  +\frac{1}{\sigma_{\beta}^2}\Biggr\}^{-1}.
\]

\subsection{Coefficient parameters (covariates)} 
The coefficient parameters of the covariates are assumed to be independent but not identically distributed. We consider the prior distribution for coefficient parameter of set of covariates $f$ being:
\[
\lambda_f \thicksim N_{\bigl[ 0, \infty \bigl]}  \bigl(\mu_{\lambda_f},\sigma_{\lambda_f}^2 \bigr)
\]
The parameters $\mu_{\lambda_f}$ and $\sigma_{\lambda_f}^2$ are nuisance parameters, distributed as:
\[
\mu_{\lambda_f} | \sigma_{\lambda_f}^2 \thicksim N_{\bigl[0, \infty \bigl]} \bigl(m_{\lambda}, \tau_{\lambda}\sigma_{\lambda_f}^2  \bigr)  \quad 
\sigma_{\lambda_f}^2 \thicksim \text{\small{Inv}} \chi_{\nu_{\lambda}}^2 
\]
Then the log-posterior distribution presented in equation in (\ref{eq:post}) has to be slightly modified with the introduction of covariates and the extra parameters:
\[
\begin{split}
\log\Bigl( P \bigl( \alpha, \beta, \mathbf{z},\mu_{\alpha}, \mu_{\beta},\sigma_{\alpha}^2 ,\sigma_{\beta}^2, \lambda, \mu_{\lambda}, \sigma_{\lambda}^2 |\mathbf{Y}, \mathbf{X} \bigr) \Bigr) \propto & 
\sum_{k = 1}^K \sum_{i = 1}^n  \sum_{j \neq i} h_{ij}^{(k)} \Bigl[ \arch \bigl(\ak  -\bk  d_{ij}  - \sum_{ f=1}^F \lambda_f x_{ijf} \bigr)\\
& - \log \Bigl( 1 + \exp\bigl\{ \ak  -\bk d_{ij} \sum_{ f=1}^F \lambda_f x_{ijf} \bigr\} \Bigr) \Bigl] \\
& -\frac{1}{2} \Biggl\{ \sum_{i = 1}^n z_i^2 + \frac{\sum_{k=1}^K (\ak -\mu_{\alpha})^2}{\sigma_{\alpha}^2} + \frac{\sum_{k=1}^K (\bk -\mu_{\beta})^2}{\sigma_{\beta}^2} +K \log(\sigma_{\alpha}^2 ) \\
&  +K \log(\sigma_{\beta}^2 ) + \log(\tau_{\alpha}\sigma_{\alpha}^2 ) +\log(\tau_{\beta}\sigma_{\beta}^2 )+\frac{\mu_{\alpha}^2}{\tau_{\alpha}\sigma_{\alpha}^2} +\frac{\mu_{\beta}^2}{\tau_{\beta}\sigma_{\beta}^2} +\frac{1}{\sigma_{\alpha}^2} + \frac{1}{\sigma_{\beta}^2}\\
& -\frac{1}{2}\sum_{f = 1}^F \Biggl( \frac{(\lambda_f - \mu_{\lambda_f})^2}{\sigma_{\lambda_f}^2}
+\log(\sigma_{\lambda_f}^2) +\log( \tau_{\lambda}\sigma_{\lambda_f}^2) + \frac{(\mu_{\lambda_f} - m_{\lambda})^2}{\tau_{\lambda}\sigma_{\lambda_f}^2} + \frac{1}{\sigma_{\lambda_f}^2}  \Biggr) \Biggr\}\\
& + \Bigl( -\frac{\nu_{\alpha} }{2} -1\Bigr)\log(\sigma_{\alpha}^2) + \Bigl( -\frac{\nu_{\beta} }{2} -1\Bigr)\log(\sigma_{\beta}^2)
+\sum_{f = 1}^F  \Bigl( -\frac{\nu_{\lambda_f} }{2} -1\Bigr)\log(\sigma_{\lambda_f}^2) .
\end{split}
\]
Given a set F covariates $f$, the corresponding nuisance parameters 
The nuisance parameters $\mu_{\lambda_f}$ and $\sigma_{\lambda_f}^2$ are update via the following full conditional distributions:
\[ 
\sigma_{\lambda_f}^2 \mid \lambda_f , \mu_{\lambda_f} , \tau_{\lambda} , \nu_{\lambda} \thicksim \text{\small{Inv}} \Gamma \Biggl( \frac{\nu_{\lambda} +2}{2} , \frac{\tau_{\lambda} + \tau_{\lambda}(\lambda_f -\mu_{\lambda_f})^2 + \mu_{\lambda_f}^2 }{2\tau_{\lambda}}\Biggr) ,
\]
\[
\mu_{\lambda_f} \mid \lambda_f , \sigma_{\lambda_f}^2, \tau_{\lambda}, m_{\lambda} \thicksim N_{\bigl[0, \infty \bigl]} \Biggl(\frac{\tau_{\lambda} \lambda_f +m_{\lambda}}{1+\tau_{\lambda}}, \frac{\tau_{\lambda} \sigma_{\lambda_f}^2}{1+\tau_{\lambda} }  \Biggr)  .
\]

The update of the parameters $\lambda_f$ is carried out with a Metropolis-Hastings algorithm, where the proposal distribution used for the parameter $f$ is derived approximating the logarithmic term in the log-posterior with its second order Taylor series expansion in $\lambda_f = \mu_{\lambda_f}$.
Then, the proposal distribution for $\lambda_f$ is:
\[
\tilde{\lambda}_f \thicksim N_{(0, \infty)}\bigl( \mu_{\tilde{\lambda}_f}, \sigma_{\tilde{\lambda}_f}^2 \bigr),
\]
where 
\[
\mu_{\tilde{\lambda}_f} = \sigma_{\tilde{\lambda}_f}^2\Biggl[ \sum_{k = 1}^K \Biggl( \sum_{i = 1}^n \sum_{j \neq i}  \frac{ x_{ijf} \exp \bigl( \ak -\bk d_{ij} -\sum_{l \neq f} \lambda_l x_{ijl} \bigr)}{1 +\exp \bigl( \ak -\bk d_{ij} -\sum_{l \neq f} \lambda_l x_{ijl} \bigr)} -y_{ij}^{(k)} x_{ijf} \Biggr)
\Biggr] +\mu_{\lambda_f},
\]
\[
\sigma_{\tilde{\lambda}_f}^2 =  \Biggl\{ \sum_{k = 1}^K \sum_{i = 1}^n \sum_{j \neq i}  \frac{ x_{ijf}^2 \exp \bigl( \ak -\bk d_{ij} -\sum_{l \neq f} \lambda_l x_{ijl} \bigr)}{\Bigl(1 +\exp \bigl( \ak -\bk d_{ij} -\sum_{l \neq f} \lambda_l x_{ijl} \bigr) \Bigr)^2} +\frac{1}{\sigma_{\lambda_f}^2} \Biggr\}^{-1}.
\]

\subsection{Other proposal distributions}
The introduction of covariates in the model requires a convenient adjustment of the proposal distribution for the latent coordinates, the intercepts and the coefficients.

The proposal distribution for latent position $i$ has the same form as the one shown in section \ref{lat_pos_prop}, but the definition of the auxiliary variables $w_{ij}^{(k)}$ changes in:
\[
w_{ij}^{(k)}= 
\begin{cases}
1 \qquad \text{if} \quad \ak  -\bk d_{ij} -\sum_{f=1}^F \lambda_f x_{ijf}  >0
  \\
 0 \qquad \text{if} \quad \ak  -\bk d_{ij} -\sum_{f=1}^F \lambda_f x_{ijf} \leq 0
   \end{cases},
\]

The mean and variance parameters of the proposal distribution for the intercept terms $\ak$ are modified by the introduction of the covariates, while the form of the distribution is left unchanged:
\[
\mu_{\tilde{\alpha}^{(k)}} = \sigma_{\tilde{\alpha}^{(k)}}^2 \Biggl\{ E^{(k)} -  \sum_{i = 1}^n  \sum_{j \neq i}  \frac{h_{ij}^{(k)} \exp\bigl\{ \mu_{\alpha} -\bk d_{ij} -\sum_{f=1}^F \lambda_f x_{ijf} \bigr\} }{1 + \exp\bigl\{ \mu_{\alpha} -\bk d_{ij} -\sum_{f=1}^F \lambda_f x_{ijf}  \bigr\}} \Biggr\} +\mu_{\alpha},
\]
\[
\sigma_{\tilde{\alpha}^{(k)}}^2 = \Biggl\{ \sum_{i = 1}^n  \sum_{j \neq i} \frac{ h_{ij}^{(k)} \exp\bigl\{ \mu_{\alpha} -\bk d_{ij} -\sum_{f=1}^F \lambda_f x_{ijf} \bigr\} }{\bigl( 1 + \exp\bigl\{ \mu_{\alpha} -\bk d_{ij}-\sum_{f=1}^F \lambda_f x_{ijf}  \bigr\}\bigr)^2}  +\frac{1}{\sigma_{\alpha}^2}\Biggr\}^{-1}
\]
The mean and variance parameters of the proposal distribution for the coefficient parameters $\bk$ changes, though the form of the distribution remains the same:
\[
 \mu_{\tilde{\beta}^{(k)}} = \sigma_{\tilde{\beta}^{(k)}}^2 \Biggl\{  \sum_{i = 1}^n  \sum_{j \neq i} h_{ij}^{(k)} d_{ij} \Biggl( \frac{\exp\bigl\{ \ak -\mu_{\beta} d_{ij} -\sum_{f=1}^F \lambda_f x_{ijf} \bigr\} }{1 + \exp\bigl\{ \ak -\mu_{\beta} d_{ij} -\sum_{f=1}^F \lambda_f x_{ijf} \bigr\}}  -\arch  \Biggr) \Biggr\} +\mu_{\beta},
\]
\[
 \sigma_{\tilde{\beta}^{(k)}}^2 = \Biggl\{ \sum_{i = 1}^n  \sum_{j \neq i} \frac{ h_{ij}^{(k)}  d_{ij}^2 \exp\bigl\{ \ak  -\mu_{\beta} d_{ij} -\sum_{f=1}^F \lambda_f x_{ijf}  \bigr\} }{\bigl( 1 + \exp\bigl\{ \ak -\mu_{\beta} d_{ij} -\sum_{f=1}^F \lambda_f x_{ijf}\bigr\}\bigr)^2}  +\frac{1}{\sigma_{\beta}^2}\Biggr\}^{-1}
\]

\clearpage
\section{ISO3 codes}
\label{app2}
\begin{table}[!h]
\centering
\label{tab:sim2-2}
\small
\begin{tabular}{@{}lclc@{}}
\toprule
Country name        & iso3 code &Country name        & iso3 code  \\
\midrule
Albania & ALB & Italy & ITA\\
Andorra & AND & Latvia & LVA\\
Armenia & ARM & Lithuania & LTU \\
Australia & AUS & Malta & MLT \\  
Austria & AUT & Moldova & MDA\\
Azerbaijan & AZE & Monaco & MCO\\
Belarus & BLR & Montenegro & MNE\\
Belgium & BEL & Norway & NOR\\
Bosnia and Herzegovina & BIH & Poland & POL\\
Bulgaria & BGR & Portugal & PRT\\
Croatia & HRV & Romania & ROU\\
Cyprus & CYP  & Russia & RUS\\
Czech Republic & CZE & San Marino & SMR\\
Denmark & DNK & Serbia & SRB\\
Estonia & EST & Serbia and Montenegro & SCG\\
Federal Republic of Macedonia & MKD & Slovakia & SVK\\
Finland & FIN  & Slovenia & SVN\\
France & FRA & Spain & ESP \\
Georgia & GEO & Sweden & SWE\\
Germany & DEU & Switzerland & CHE\\
Greece & GRC & Netherlands & NLD\\
Hungary & HUN   & Turkey & TUR\\
Iceland & ISL  &  Ukraine & UKR\\
Ireland & IRL & United Kingdom & GBR\\    
Israel & ISR  & &\\
\bottomrule
\end{tabular}
\end{table}

\clearpage
\section{Results for the Eurovision sub-periods}
\label{app3}
\begin{table}[!h]
\centering
\label{tab:par-98-07}
\small
\begin{tabular}{@{}lllllllllllll@{}}
\toprule
Year  & $\hat{\alpha}$ & $sd(\alpha)$ & $\hat{\beta}$ & $sd(\beta)$ &&
Year  & $\hat{\alpha}$ & $sd(\alpha)$ & $\hat{\beta}$ & $sd(\beta)$  \\
\midrule
1998 & $0$    &   -    & $1$    & -      && 2003 & $0.25$ & $0.17$ & $0.79$ & $0.19$ \\
1999 & $0.19$ & $0.17$ & $0.37$ & $0.15$ && 2004 & $0.31$ & $0.19$ & $0.67$ & $0.21$ \\
2000 & $0.32$ & $0.18$ & $0.62$ & $0.17$ && 2005 & $0.33$ & $0.16$ & $0.71$ & $0.17$ \\
2001 & $0.01$ & $0.15$ & $0.19$ & $0.12$ && 2006 & $0.40$ & $0.18$ & $0.76$ & $0.19$ \\
2002 & $0.15$ & $0.19$ & $0.40$ & $0.17$ && 2007 & $0.63$ & $0.15$ & $1.10$ & $0.18$ \\
\bottomrule
\end{tabular}
\caption{Estimated averages and standard deviations for the network parameters in the multiplex 1998-2007. }
\end{table}

\begin{table}[!h]
\centering
\label{tab:par-08-15}
\small
\begin{tabular}{@{}lllllllllll@{}}
\toprule
Year  & $\hat{\alpha}$ & $sd(\alpha)$ & $\hat{\beta}$ & $sd(\beta)$ &&
Year  & $\hat{\alpha}$ & $sd(\alpha)$ & $\hat{\beta}$ & $sd(\beta)$
 \\
\midrule
2008 & $0$    &   -    & $1$    & -      && 2012 & $0.44$ & $0.14$ & $0.96$ & $0.15$ \\
2009 & $0.21$ & $0.14$ & $0.60$ & $0.13$ && 2013 & $0.40$ & $0.13$ & $0.93$ & $0.14$ \\
2010 & $0.40$ & $0.14$ & $0.80$ & $0.13$ && 2014 & $0.28$ & $0.16$ & $0.80$ & $0.16$ \\
2011 & $0.35$ & $0.13$ & $0.77$ & $0.13$ && 2015 & $0.35$ & $0.12$ & $1.01$ & $0.14$ \\
\bottomrule
\end{tabular}
\caption{Estimated averages and standard deviations for the network parameters in the multiplex 2008-2015. }
\end{table}

\begin{figure}[!h]%
    \centering
    \subfloat[1998-2007.]{{\includegraphics[width=8.2cm]{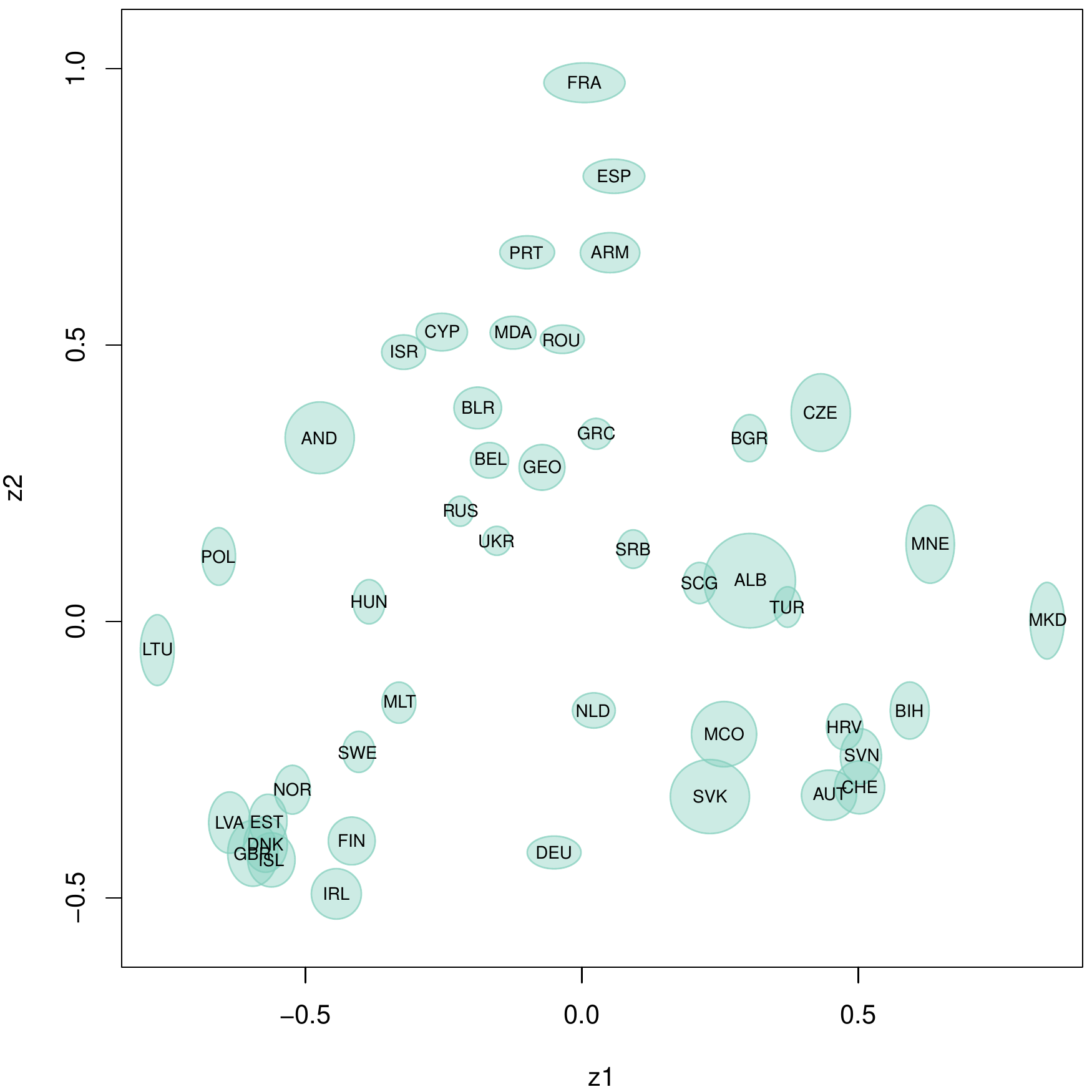} }}%
    \quad 
    \subfloat[2008-2015.]{{\includegraphics[width=8.2cm]{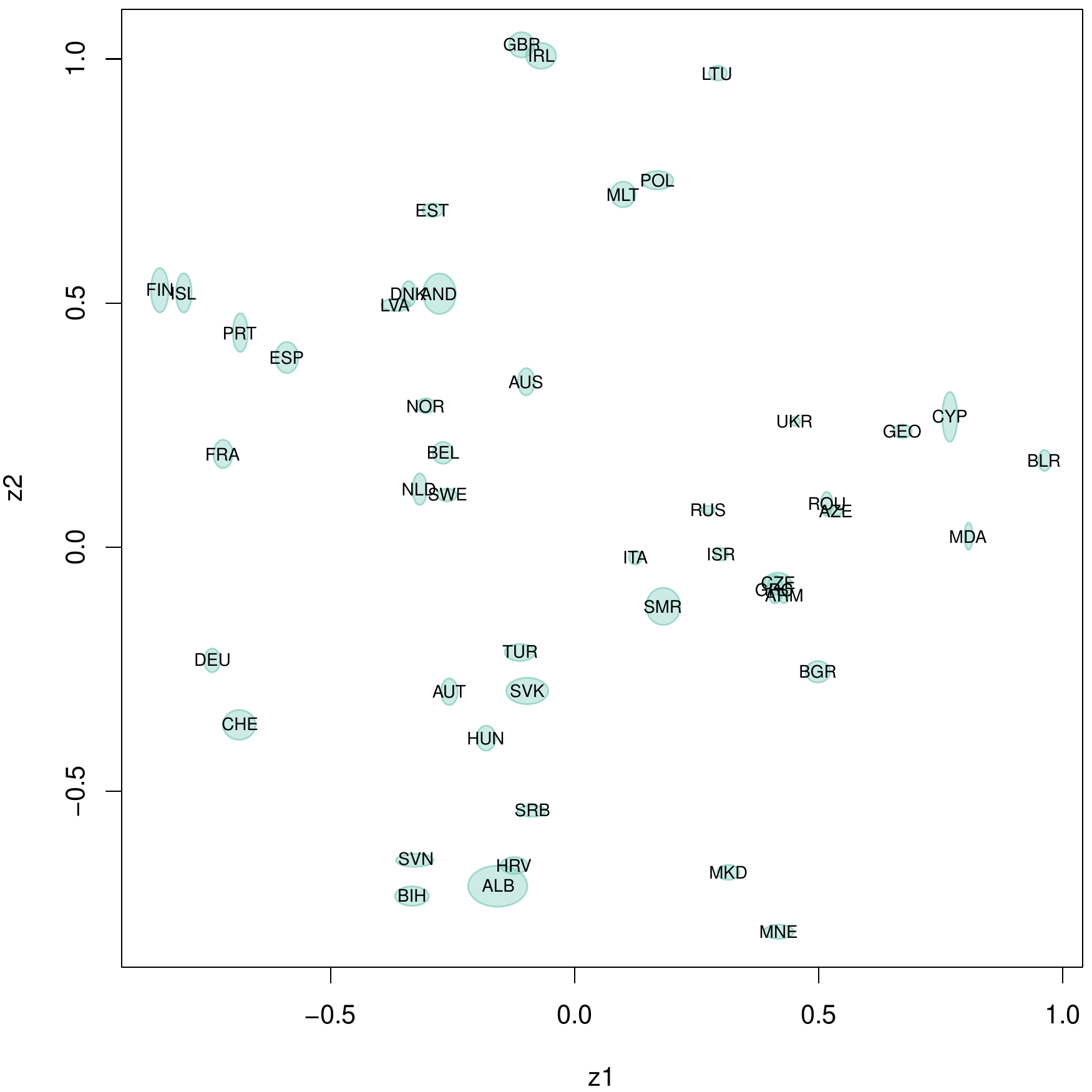} }}%
     \caption{Estimated latent positions.}
\end{figure}

\begin{figure}[!h]%
    \centering
    \subfloat[1998-2007.]{{\includegraphics[width=8.2cm]{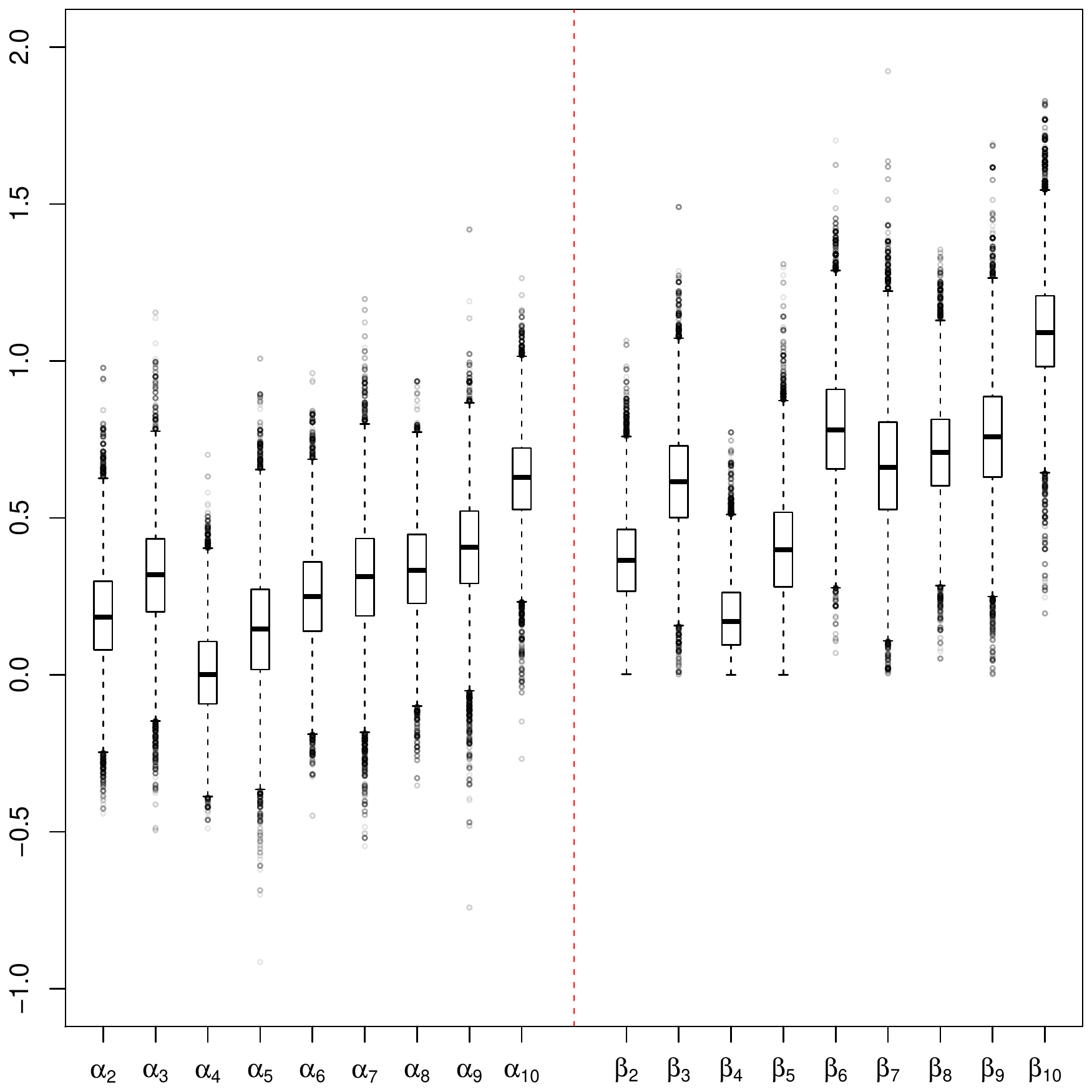} }}%
    \quad 
    \subfloat[2008-2015]{{\includegraphics[width=8.2cm]{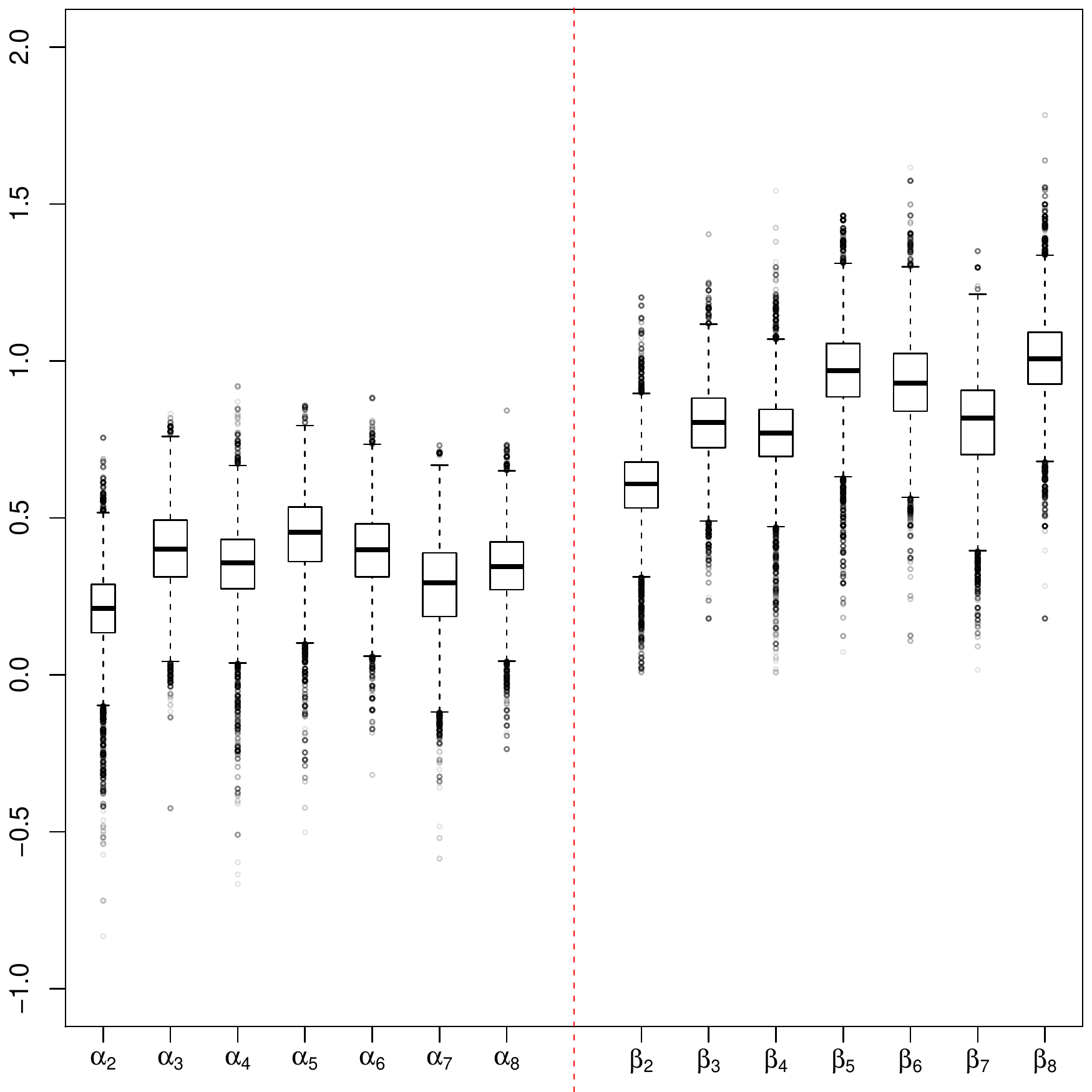} }}
     \caption{Boxplots for the estimates of the logistic parameters.}
\end{figure}

\begin{figure}%
    \centering
{{\includegraphics[width=18.1cm]{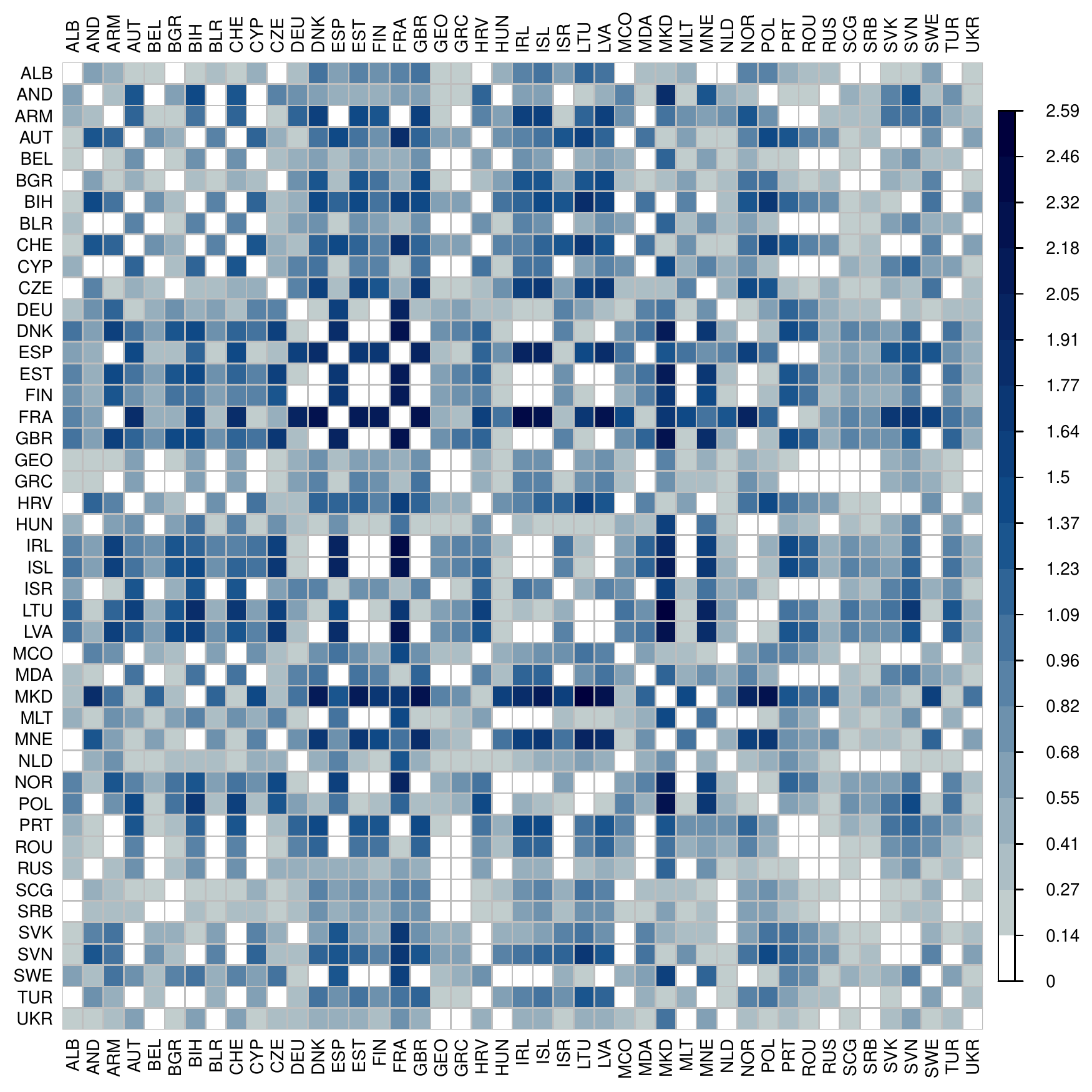} }}%
    \caption{Estimated distances between couple of countries for the period 1998-2007.}
\end{figure}

\begin{figure}%
    \centering
{{\includegraphics[width=18.1cm]{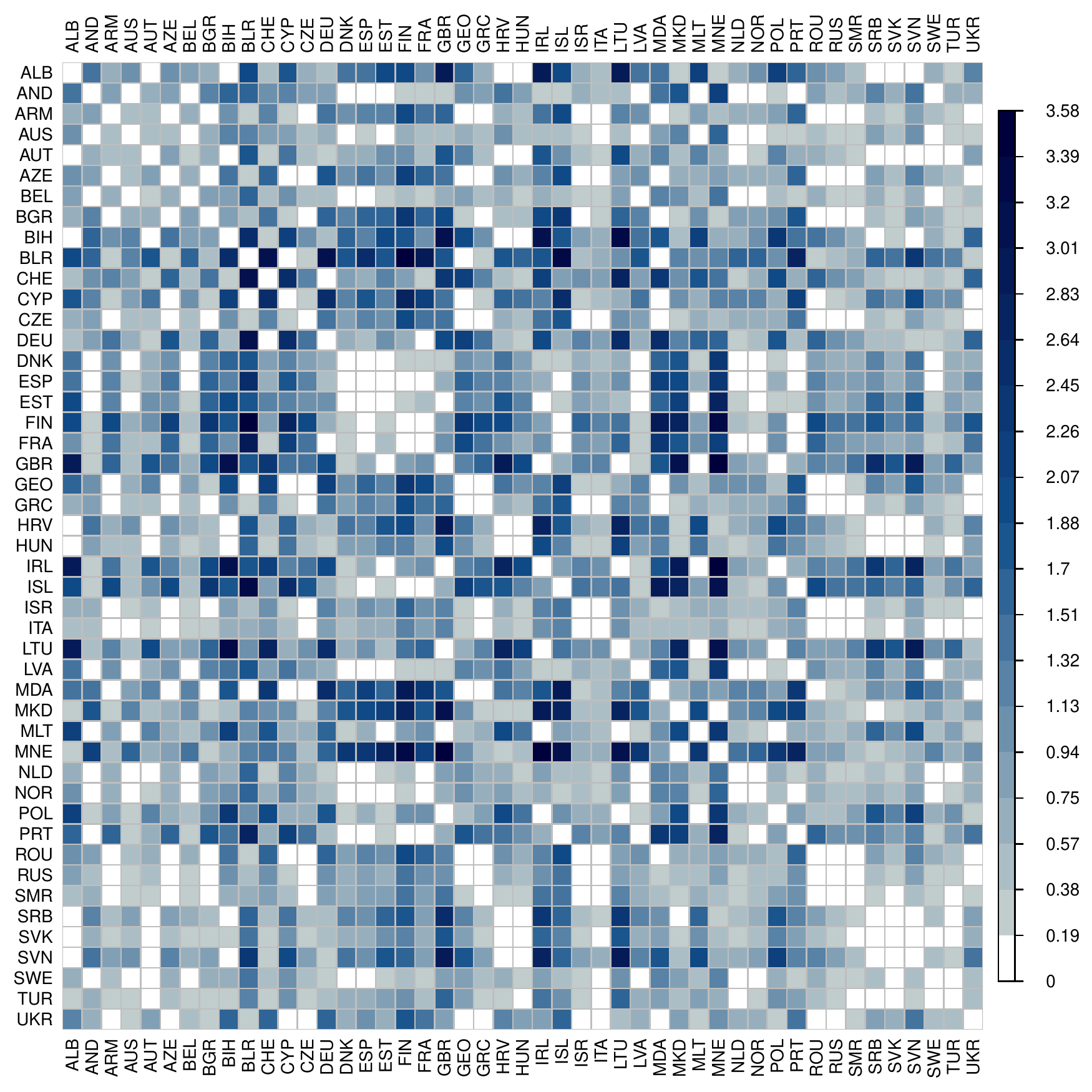} }}%
    \caption{Estimated distances between couple of countries for the period 2008-2015.}
\end{figure}

\begin{figure}[!h]%
    \centering
    \subfloat[$neigh = 1$]{{\includegraphics[width=7cm]{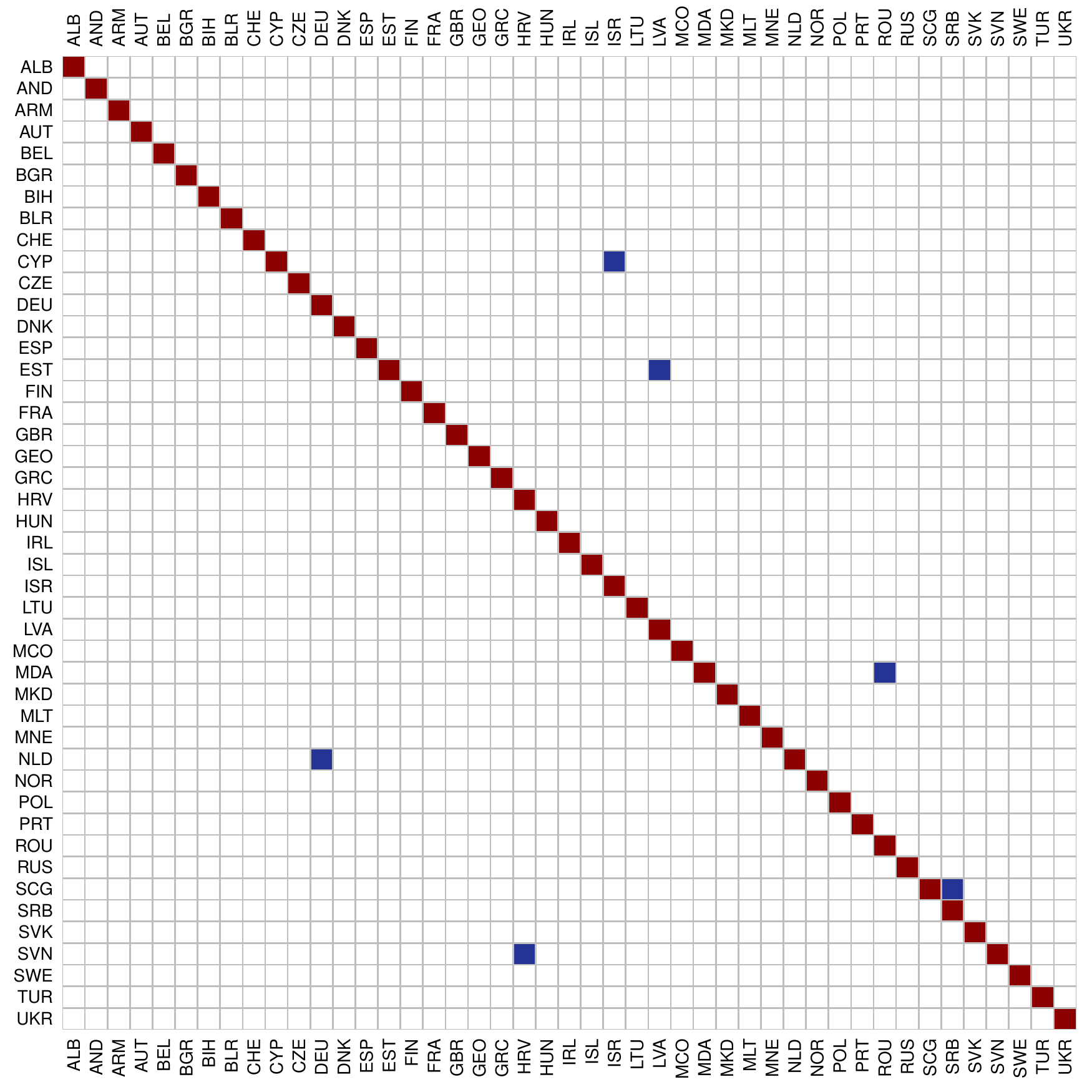} }}%
    \quad 
    \subfloat[$neigh = 2$]{{\includegraphics[width=7cm]{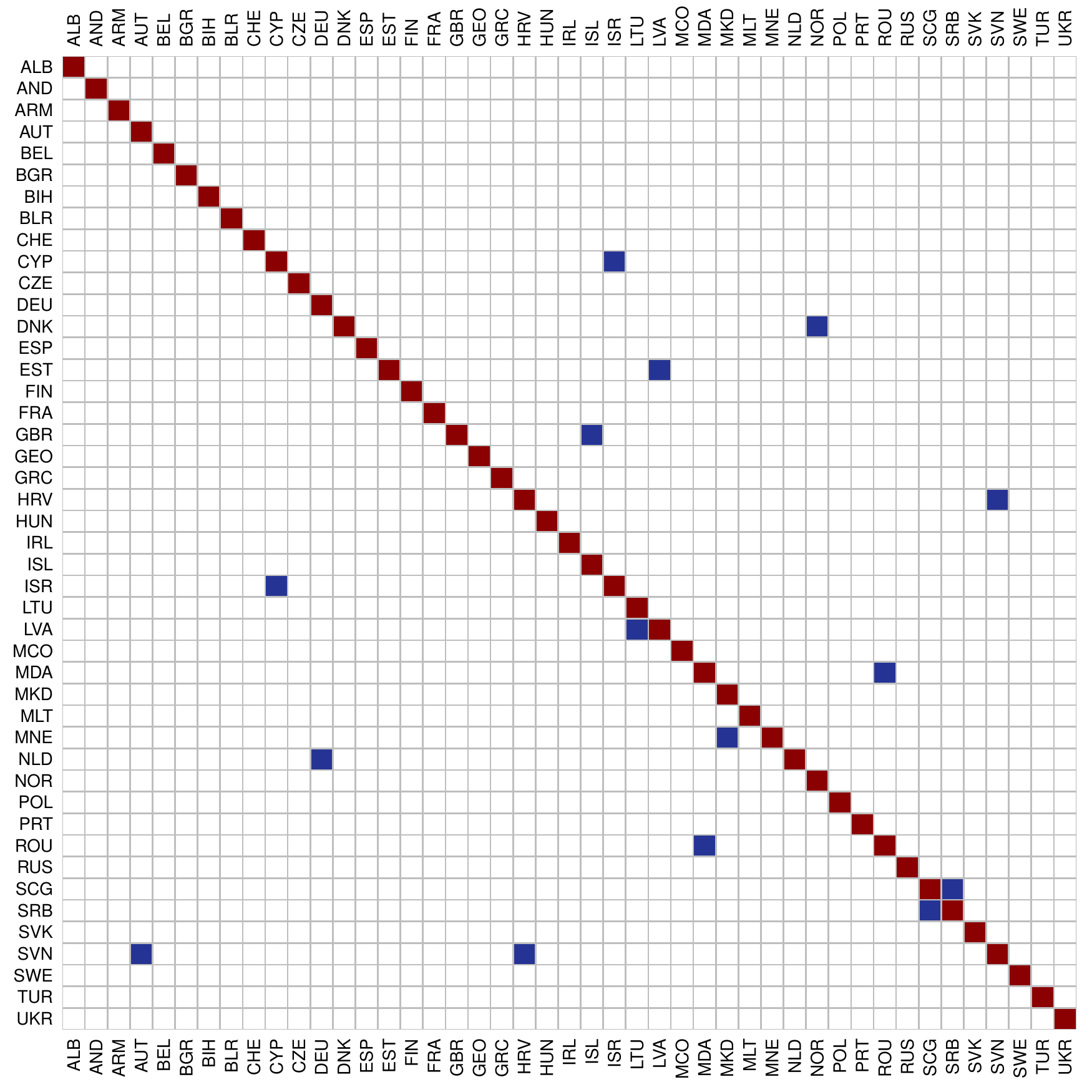} }}%
    \qquad
    \subfloat[$neigh = 3$]{{\includegraphics[width=7cm]{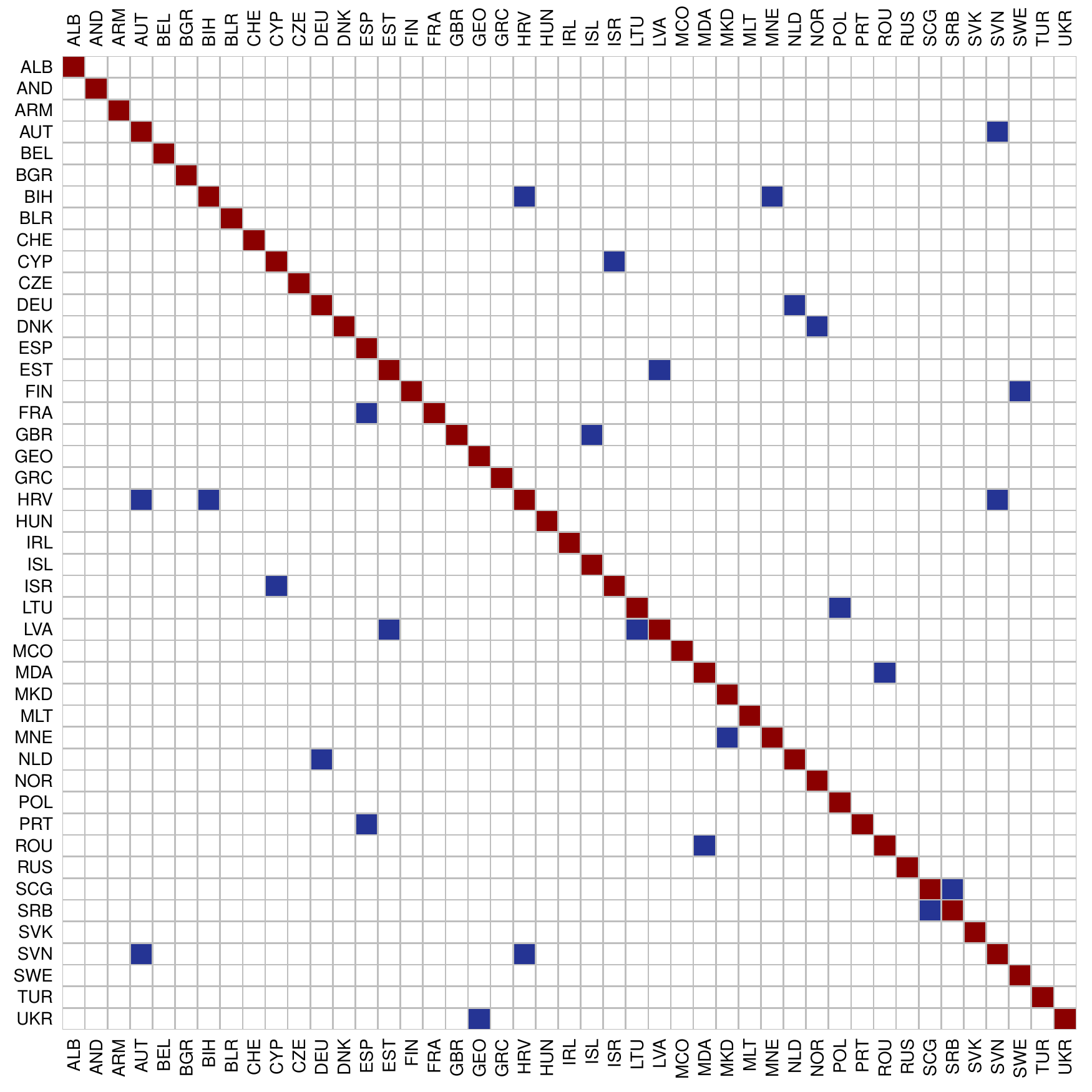} }}%
    \quad 
    \subfloat[$neigh = 5$]{{\includegraphics[width=7cm]{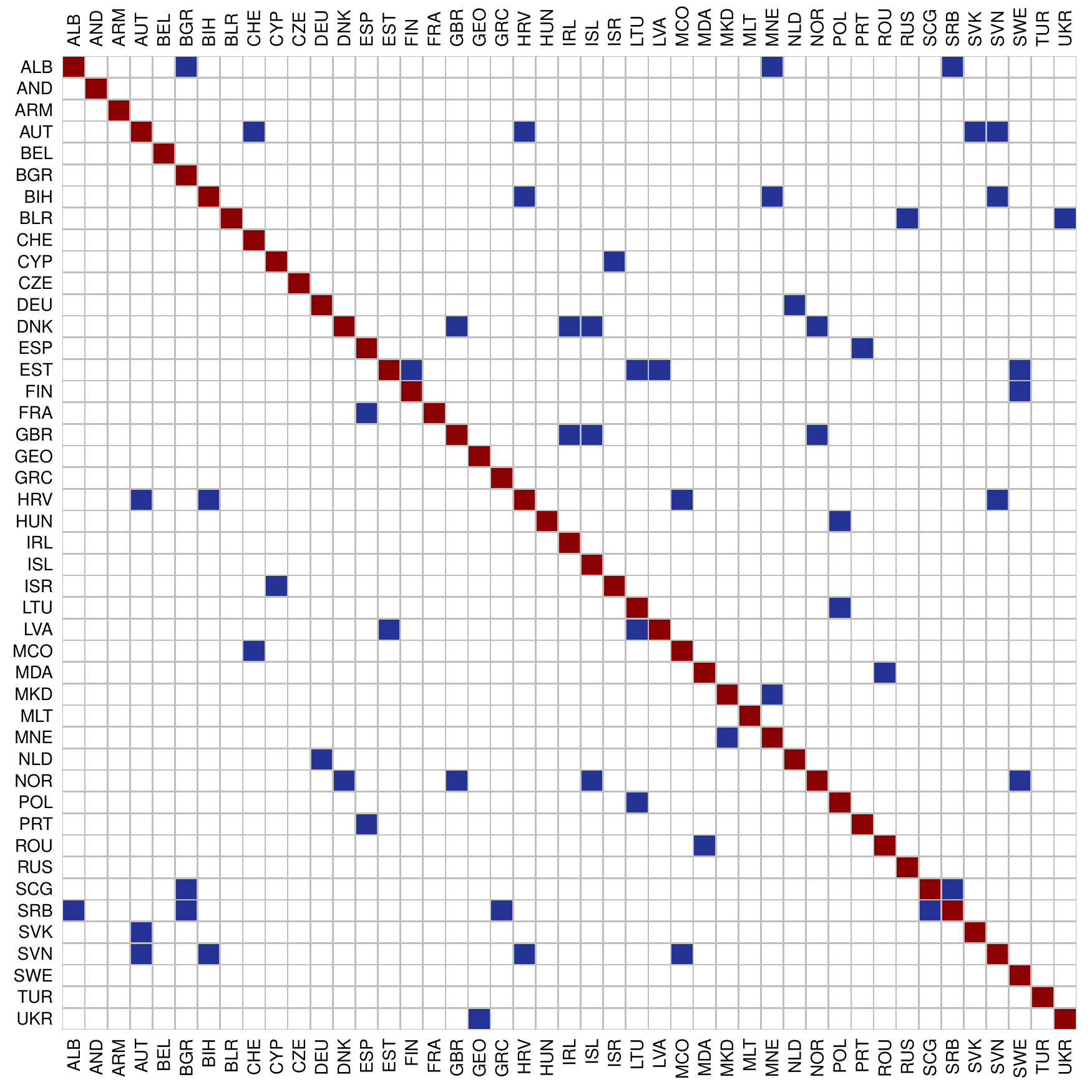} }}%
    \qquad
    \subfloat[$neigh = 10$]{{\includegraphics[width=7cm]{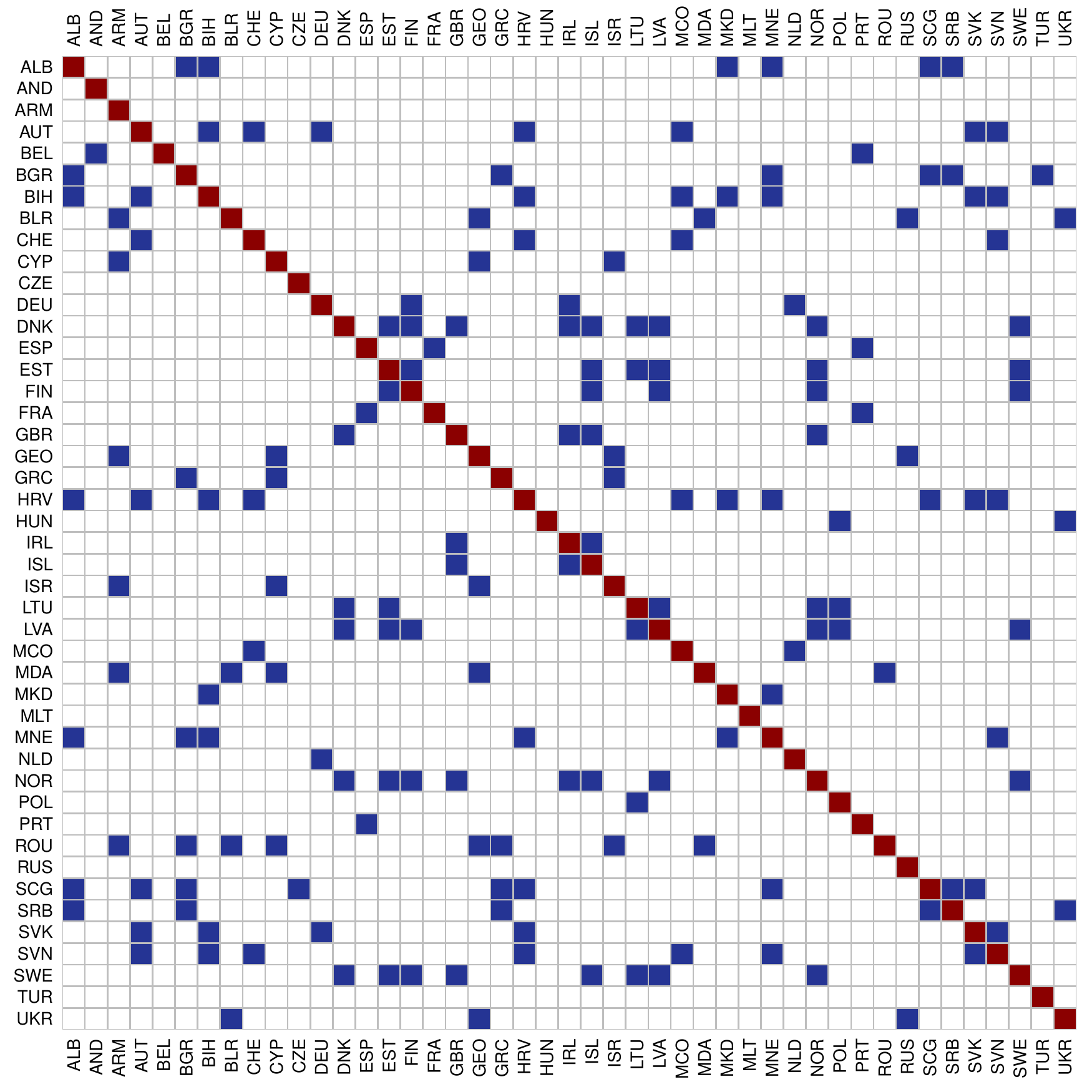} }}%
    \quad 
    \subfloat[$neigh = 15$]{{\includegraphics[width=7cm]{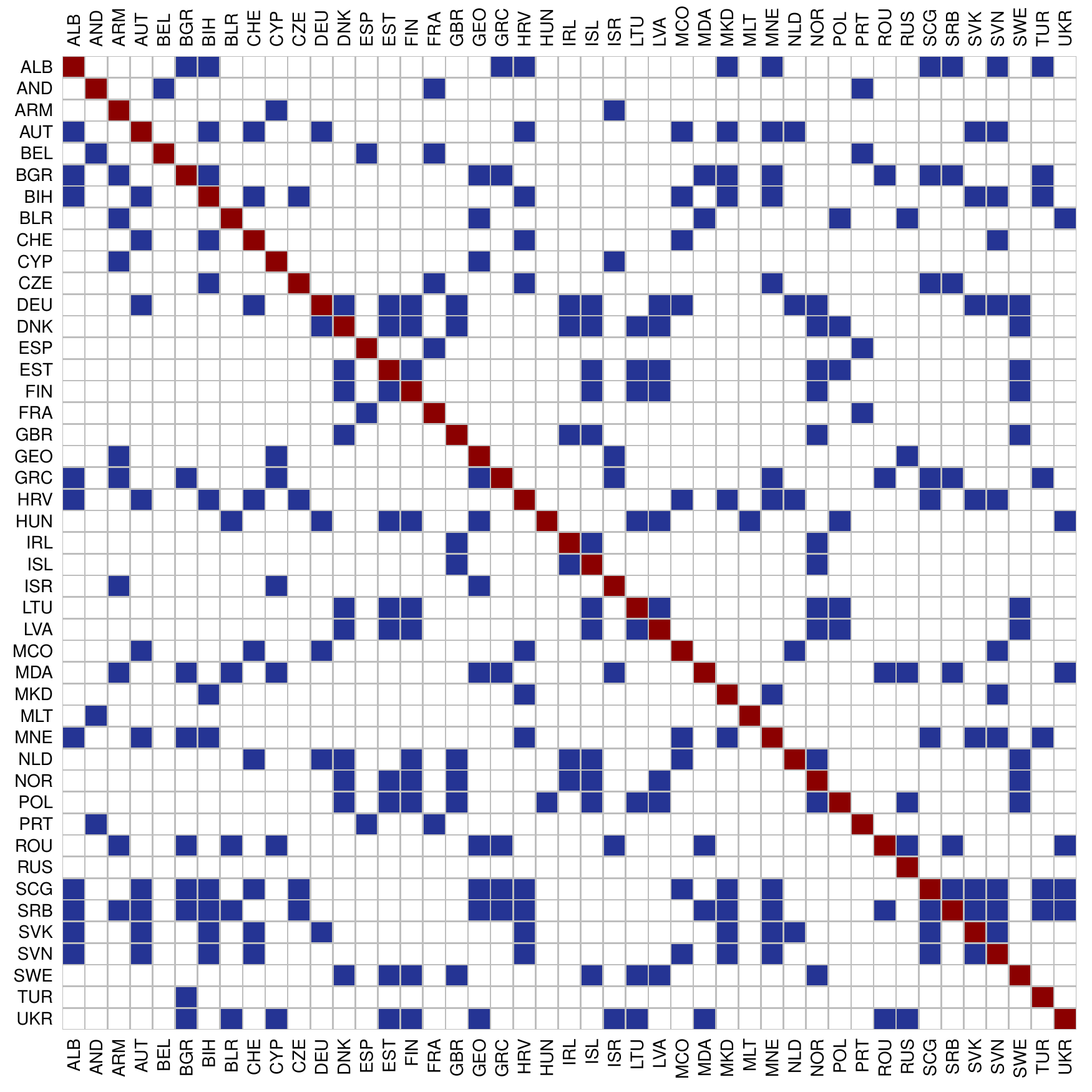} }}%
    \caption{neigh plot dist lat e go 1998-2007.}
\end{figure}

\begin{figure}%
    \centering
{{\includegraphics[width=18.1cm]{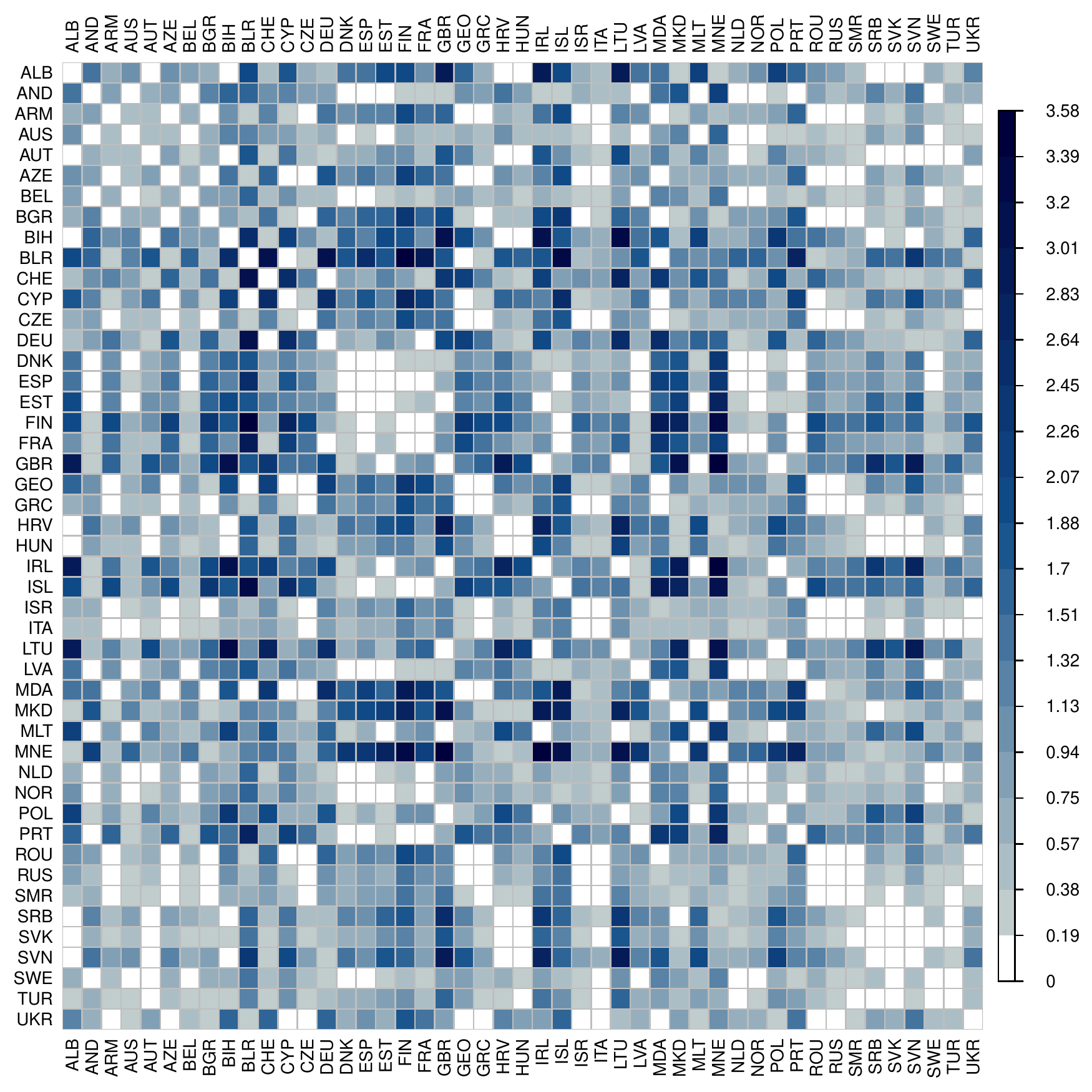} }}%
    \caption{Estimated distances between couple of countries for the period 2008-2015.}
\end{figure}

\begin{figure}[!h]%
    \centering
    \subfloat[$neigh = 1$]{{\includegraphics[width=7cm]{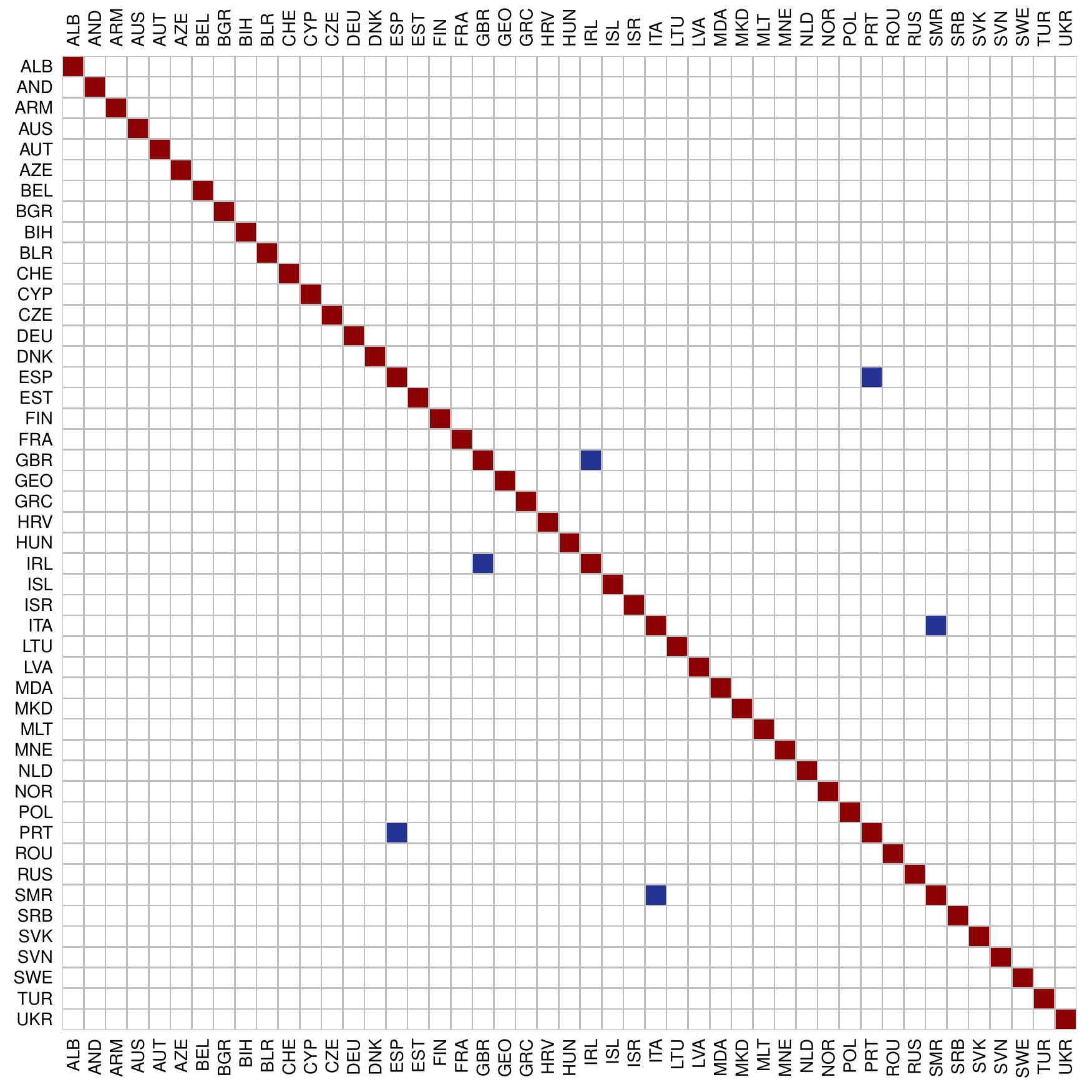} }}%
    \quad 
    \subfloat[$neigh = 2$]{{\includegraphics[width=7cm]{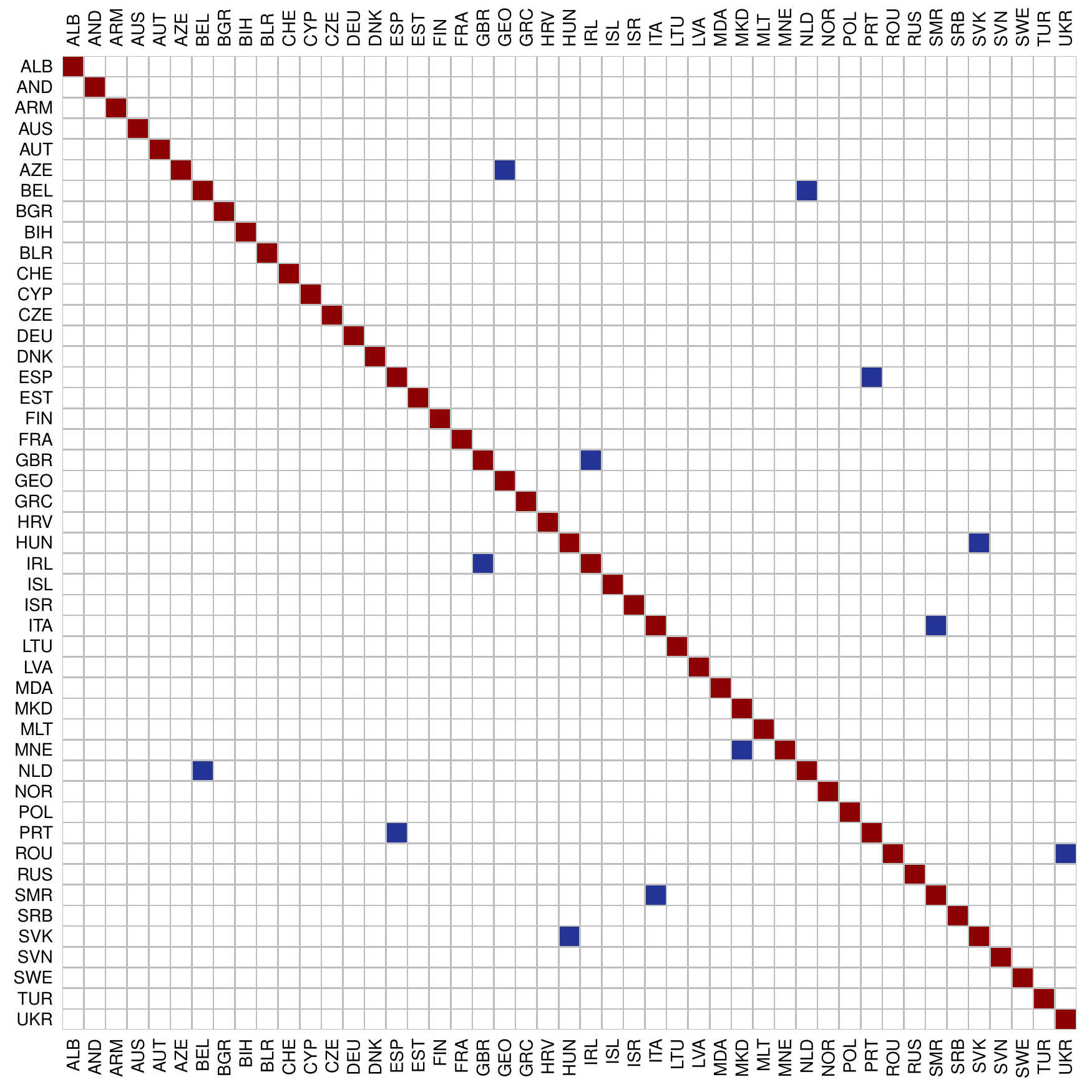} }}%
    \qquad
    \subfloat[$neigh = 3$]{{\includegraphics[width=7cm]{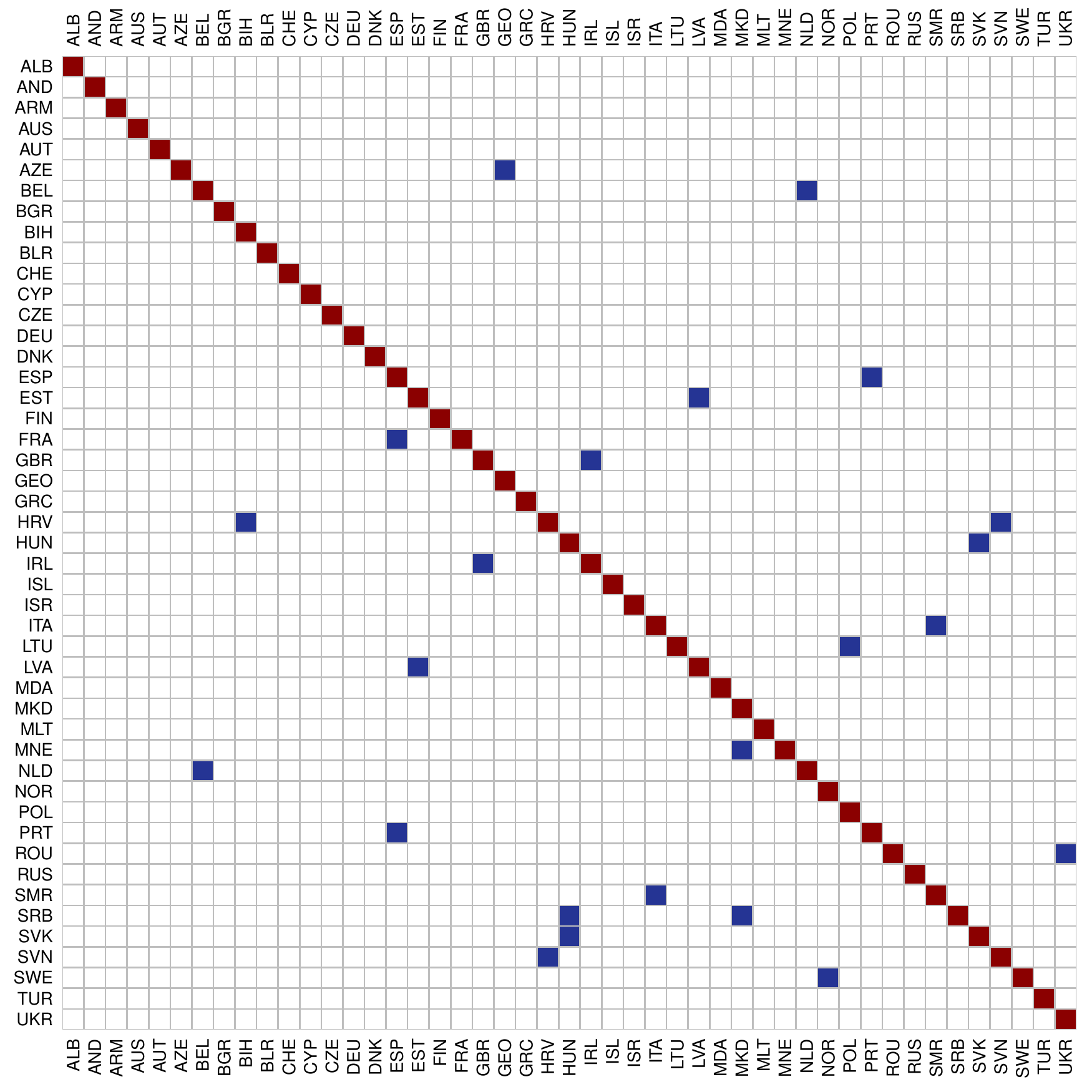} }}%
    \quad 
    \subfloat[$neigh = 5$]{{\includegraphics[width=7cm]{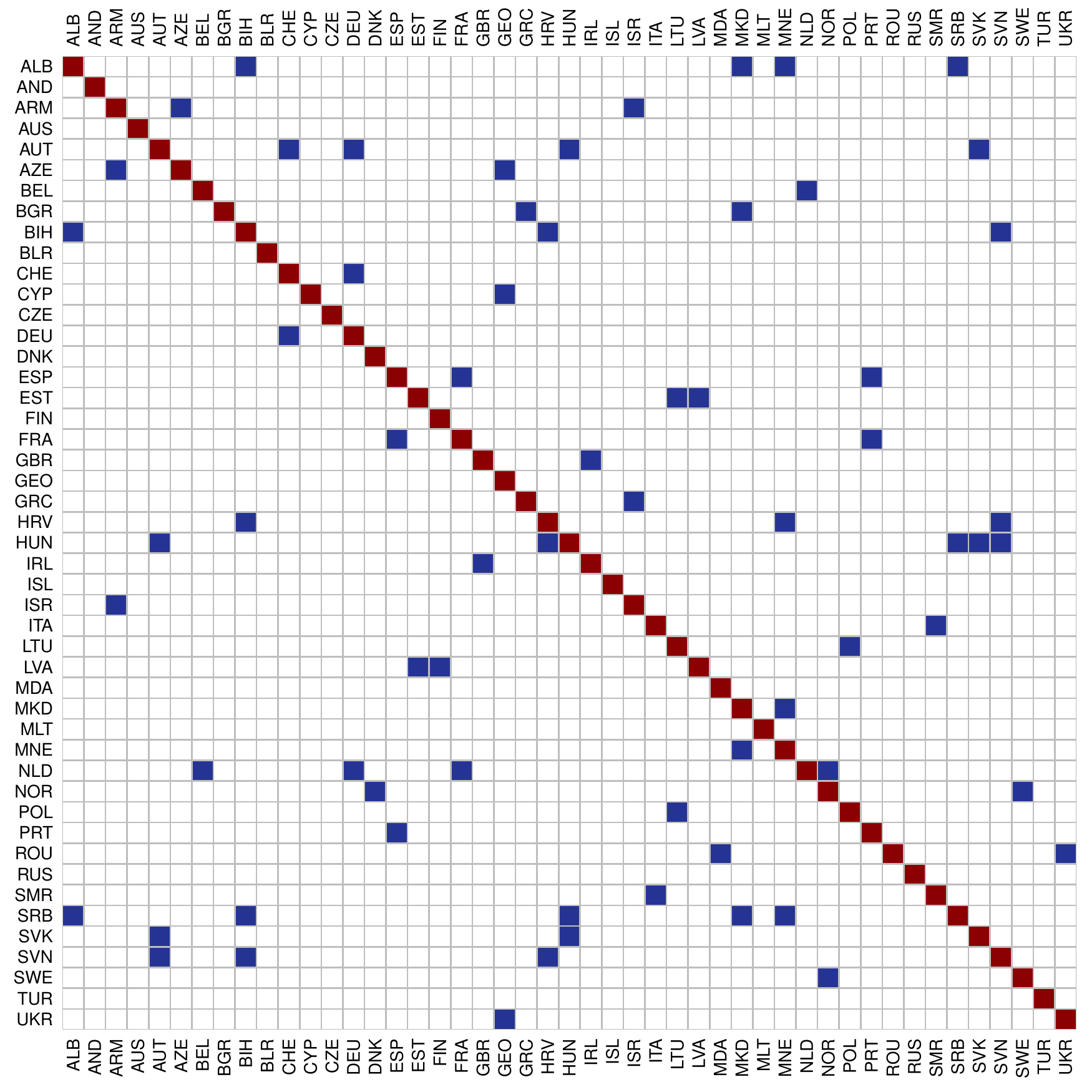} }}%
    \qquad
    \subfloat[$neigh = 10$]{{\includegraphics[width=7cm]{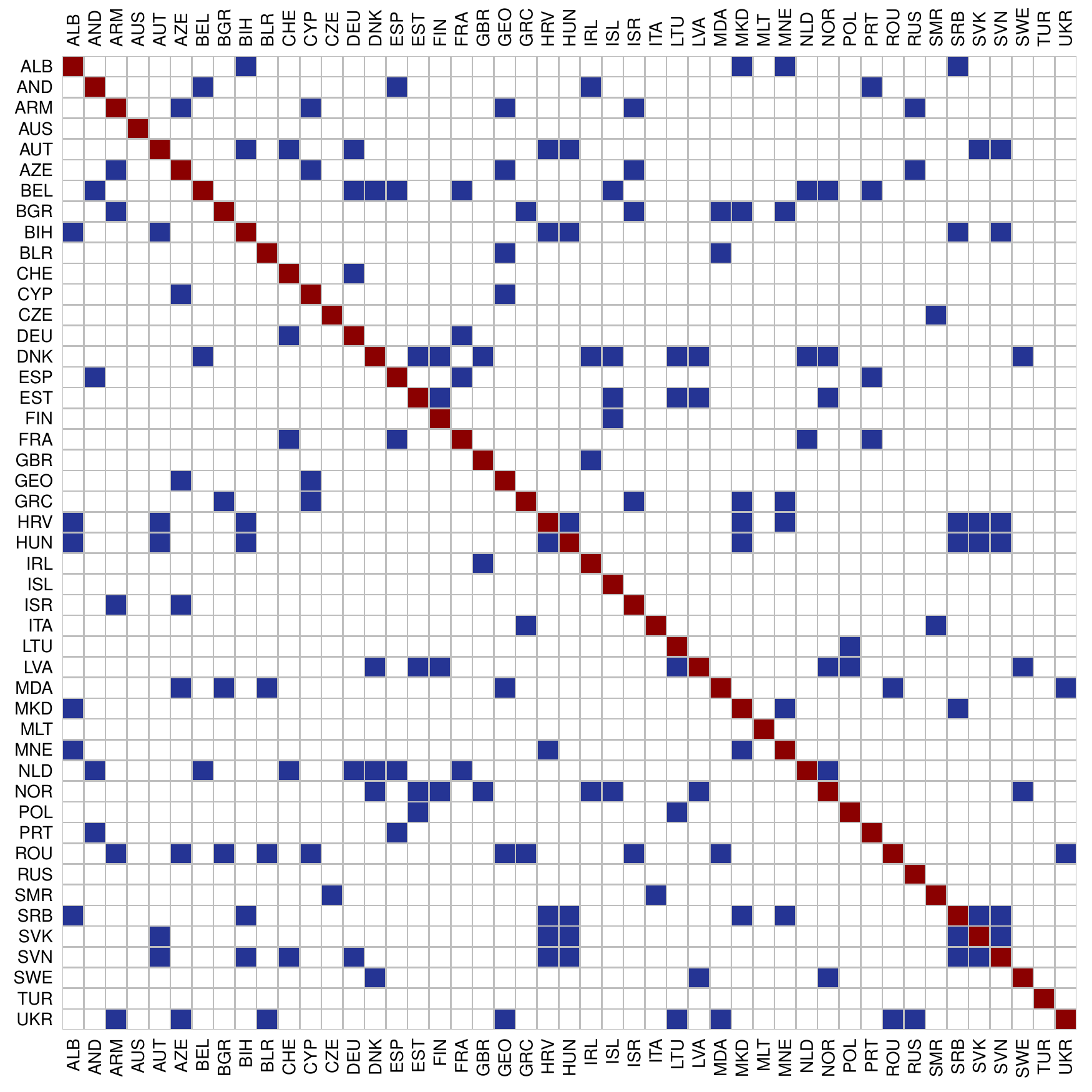} }}%
    \quad 
    \subfloat[$neigh = 15$]{{\includegraphics[width=7cm]{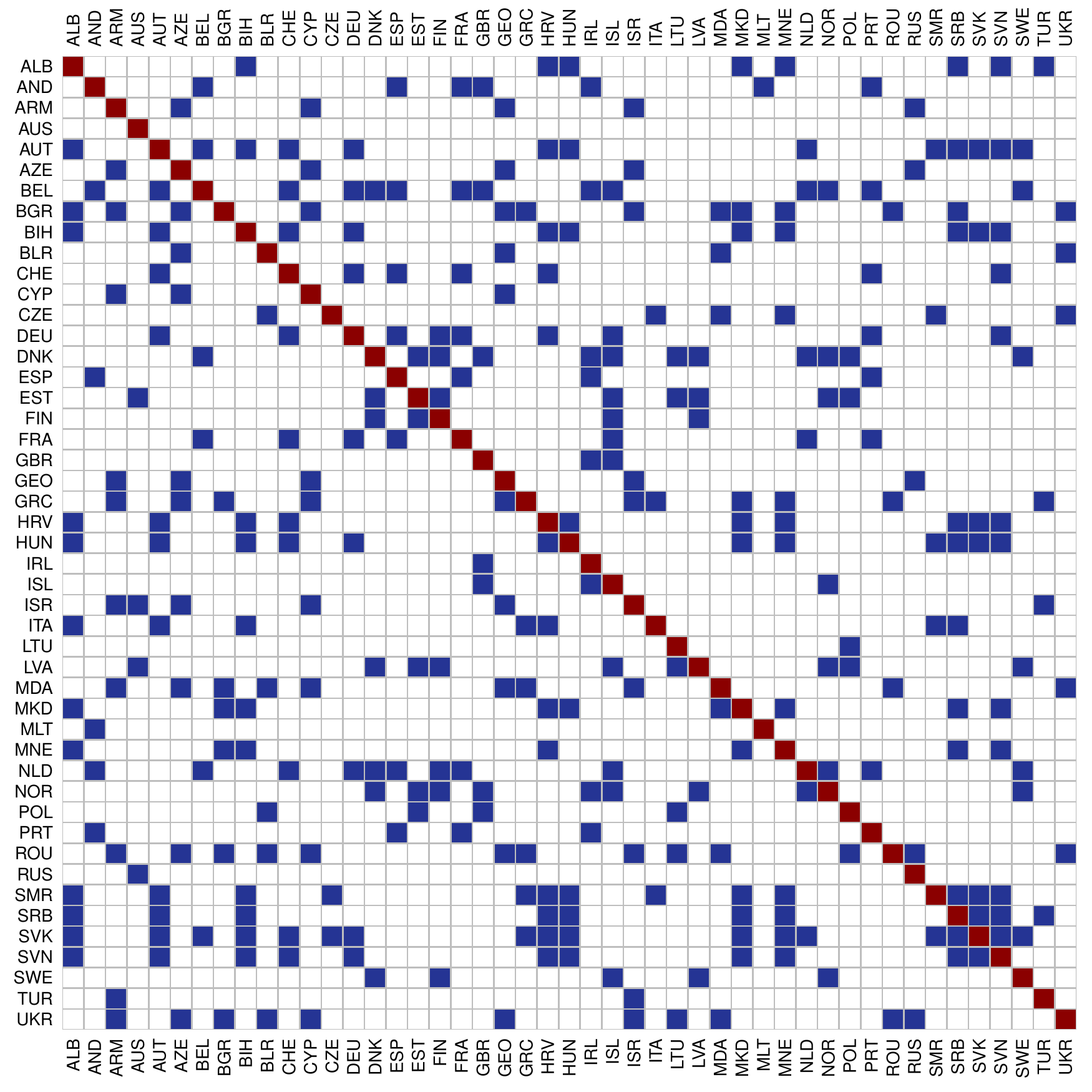} }}%
    \caption{neigh plot dist lat e go 2008-2015.}
\end{figure}




\clearpage
\section{Simulations results}
\label{app4}
\subsection{Results for block I}
\begin{table}[!h]
\centering
\caption{Simulated values for the intercept and the coefficient terms in the multidimensional networks considered in scenarios I-III.}
\label{tab:val_veri_b1}
\small
\scalebox{.8}{
\begin{tabular}{@{}lllllllll@{}}
\toprule   
 & \multicolumn{2}{c}{$K =2$} & \multicolumn{2}{c}{$K =3$} &\multicolumn{2}{c}{$K =4$} & \multicolumn{2}{c}{$K =5$} \\
Multiplex & $\alpha$ & $\beta$ & $\alpha$ & $\beta$ & $\alpha$ & $\beta$ & $\alpha$ & $\beta$ \\
\midrule 
$n= 25,K =3$ & $-0.22$ & $0.91$ &  $0.69$ & $0.22$ & - & - & - & -\\
$n= 50,K =3$ & $0.51$  & $0.68$ & $-0.83$ & $0.12$ & - & - & - & -\\
$n= 100, K =3$& $0.21$ & $0.70$ & $-0.74$ & $1.09$ & - & - & - & -\\
$n= 50, K =5$ & $1.10$  & $1.36$ & $0.23$ & $0.45$  & $0.47$ & $0.07$ & $-0.52$ & $0.95$\\
\bottomrule
\end{tabular}
}
\end{table}
\subsubsection{First scenario}
\label{app3_sim}
\begin{table}[!h]
\centering
\caption{Multivariate Gaussian latent coordinates. Averages for the estimated logistic parameters and the procrustes correlation between true and estimated latent spaces.}
\small
\begin{tabular}{@{}lllllllllllllll@{}}
\toprule   
        & $\alpha^{(1)}$  & $\hat{\alpha}^{(2)}$ & $\hat{\alpha}^{(3)}$& $\hat{\alpha}^{(4)}$   & $\hat{\alpha}^{(5)}$&  & $\beta^{(1)}$  & $\hat{\beta}^{(2)}$  & $\hat{\beta}^{(3)}$ & $\hat{\beta}^{(4)}$  & $\hat{\beta}^{(5)}$ & & PC &                
         \\
\midrule
\multirow{2}{*}{$n= 25$, $K =3$}  & 0 & $-0.19$ & $0.63$ & - & - &  & 1 & $1.00$ & $0.22$ & - & - & & $0.92$ &\texttt{P}\\
  & 0 & $-0.25$ & $0.78$ & - & - &  & 1 & $1.00$ & $0.27$ & - & - & & $0.91$ & \texttt{A}                        \\
\multirow{2}{*}{$n= 50$, $K =3$}  & 0 & $0.18$ & $-0.84$ & - & - &  & 1 & $0.58$ & $0.12$ & - & - & & $0.91$ & \texttt{P}\\
  & 0 & $0.39$ & $-0.72$ & - & - &  & 1 & $0.65$ & $0.14$ & - & - & & $0.96$ & \texttt{A}                        \\
\multirow{2}{*}{$n= 100$, $K =3$}  & $0$ & $0.09$ & $-0.97$ & - & - &  & 1 & $0.90$ & $1.32$ & - & - & & $0.95$  & \texttt{P}\\
  & 0 & $0.16$ & $-0.95$ & - & - &  & 1 & $0.60$ & $0.85$ & - & - & & $0.96$ & \texttt{A}                        \\
\multirow{2}{*}{$n= 50$, $K =5$}  & 0 & $0.83$ & $0.16$ & $0.60$ & $-0.71$ &  & 1 & $1.35$ & $0.48$ & $0.09$ & $1.05$ & & $0.97$ & \texttt{P}\\
  & 0 & $0.85$ & $0.15$ & $0.64$ & $-0.73$ &  & 1 & $1.39$ & $0.49$ & $0.09$ & $1.01$ & & $0.96$ & \texttt{A}                        \\
\bottomrule
\end{tabular}
\end{table}
\begin{table}[!h]
\centering
\caption{Multivariate Gaussian latent coordinates. Standard deviations for the estimated logistic parameters and the procrustes correlation between true and estimated latent spaces.}
\small
\begin{tabular}{@{}lllllllllllllll@{}}
\toprule
    & $sd(\hat{\alpha}^{(2)})$ & $sd(\hat{\alpha}^{(3)})$& $sd(\hat{\alpha}^{(4)})$ & $sd(\hat{\alpha}^{(5)})$&  & $sd(\hat{\beta}^{(2)})$  & $sd(\hat{\beta}^{(3)})$ & $sd(\hat{\beta}^{(4)})$  & $sd(\hat{\beta}^{(5)})$ & & sd(PC) &                
         \\
\midrule
\multirow{2}{*}{$n= 25$, $K =3$}   & $0.06$ & $0.04$ & - & - &  & $0.05$ & $0.03$ & - & - & & $0.02$ &\texttt{P}\\
   & $0.02$ & $0.01$ & - &   & - & $0.03$ & $0.02$ & - & - & & $0.05$ & \texttt{A}                        \\
\multirow{2}{*}{$n= 50$, $K =3$}  & $0.28$ & $0.04$ & - & - &  & $0.11$ & $0.01$ & - & - & & $0.06$ & \texttt{P}\\
  & $0.05$ & $0.02$ & - & - &   & $0.04$ & $0.01$ & - & - & & $0.01$ & \texttt{A}                        \\
\multirow{2}{*}{$n= 100$, $K =3$}   & $0.07$ & $0.09$ & - & - &   & $0.06$ & $0.10$ & - & - & & $0.02$ & \texttt{P}\\
  & $0.05$ & $0.06$ & - &  & - & $0.05$ & $0.08$ & - & - & & $0.02$ & \texttt{A}                        \\
\multirow{2}{*}{$n= 50$, $K =5$}   & $0.08$ & $0.03$ & $0.03$ & $0.04$ &  & $0.16$ & $0.08$ & $0.01$ & $0.11$ & & $0.01$ & \texttt{P}\\
  & $0.11$ & $0.04$ & $0.02$ & $0.09$ &   & $0.16$ & $0.04$ & $0.01$ & $0.13$ & & $0.01$ & \texttt{A}                        \\
\bottomrule
\end{tabular}
\end{table}
\clearpage

\begin{figure}%
    \centering
    \subfloat[$n=25, K=3$]{{\includegraphics[width=8cm]{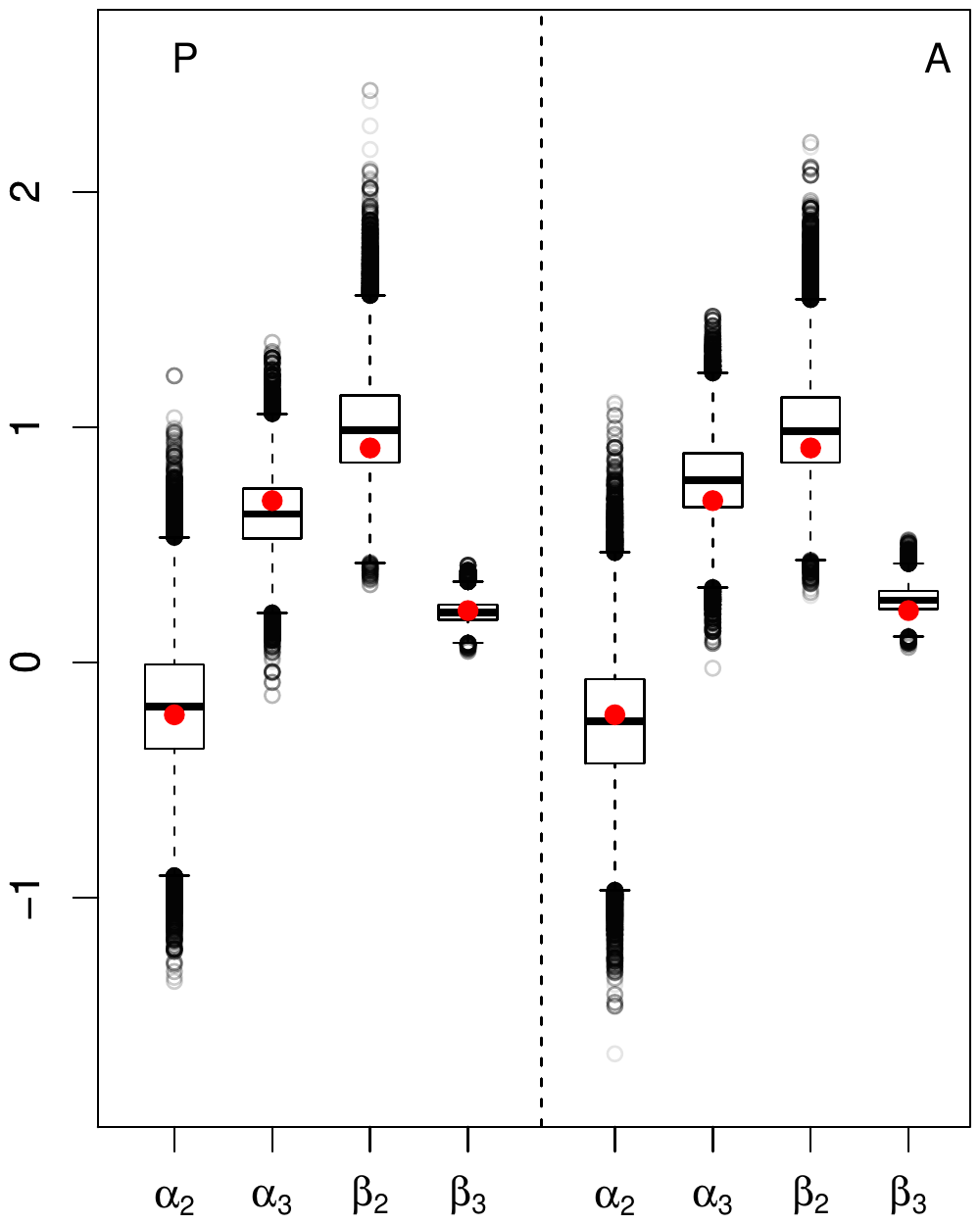} }}%
    \qquad 
    \subfloat[$n=50, K=3$]{{\includegraphics[width=8cm]{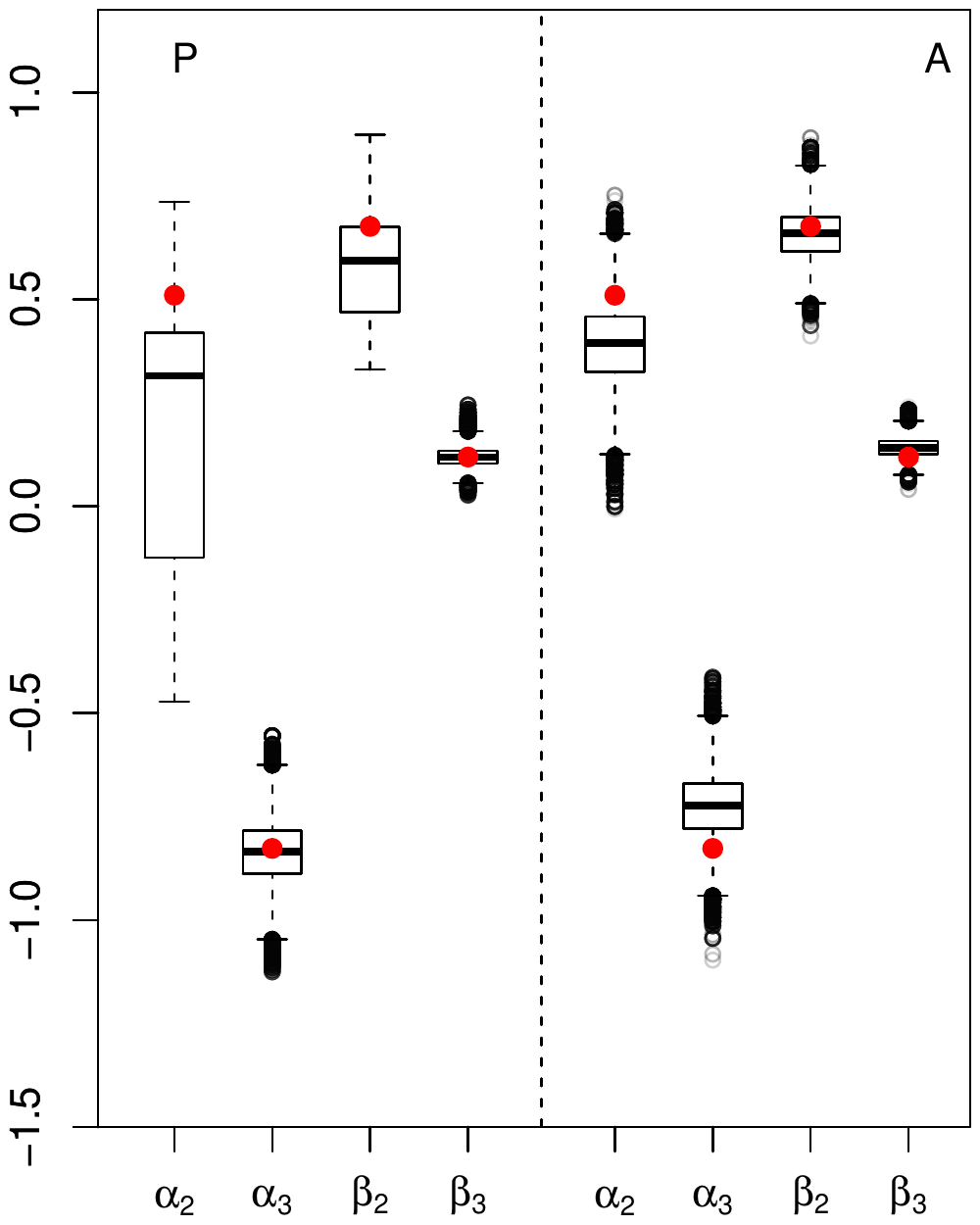} }}%
    \qquad
    \subfloat[$n=100, K=3$]{{\includegraphics[width=8cm]{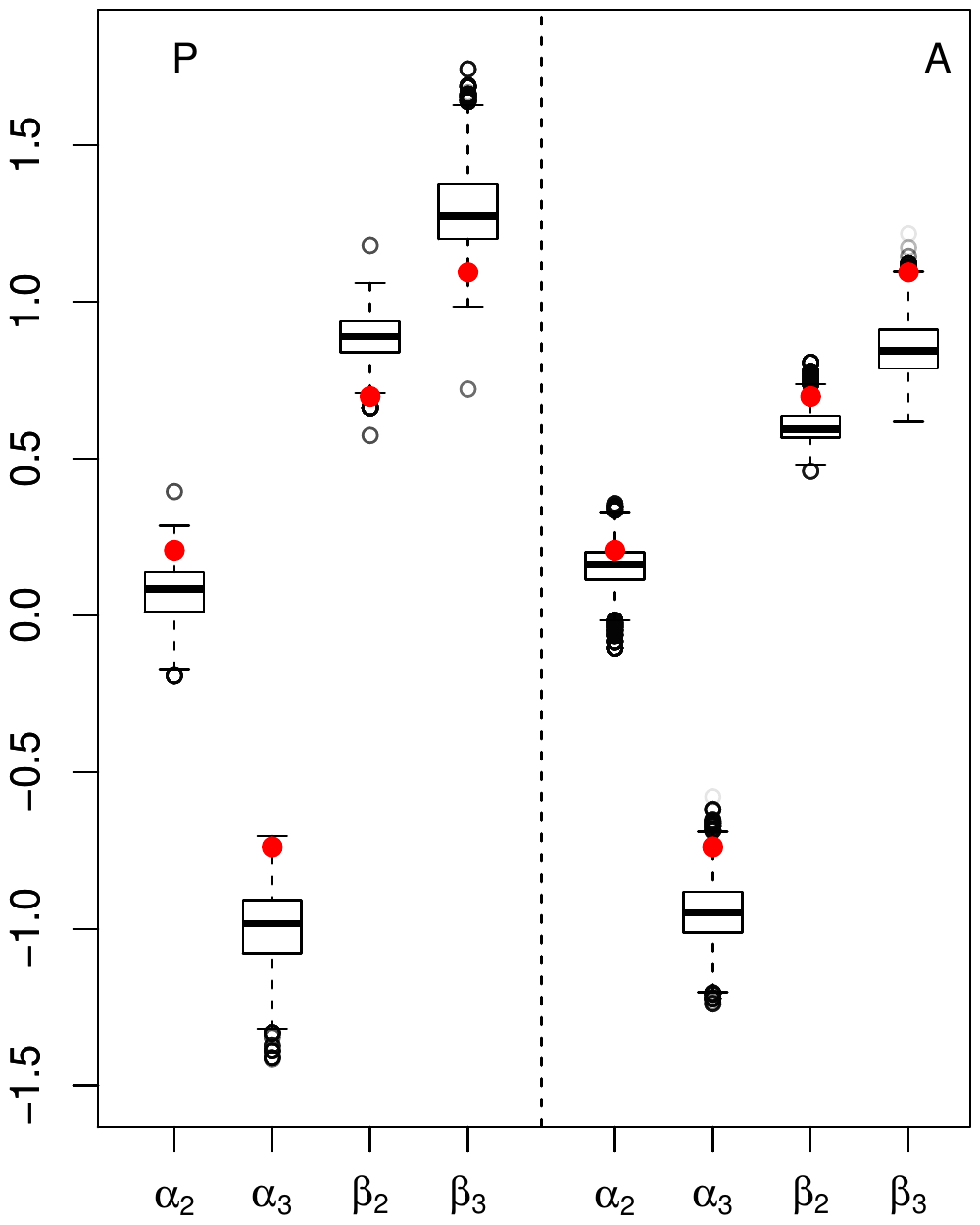} }}
    \qquad 
    \subfloat[$n=50, K=5$]{{\includegraphics[width=8cm]{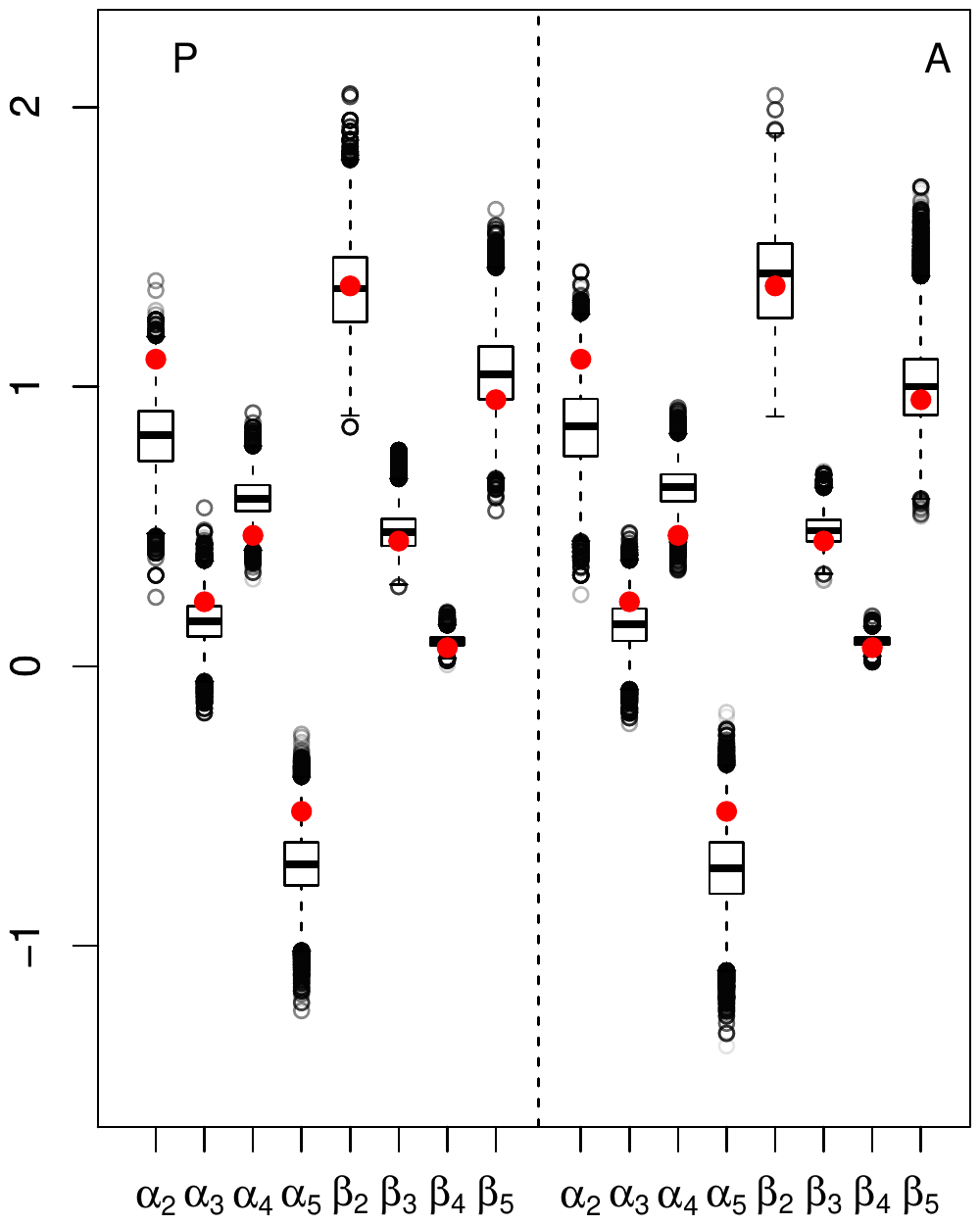} }}%
    \caption{First scenario.}%
\end{figure}
\clearpage

\subsubsection{Second scenario}
\begin{table}[!h]
\centering
\caption{Mixture of multivariate Gaussian distributions latent coordinates. Averages for the estimated logistic parameters and the procrustes correlation between true and estimated latent spaces.}
\small
\begin{tabular}{@{}lllllllllllllll@{}}
\toprule
     & $\alpha^{(1)}$  & $\hat{\alpha}^{(2)}$ & $\hat{\alpha}^{(3)}$& $\hat{\alpha}^{(4)}$   & $\hat{\alpha}^{(5)}$&  & $\beta^{(1)}$  & $\hat{\beta}^{(2)}$  & $\hat{\beta}^{(3)}$ & $\hat{\beta}^{(4)}$  & $\hat{\beta}^{(5)}$ & & PC &                
         \\
\midrule
\multirow{2}{*}{$n= 25$, $K =3$}  & 0 & $-0.39$ & $0.58$ & - & - &  & 1 & $1.24$  & $0.41$ & - & - &  & $0.85$ &\texttt{P}\\
  & 0 & $-0.38$ & $0.60$ & - & - &  & 1 & $1.26$ & $0.40$ & - & - & & $0.86$ & \texttt{A}                        \\
\multirow{2}{*}{$n= 50$, $K =3$}  & 0 & $0.55$ & $-0.79$ & - & - &  & 1 & $0.63$ & $0.11$ & - & - & & $0.93$ & \texttt{P}\\
  & 0 & $0.59$ & $-0.80$ & - & - &  & 1 & $0.63$ & $0.10$ & - & - & & $0.93$ & \texttt{A}                        \\
\multirow{2}{*}{$n= 100$, $K =3$}  & 0 & $0.10$ & $-0.79$ & - & - &  & 1 & $0.54$ & $0.91$ & - & - & & $0.95$ & \texttt{P}\\
  & 0 & $0.12$ & $-0.81$ & - & - &  & 1 & $0.53$ & $0.91$ & - & - & & $0.95$ & \texttt{A}                        \\
\multirow{2}{*}{$n= 50$, $K =5$}  & 0 & $1.03$ & $0.29$ & $0.56$ & $-0.64$ &  & 1 & $1.10$ & $0.39$ & $0.04$ & $0.70$ & & $0.95$ & \texttt{P}\\
  & 0 & $1.03$ & $0.24$ & $0.57$ & $-0.7$ &  & 1 & $1.13$ & $0.37$ & $0.04$ & $0.68$ & & $0.95$ & \texttt{A}                        \\
\bottomrule
\end{tabular}
\end{table}
\begin{table}[!h]
\centering
\caption{Mixture of multivariate Gaussian distributions latent coordinates. Standard deviations for the estimated logistic parameters and the procrustes correlation between true and estimated latent spaces.}
\small
\begin{tabular}{@{}lllllllllllllll@{}}
\toprule
           & $sd(\hat{\alpha}^{(2)})$ & $sd(\hat{\alpha}^{(3)})$& $sd(\hat{\alpha}^{(4)})$   & $sd(\hat{\alpha}^{(5)})$&  & $sd(\hat{\beta}^{(2)})$  & $sd(\hat{\beta}^{(3)})$ & $sd(\hat{\beta}^{(4)})$  & $sd(\hat{\beta}^{(5)})$ & & sd(PC) &                
         \\
\midrule
\multirow{2}{*}{$n= 25$, $K =3$}   & $0.07$  & $0.15$ & - & - &  & $0.06$ & $0.07$ & - & - &  & $0.17$ & \texttt{P}\\
   & $0.08$ & $0.17$  & - & - &   & $0.08$ & $0.6$ & - & - & & $0.18$ & \texttt{A}                        \\
\multirow{2}{*}{$n= 50$, $K =3$}   & $0.02$ & $0.01$ & - & - &   & $0.04$ & $0.01$ & - & - & & $0.02$ & \texttt{P}\\
   & $0.06$ & $0.01$ & - & - &   & $0.05$ & $0.01$ & - & - & & $0.02$ & \texttt{A}                        \\
\multirow{2}{*}{$n= 100$, $K =3$}  & $0.02$ & $0.03$ & - & - &  & $0.12$ & $0.18$ & - & - & & $0.01$ & \texttt{P}\\
  & $0.02$ & $0.03$ & - & - &   & $0.03$ & $0.05$ & - & - & & $0.01$ & \texttt{A}                        \\
\multirow{2}{*}{$n= 50$, $K =5$}  & $0.04$ & $0.02$ & $0.01$ & $0.02$ &   & $0.09$ & $0.03$ & $0.00$ & $0.06$ & & $0.01$ & \texttt{P}\\
   & $0.04$ & $0.02$ & $0.01$ & $0.02$ &   & $0.08$ & $0.03$ & $0.01$ & $0.04$ & & $0.02$ & \texttt{A}                        \\
\bottomrule
\end{tabular}
\end{table}
\begin{figure}%
    \centering
    \subfloat[$n=25, K=3$]{{\includegraphics[width=8cm]{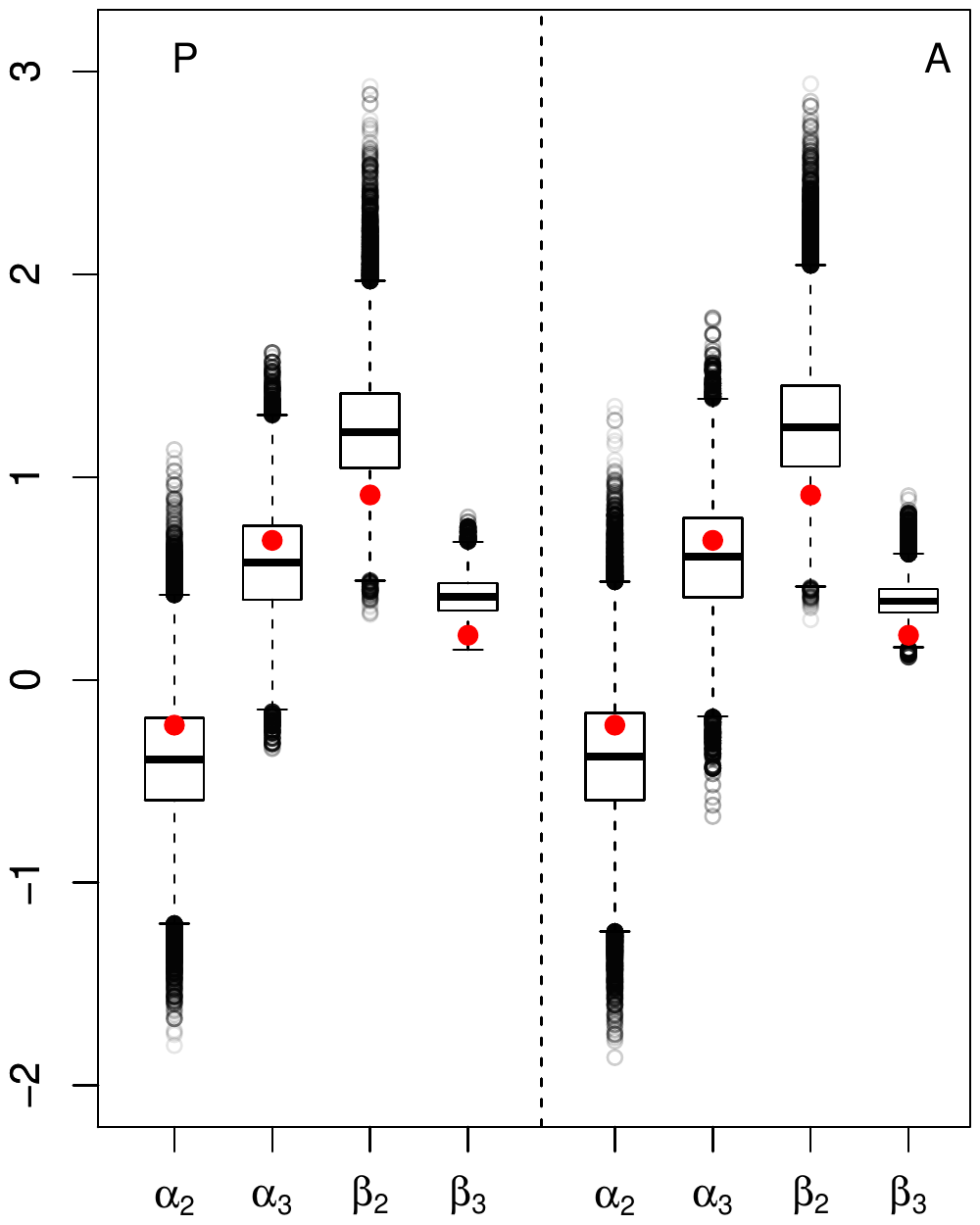} }}%
    \qquad 
    \subfloat[$n=50, K=3$]{{\includegraphics[width=8cm]{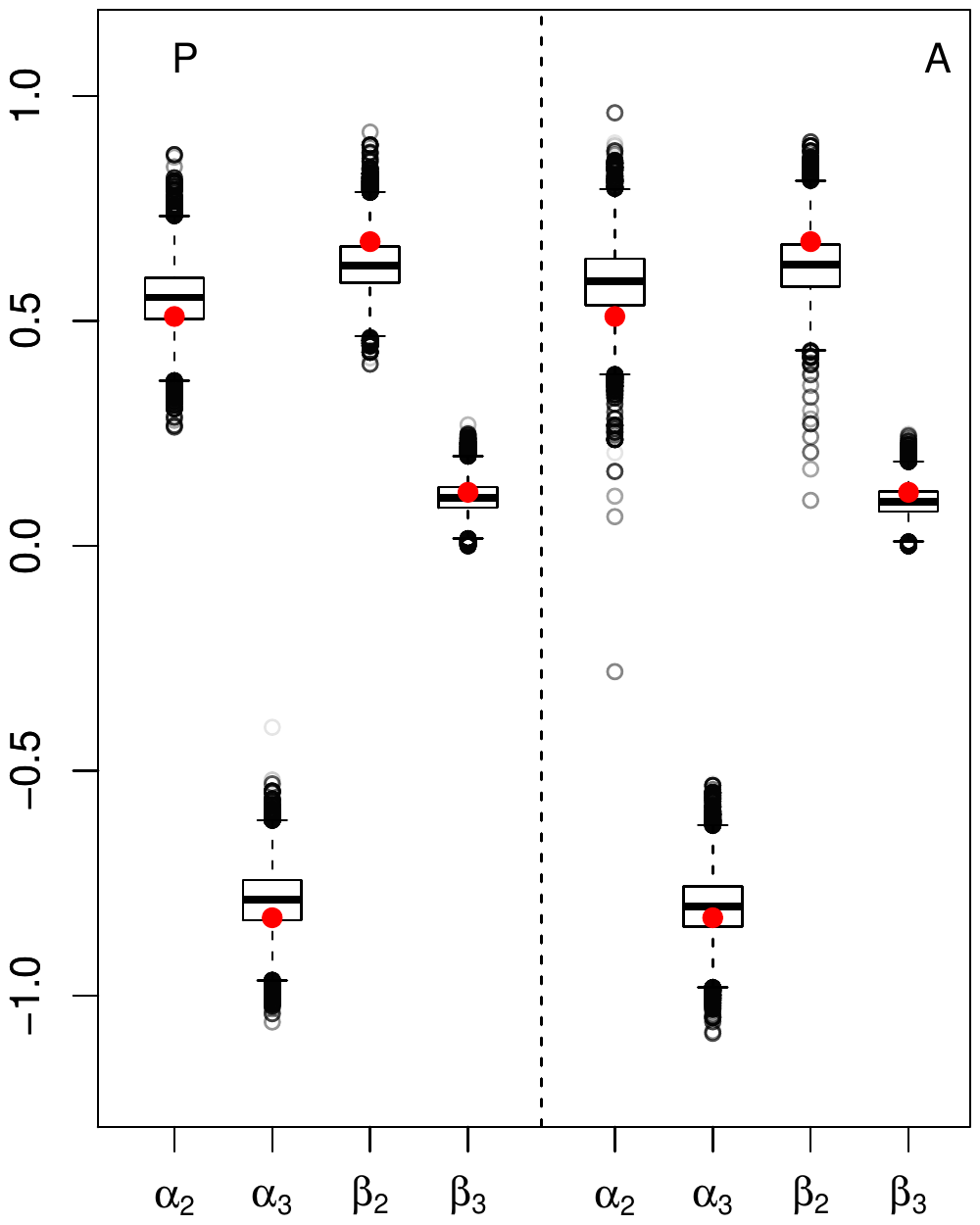} }}%
    \qquad
    \subfloat[$n=100, K=3$]{{\includegraphics[width=8cm]{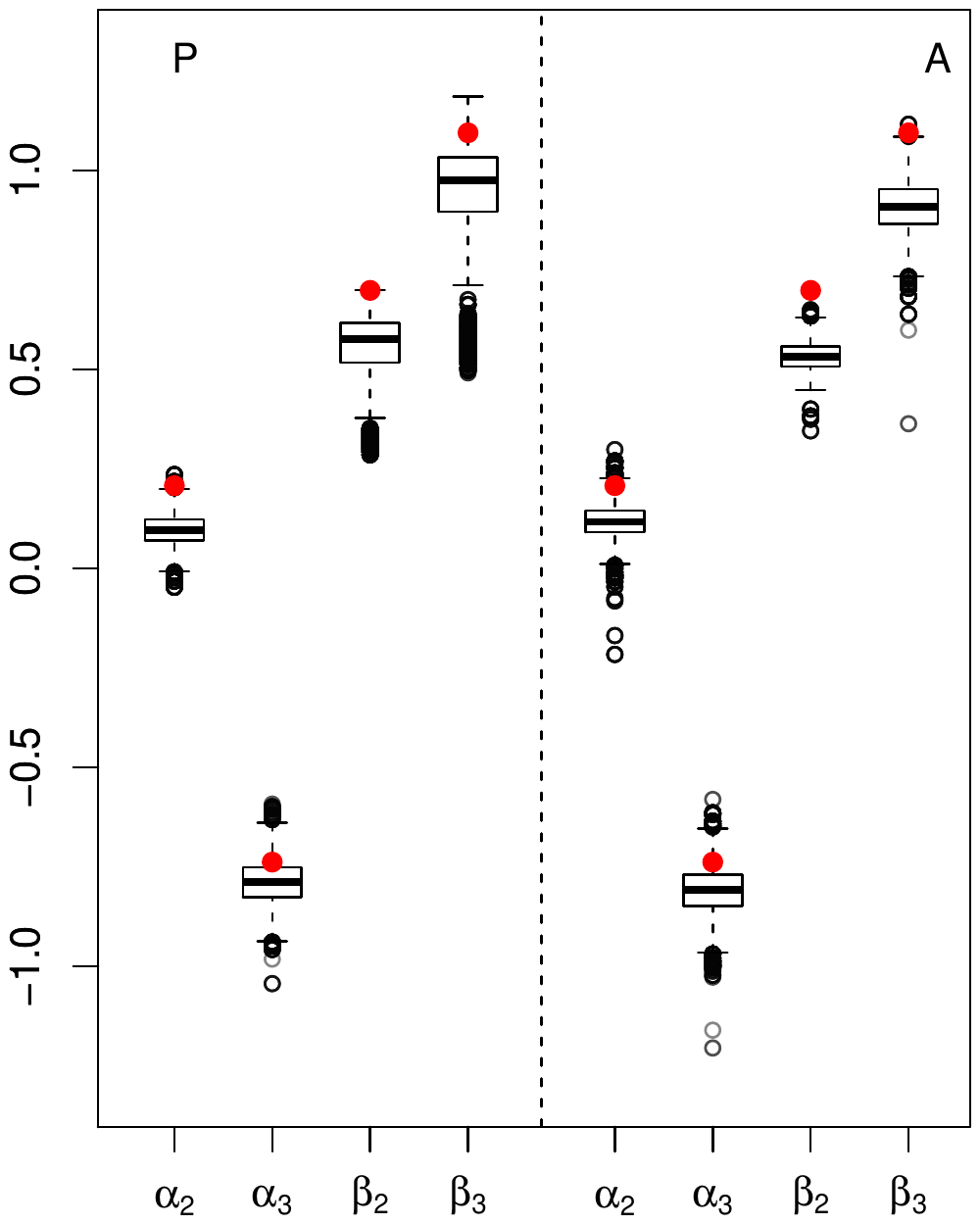} }}
    \qquad 
    \subfloat[$n=50, K=5$]{{\includegraphics[width=8cm]{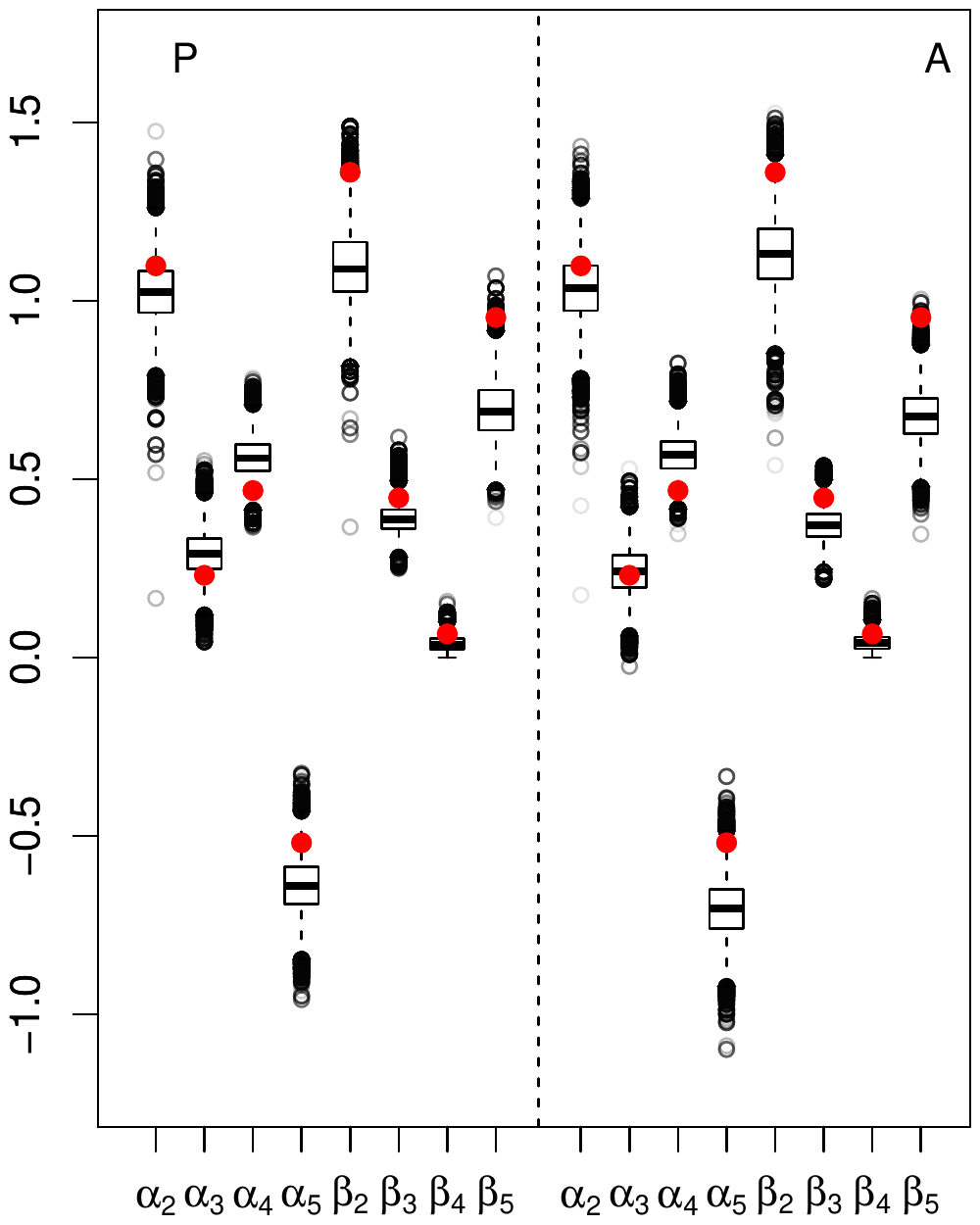} }}%
    \caption{Second scenario.}%
\end{figure}
\clearpage

\subsection{Third scenario}
\begin{table}[!h]
\centering
\caption{Hotelling T squared latent coordinates. Averages for the estimated logistic parameters and the procrustes correlation between true and estimated latent spaces.}
\small
\begin{tabular}{@{}lllllllllllllll@{}}
\toprule
        & $\alpha^{(1)}$  & $\hat{\alpha}^{(2)}$ & $\hat{\alpha}^{(3)}$& $\hat{\alpha}^{(4)}$   & $\hat{\alpha}^{(5)}$&  & $\beta^{(1)}$  & $\hat{\beta}^{(2)}$  & $\hat{\beta}^{(3)}$ & $\hat{\beta}^{(4)}$  & $\hat{\beta}^{(5)}$ & & PC &                
         \\
\midrule
\multirow{2}{*}{$n= 25$, $K =3$}  & 0 & $-0.69$ & $0.54$ & - & - &  & 1 & $1.04$ & $0.27$ & - & - & & $0.92$ &\texttt{P}\\
  & 0 & $-0.69$ & $0.56$ & - & - &  & 1 & $1.02$ & $0.26$ & - & - & & $0.90$ & \texttt{A}                        \\
\multirow{2}{*}{$n= 50$, $K =3$} & 0 & $0.18$ & $-0.82$ & - & - &  & 1 & $0.51$ & $0.12$ & - & - & & $0.95$ &\texttt{P}\\
  & 0 & $0.29$ & $-0.74$ & - & - &  & 1 & $0.59$ & $0.14$ & - & - & & $0.95$ &\texttt{A}                        \\
  \multirow{2}{*}{$n= 100$, $K =3$}  & $0$ & $0.02$ & $-0.93$ & - & - &  & 1 & $0.93$ & $1.39$ & - & - & & $0.80$ &\texttt{P}\\
  & $0$ & $0.04$ & $-0.87$ & - & -&  & 1 & $0.90$ & $1.40$ & - & - & & $0.80$ & \texttt{A}                        \\
  \multirow{2}{*}{$n= 50$, $K =5$}  & $0$ & $0.47$ & $0.12$ & $0.60$ & $-0.65$ &  & 1 & $1.26$ & $0.60$ & $0.11$ & $1.16$ & & $0.97$ &\texttt{P}\\
 & $0$ & $0.53$ & $0.13$ & $0.58$ & $-0.72$ &  & 1 & $1.23$ & $0.56$ & $0.10$ & $1.06$ & & $0.96$ &\texttt{A}                        \\
\bottomrule
\end{tabular}
\end{table}
\begin{table}[!h]
\centering
\caption{Hotelling T squared latent latent coordinates. Standard deviations for the estimated logistic parameters and the procrustes correlation between true and estimated latent spaces.}
\small
\begin{tabular}{@{}lllllllllllllll@{}}
\toprule
      & $sd(\hat{\alpha}^{(2)})$ & $sd(\hat{\alpha}^{(3)})$& $sd(\hat{\alpha}^{(4)})$   & $sd(\hat{\alpha}^{(5)})$&  & $sd(\hat{\beta}^{(2)})$  & $sd(\hat{\beta}^{(3)})$ & $sd(\hat{\beta}^{(4)})$  & $sd(\hat{\beta}^{(5)})$ & & sd(PC) &                
         \\
\midrule
\multirow{2}{*}{$n= 25$, $K =3$}   & $0.04$ & $0.03$ & - & - &  & $0.02$ & $0.02$ & - & - &  & $0.02$ & \texttt{P}\\
   & $0.04$ & $0.02$ & - & - &   & $0.03$ & $0.01$ & - & - & & $0.02$ & \texttt{A}                        \\
\multirow{2}{*}{$n= 50$, $K =3$}   & $0.31$ & $0.10$ & - & - &  & $0.06$ & $0.02$ & - & - &  & $0.02$ & \texttt{P}\\
   & $0.04$ & $0.01$ & - & - &   & $0.04$ & $0.01$ & - & - & & $0.01$ & \texttt{A}                        \\  
\multirow{2}{*}{$n= 100$, $K =3$}   & $0.01$ & $0.01$  & - & - &  & $0.02$ & $0.02$ &  & - & - & $0.02$ & \texttt{P}\\
   & $0.01$ & $0.01$ & - & - &   & $0.01$ & $0.02$ & - & - & & $0.02$ & \texttt{A}                        \\
\multirow{2}{*}{$n= 50$, $K =5$}   & $0.10$ & $0.03$ & $0.01$ & $0.07$ &  & $0.08$ & $0.04$ & $0.01$ & $0.08$ &  & $0.01$ & \texttt{P}\\
   & $0.08$ & $0.03$  & $0.02$ & $0.05$ &   & $0.12$ & $0.06$ & $0.01$ & $0.08$ & & $0.01$ & \texttt{A}                        \\
\bottomrule
\end{tabular}
\end{table}
\begin{figure}%
    \centering
    \subfloat[$n=25, K=3$]{{\includegraphics[width=8cm]{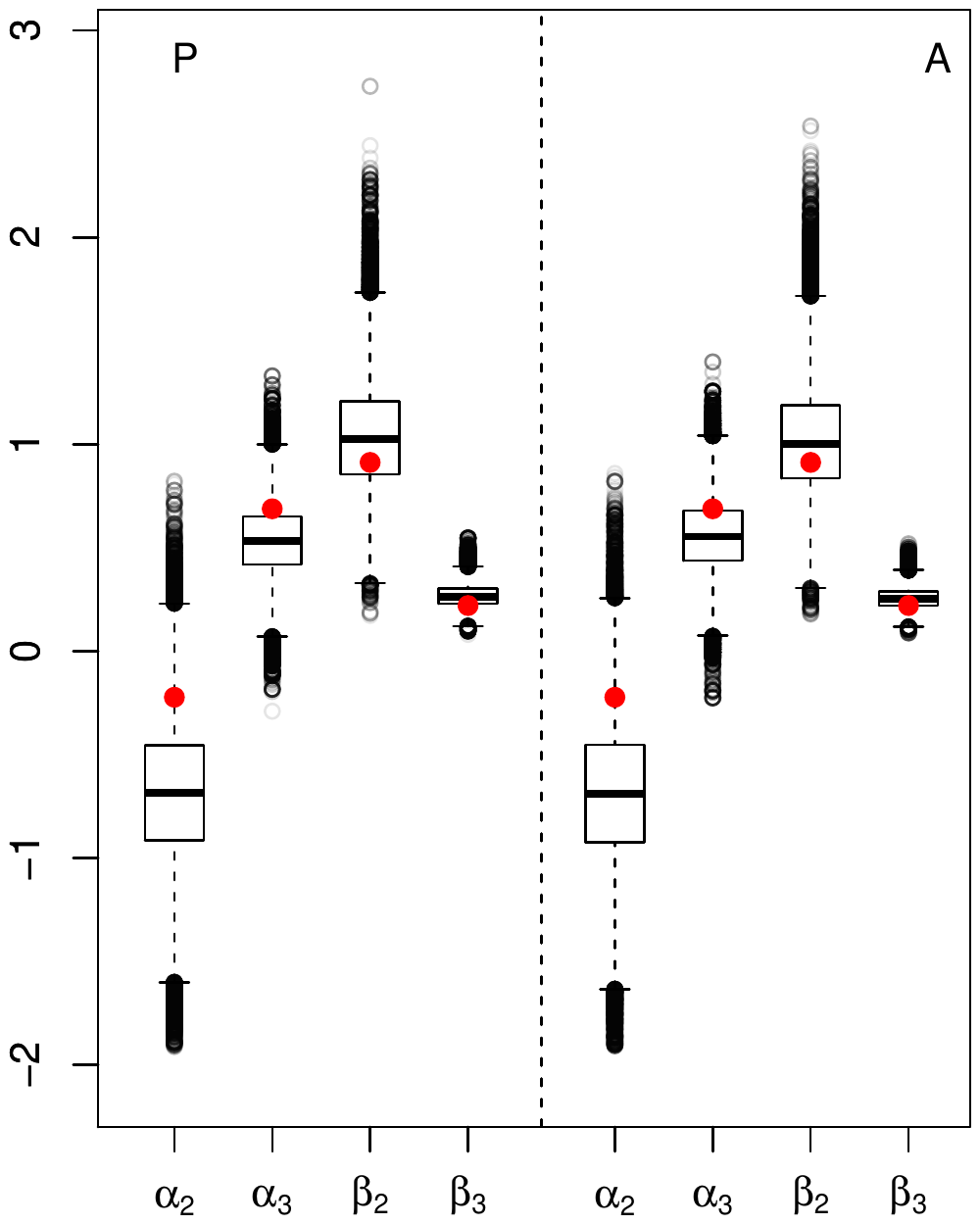} }}%
    \qquad 
    \subfloat[$n=50, K=3$]{{\includegraphics[width=8cm]{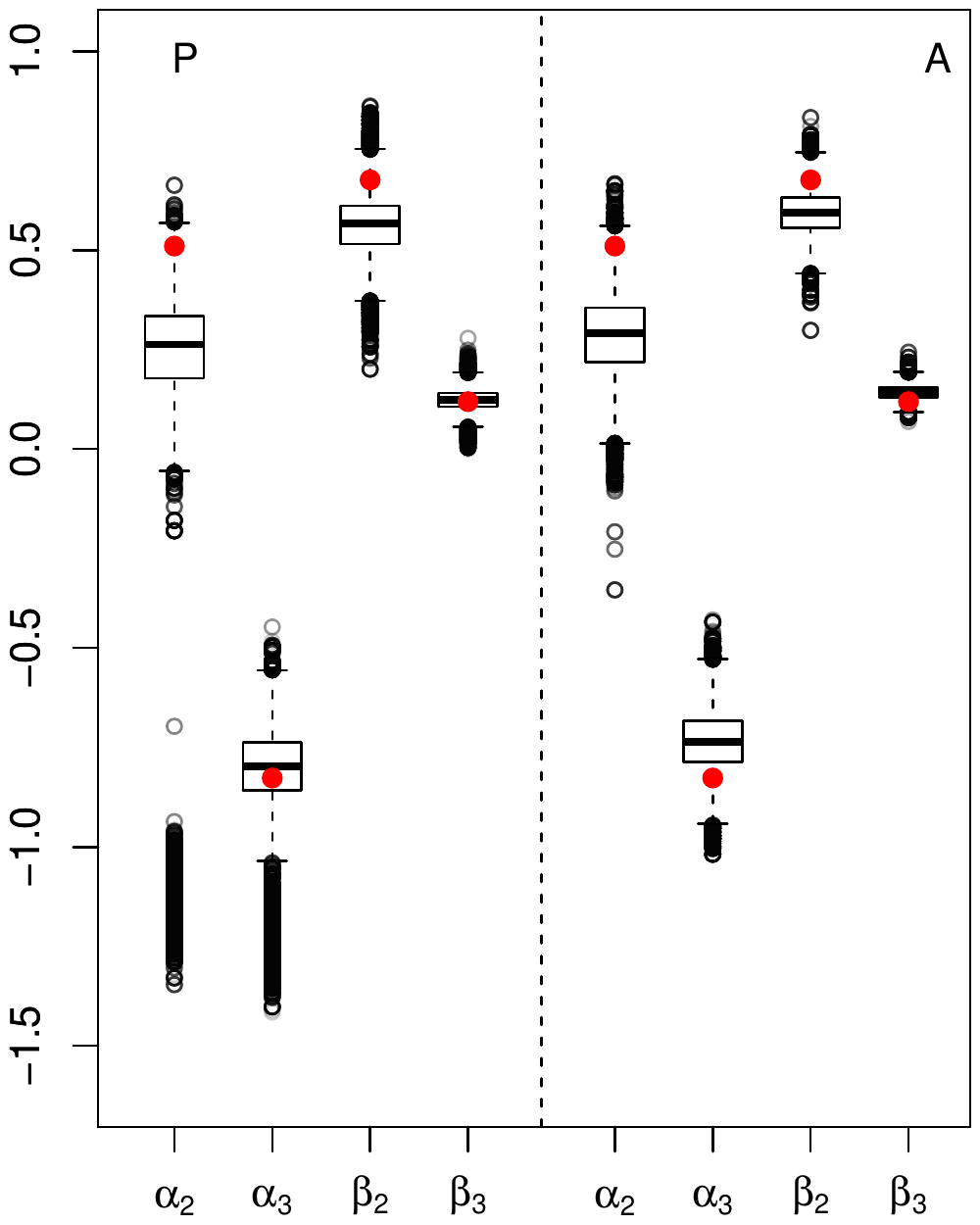} }}%
    \qquad
    \subfloat[$n=100, K=3$]{{\includegraphics[width=8cm]{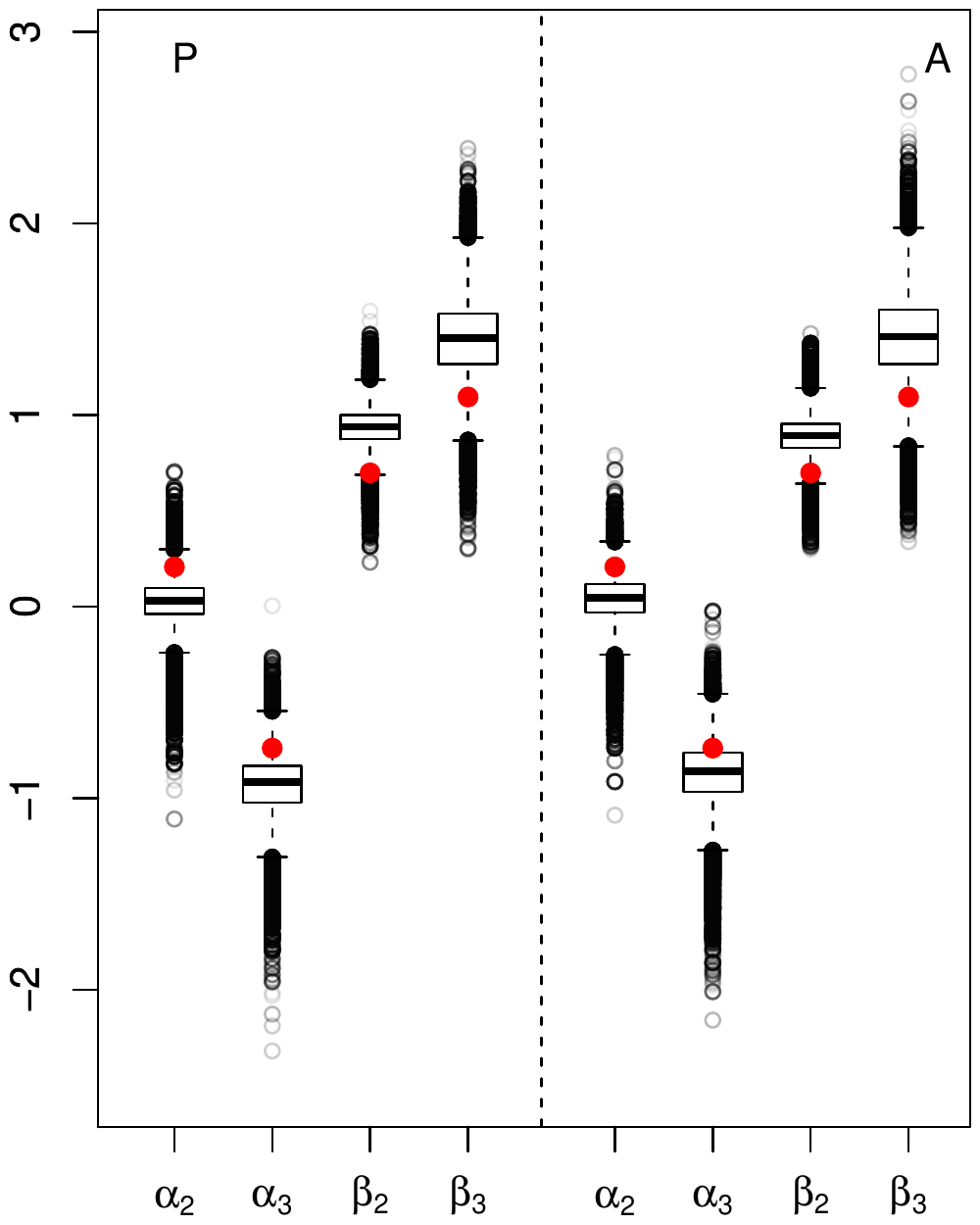} }}
    \qquad 
    \subfloat[$n=50, K=5$]{{\includegraphics[width=8cm]{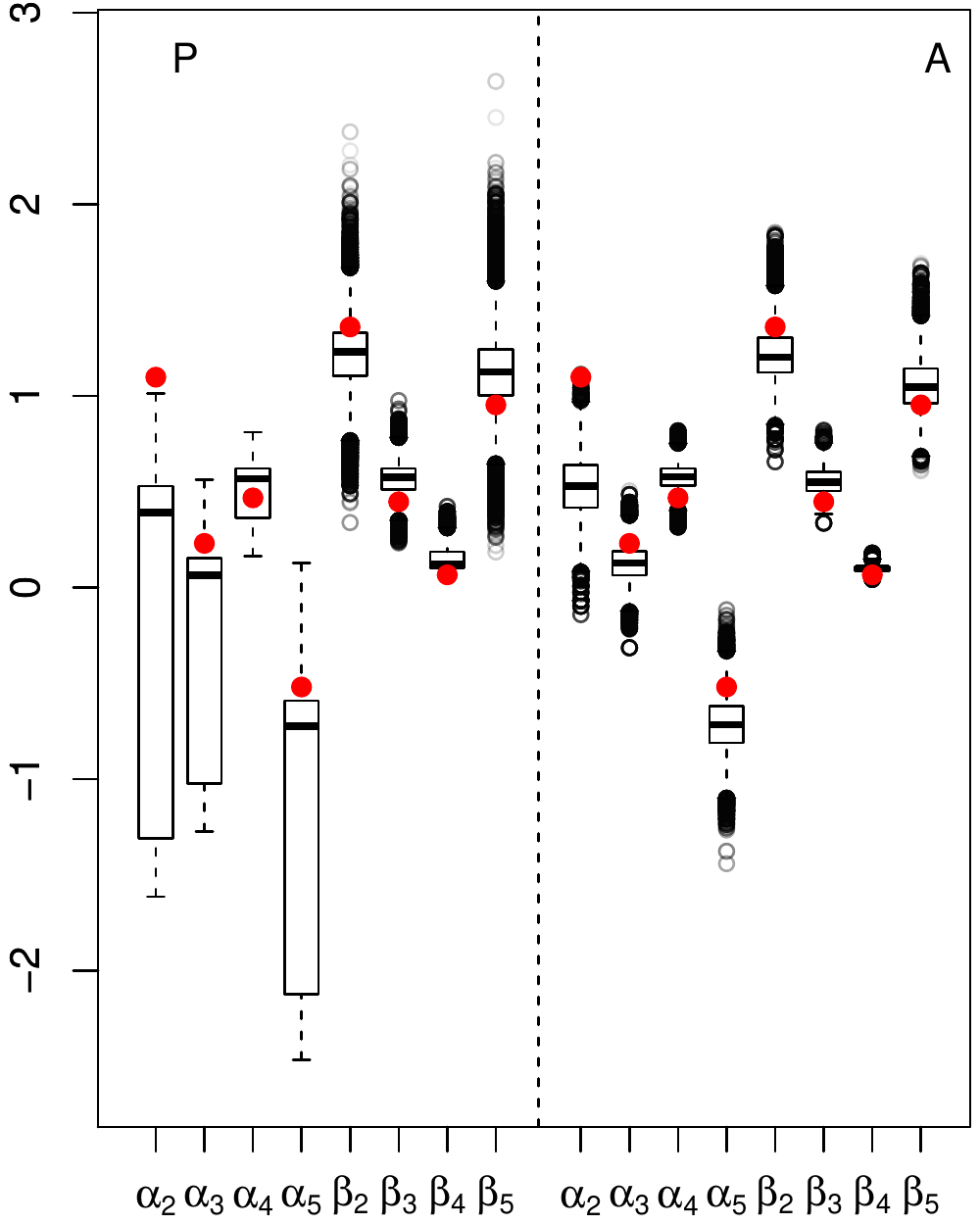} }}%
    \caption{Third scenario.}%
  \end{figure}
\clearpage

\subsection{Results for block II}
\subsubsection{Fourth scenario}
\begin{table}[!h]
\centering
\caption{Procrustes correlation, second block}
\begin{tabular}{@{}ccccccccc@{}}
\toprule
 & \multicolumn{2}{c}{{\large $K = 10$} } &  \multicolumn{2}{c}{{\large $K = 20$} }& \multicolumn{2}{c}{{\large $K = 30$}} \\
     & {\footnotesize P} & {\footnotesize A} & {\footnotesize P} & {\footnotesize A} & {\footnotesize P} & {\footnotesize A} \\
\midrule
PC & $0.97$ & $0.96$ & $0.97$ & $0.97$ & $0.97$ & $0.97$ \\
sd(PC)  & $0.01$ & $0.01$ & $0.01$ & $0.01$ & $0.01$ & $0.01$ \\
\bottomrule
\end{tabular}
\end{table}
\begin{figure}[!h]%
    \centering
    \subfloat[$\alpha$ parameters.]{{\includegraphics[width=8.1cm]{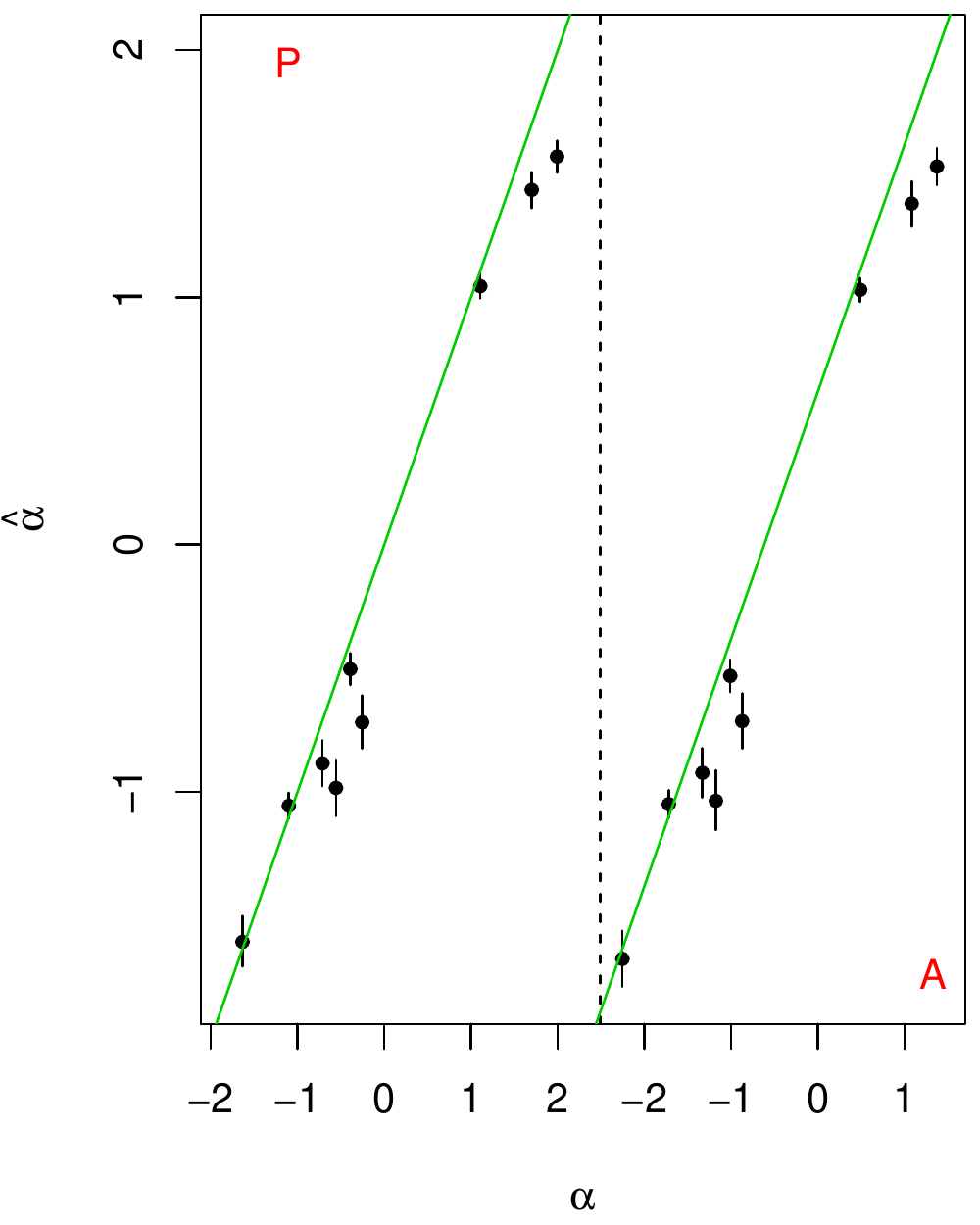} }}%
        \subfloat[$\beta$ parameters.]{{\includegraphics[width=8.1cm]{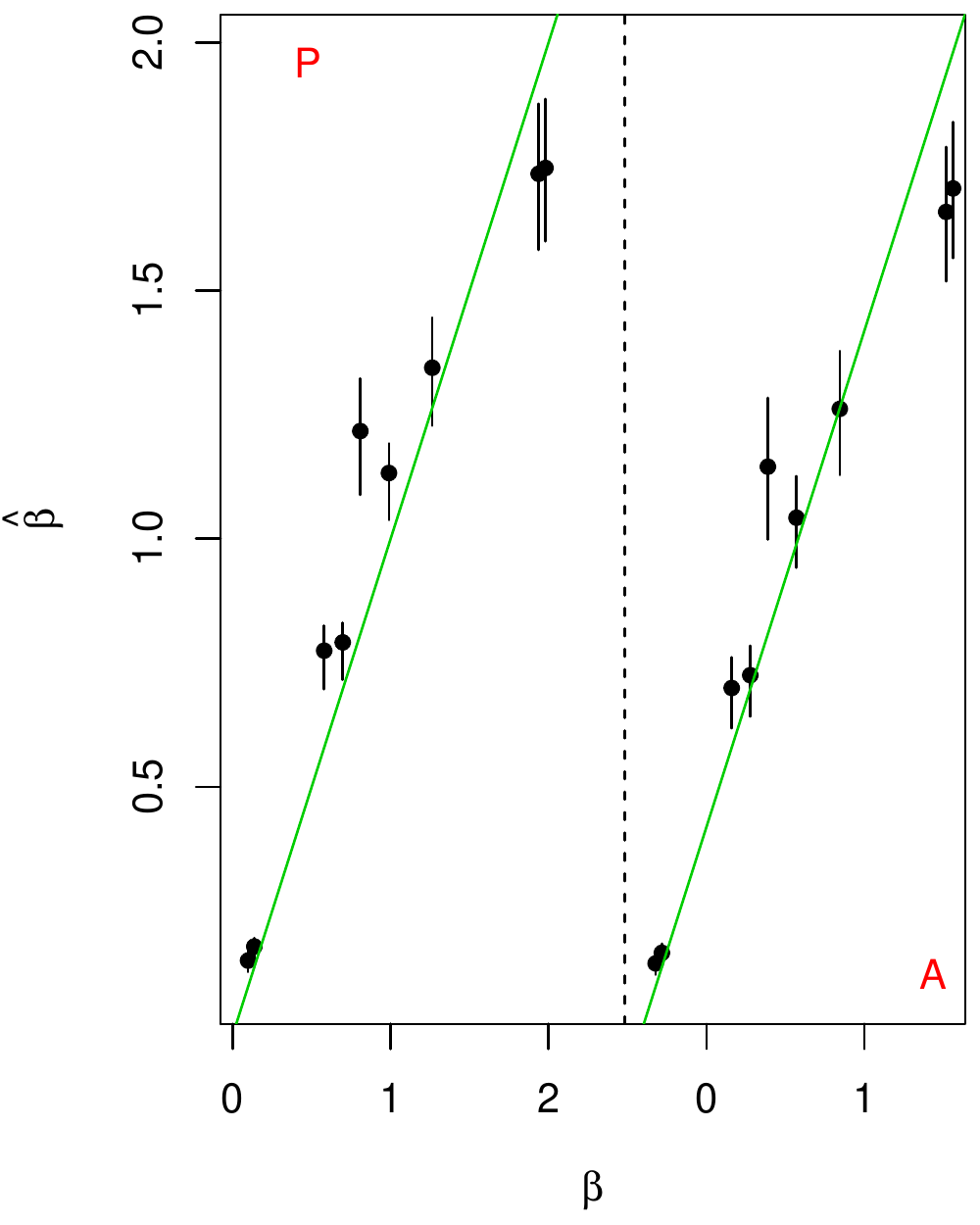} }}%
    \caption{Second block. Multiplex with $n=50$ and $K =10$.}%
    \end{figure}
\begin{figure}[!h]%
    \centering
    \subfloat[$\alpha$ parameters.]{{\includegraphics[width=8.1cm]{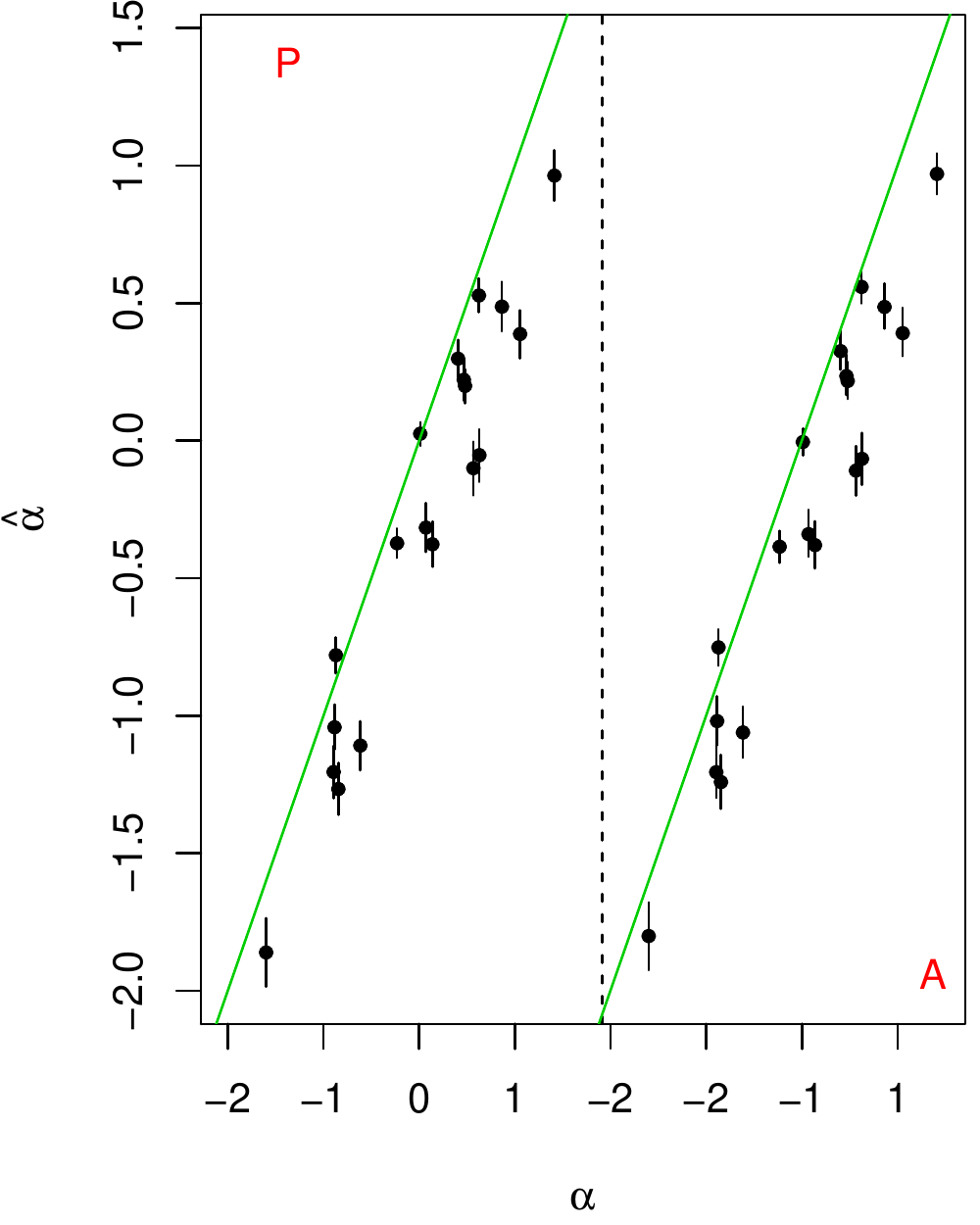} }}%
    \subfloat[$\beta$ parameters.]{{\includegraphics[width=8.1cm]{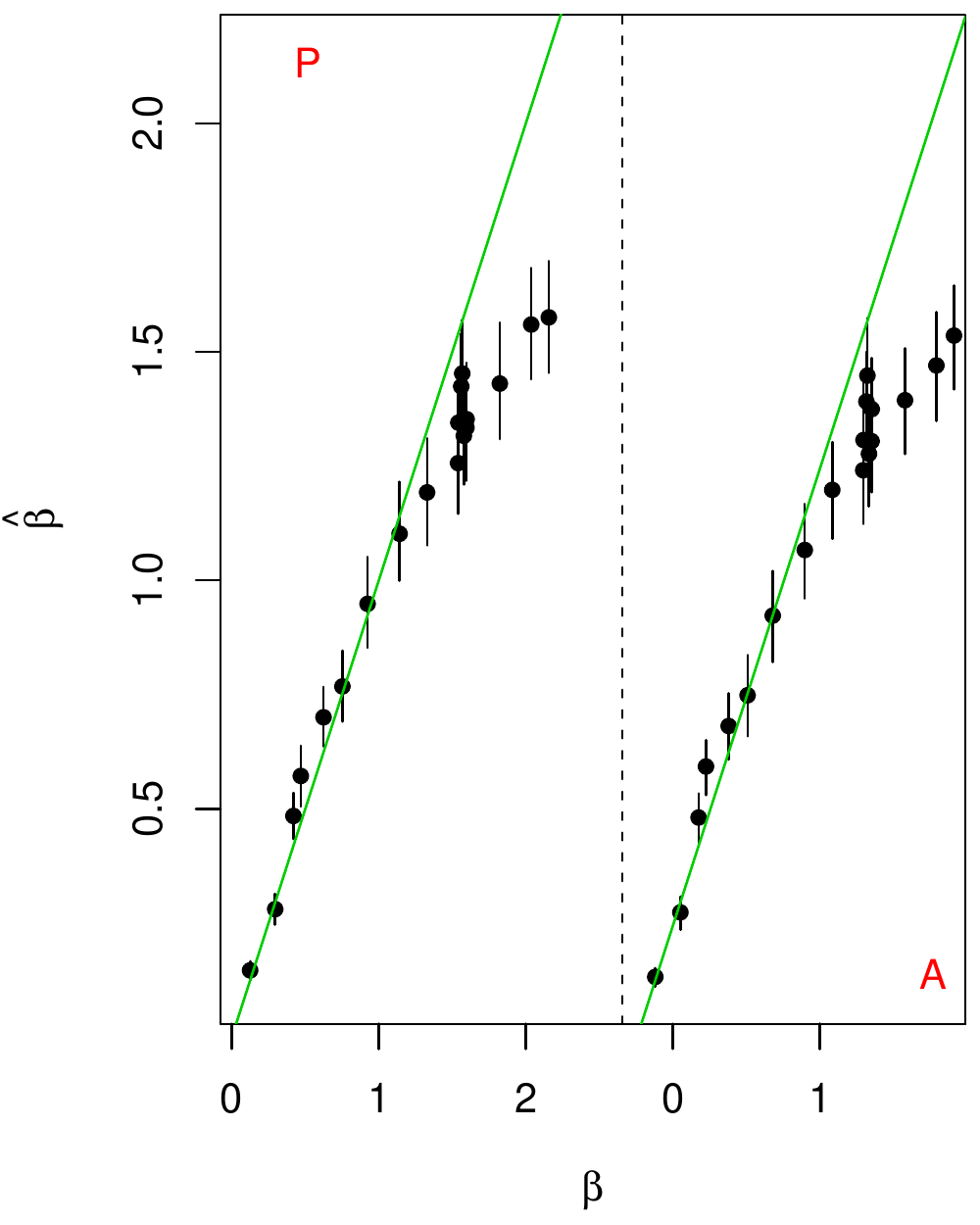} }}%
    \caption{Second block. Multiplex with $n=50$ and $K =20$.}%
  \end{figure}
\begin{figure}[!h]%
    \centering
    \subfloat[$\alpha$ parameters.]{{\includegraphics[width=8.1cm]{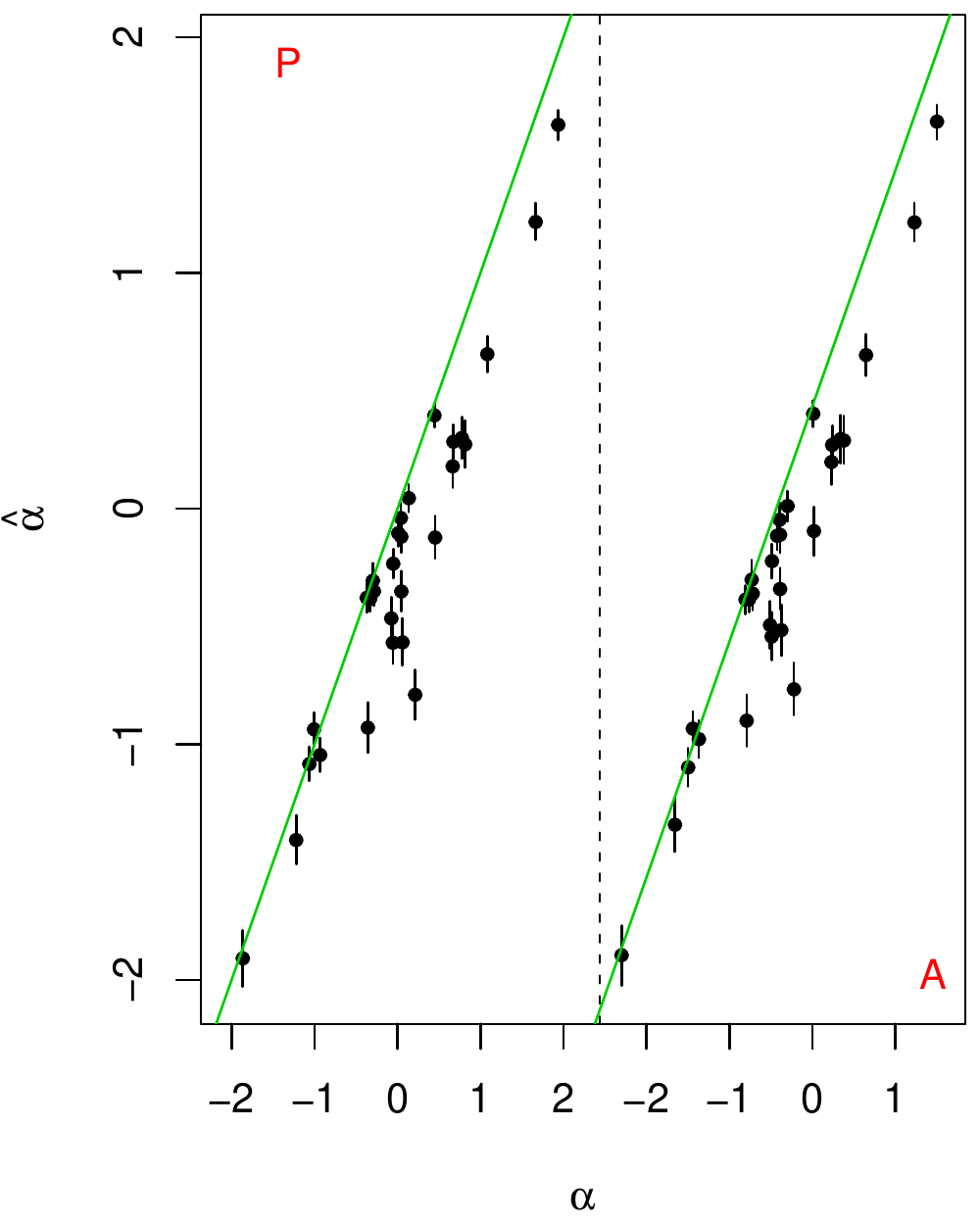} }}%
   \subfloat[$\beta$ parameters.]{{\includegraphics[width=8.1cm]{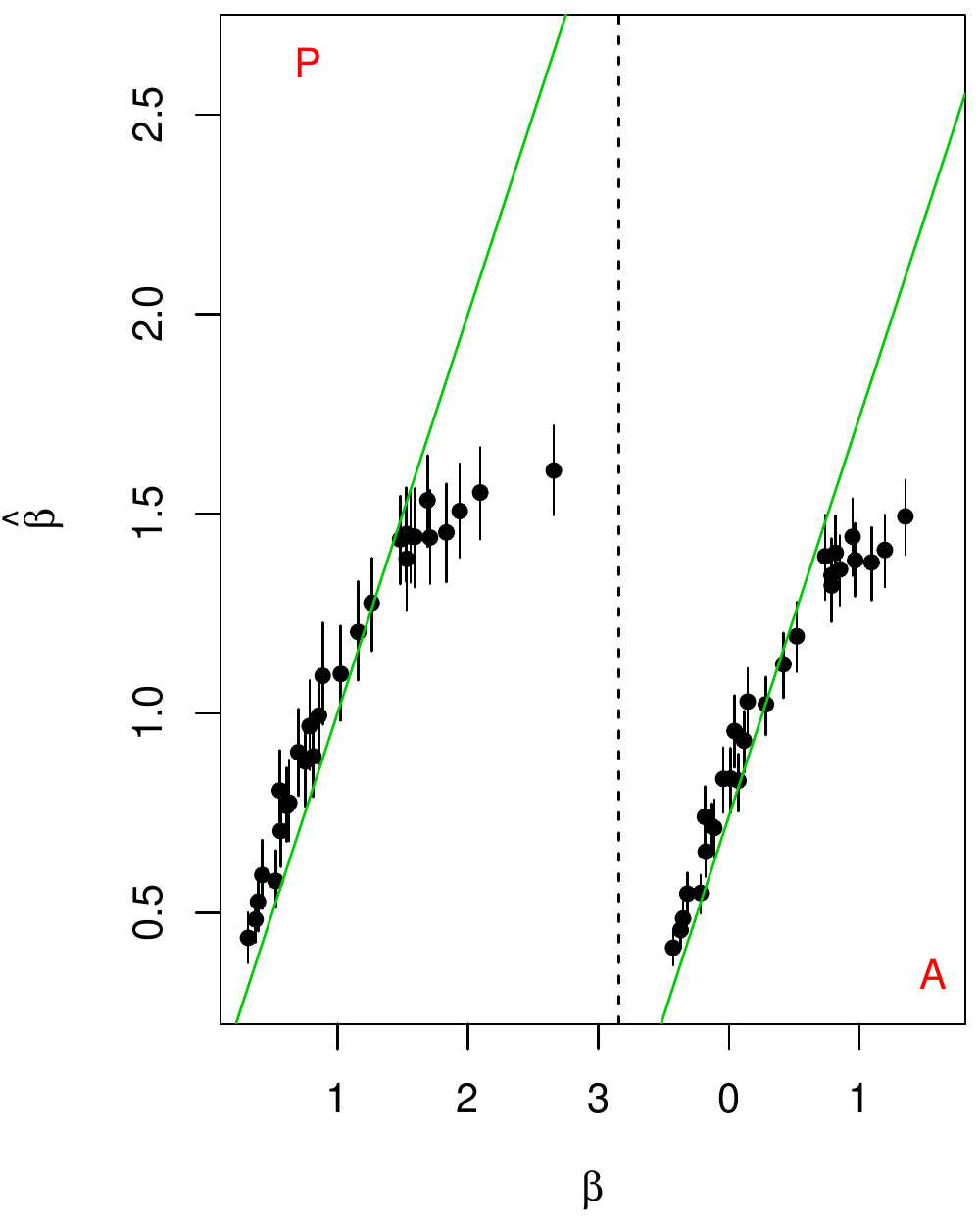} }}%
    \caption{Second block. Multiplex with $n=50$ and $K =30$.}%
   \end{figure}
\clearpage

\subsection{Comparison with the \emph{lsjm} model. Results}
\label{app3_isa}
\begin{table}[!h]
\centering
\caption{Results of the comparison with the \emph{lsjm} model. Averages and standard deviations for the estimated intercepts and the procrustes correlation between true and estimated latent spaces. The acronym \emph{lsmmn} is used to indicate the model presented in this work. }
\small
\begin{tabular}{@{}llllllllllllll@{}}
\toprule
   
        & $\hat{\alpha}^{(1)}$  & $\hat{\alpha}^{(2)}$ & $\hat{\alpha}^{(3)}$& & $sd(\hat{\alpha}^{(1)})$  & $sd(\hat{\alpha}^{(2)})$ & $sd(\hat{\alpha}^{(3)})$& & PC & & sd(PC) &                
         \\
\midrule
\multirow{2}{*}{$n= 50$, $K =3$}  & $-201.89$ & $-211.10$ & $-193.75$ &  & $1501.55$ & $1500.95$ & $ 1496.83$ & & $0.46$ & & $0.27$ &\texttt{lsjm}\\
  & $0$ & $-0.73$ & $-0.64$ &  & - & $0.04$ &  $0.07$ & & $0.96$ & & $0.01$ & \texttt{lsmmn}    \\
\multirow{2}{*}{$n= 70$, $K =2$}  & $-0.84$ & $-1.30$ & - &  & $0.06$ & $0.06$ & - & & $0.79$ & & $0.05$ &\texttt{lsjm}\\
  & $0$ & $-1.36$ & - &  & - & $0.01$ & - & & $0.85$ & & $0.05$ & \texttt{lsmmn}    \\
\multirow{2}{*}{$n= 25$, $K =3$}  & $ -44.38$ & $15.40$ & $-16.80$ &  & $360.49$ & $353.41$ & $356.76$ & & $0.57$ & & $0.31$ &\texttt{lsjm}\\
  & $0$ & $0.69$ & $0.26$ &  & - & $0.08$ & $0.06$ & & $0.95$ & & $0.05$ & \texttt{lsmmn}    \\
\bottomrule
\end{tabular}
\end{table}
Concerning the \emph{lsjm} estimates, the instability in the estimation of the intercept parameters in the first and third scenario could be caused by similar identifiability issues as the one described in section \ref{eq:edge_prob} and in section \ref{identif}. The second scenario returns stable and truthful estimates of the intercept parameters and high correlation on average between the true and the estimated latent space. 
The \emph{lsmmn} model always recovers a set of latent coordinates which is highly correlated with the simulated one. The average value for this correlation in the estimates for the second multiplex is lower than the values estimated for multidimensional network one and three. This could be due to the presence of only $2$ views in the second multiplex, meaning that there are only two replications of the latent space in the data.
The \emph{lsmmn} model estimated also the coefficient parameters $\bk$, which in the simulations have all been set to $1$. The estimates produced for all the three multiplex returned faithful values for the coefficients, as in all the cases $.93 \leq \hat{\beta}^{(k)} \leq 1.05$, $\forall k$. The intercept coefficient in the first network as been fixed $\alpha^{(1)}=0$ to estimate all the three multiplex.
\clearpage

\clearpage
\section{Pseudo-code of the mcmc algorithm}
\label{app5}
\begin{algorithm}[!h]
\caption{MCmc for LSM-MN}\label{mcmc}
\begin{algorithmic}[]
\Procedure{}{}
\State $\texttt{Given:} \quad  \mathbf{Y}, K, n, LB(\alpha), \mathbf{H}$
\State $\texttt{Fix:} \quad  \eta$
\State $\texttt{Initialize:} \quad  \mathbf{z}^{\{0\}} ,\mathbf{D}^{\{0\}}, \mathbf{\alpha}^{\{0\}}, \mathbf{\beta}^{\{0\}}, \mu_{\alpha}^{\{0\}}, \mu_{\beta}^{\{0\}}, \sigma_{\alpha}^{2 {\{0\}}}, \sigma_{\beta}^{2 {\{0\}}}$
\State $\texttt{Fix: } \quad \alpha^{(1)\{0\}} = 0 , \quad\beta^{(1)\{0\}} = 1$
\For{\texttt{iter in 1:niter}}
 \State $\texttt{Simulate} \quad \sigma_{\alpha}^{2 {\{iter\}}} \quad \texttt{from} \quad g \Bigl(\sigma_{\alpha}^{2} \mid \mathbf{\alpha}^{\{iter -1\}}, \mu_{\alpha}^{\{iter -1\}}, \tau_{\alpha}, \nu_{\alpha} , K \Bigr) \texttt{;}$ 
 \State $\texttt{Simulate} \quad \sigma_{\beta}^{2 {\{iter\}}} \quad \texttt{from} \quad g\Bigl(\sigma_{\beta}^{2} \mid \mathbf{\beta}^{\{iter -1\}}, \mu_{\beta}^{\{iter -1\}}, \tau_{\beta}, \nu_{\beta}, K \Bigr) \texttt{;}$
 \State $\texttt{Simulate} \quad \mu_{\alpha}^{\{iter\}} \quad \texttt{from} \quad  g\Bigl(\mu_{\alpha} \mid  \mathbf{\alpha}^{\{iter -1\}} ,  \sigma_{\alpha}^{2 {\{iter\}}} , \tau_{\alpha},K \Bigr)  \texttt{;}$ 
  \State $\texttt{Simulate} \quad \mu_{\beta}^{\{iter\}} \quad \texttt{from} \quad g\Bigl(\mu_{\beta} \mid \mathbf{\beta}^{\{iter -1\}} ,  \sigma_{\beta}^{2 {\{iter\}}} , \tau_{\beta},K  \Bigr)\texttt{;}$ 
  \For{\texttt{k in 2:K}}
  \State $\texttt{Propose} \quad  \tilde{\alpha}^{(k) } \quad \texttt{from} \quad 
  g \Bigl( \ak \mid \mathbf{Y}, \mathbf{D}^{\{iter -1\} }, \mathbf{H} , \beta^{(k)\{iter-1\}}, \mu_{\alpha}^{\{iter \} }, \sigma_{\alpha}^{2\{iter\} },n \Bigr)\texttt{;}$
\State $\texttt{Propose} \quad  \tilde{\beta}^{(k) } \quad \texttt{from} \quad 
  g \Bigl( \bk \mid \mathbf{Y}, \mathbf{D}^{\{iter -1\} }, \mathbf{H} , \alpha^{(k)\{iter-1\}}, \mu_{\beta}^{\{iter \} }, \sigma_{\beta}^{2\{iter\} },n \Bigr)\texttt{;}$  
  
 \State $\texttt{Accept} \quad \Bigl( \tilde{\alpha}^{(k) }, \tilde{\beta}^{(k)} \Bigr) \quad \texttt{with probability} \quad A1 $
  \If{Accept}
 \State $\alpha^{(k)\{iter \} }, \beta^{(k)\{iter \}} \leftarrow \tilde{\alpha}^{(k) }, \tilde{\beta}^{(k)} $ 
 \Else
 \State $\alpha^{(k)\{iter \} }, \beta^{(k)\{iter \}} \leftarrow \alpha^{(k)\{iter -1\} }, \beta^{(k)\{iter -1\}} $ 
 \EndIf
 \EndFor
 \State $\texttt{Assign} \quad \mathbf{z}^{\{old\}} \leftarrow \mathbf{z}^{\{iter -1\}} $
 \For{\texttt{i in 1:n}}
 \State  $\texttt{Propose} \quad  \tilde{z}_i \quad \texttt{from} \quad 
  g \Bigl( z_i \mid \mathbf{Y},\alpha^{\{iter\}}, \beta^{\{iter\}}, \mathbf{D}^{\{iter-1\}}, K \Bigr)\texttt{;}$  
  \State  $\texttt{Accept} \quad \tilde{z}_i \quad \texttt{with probability} \quad A2$
   \If{Accept}
  \State $ z_i^{\{iter \}} \leftarrow \tilde{z}_i $
  \Else
  \State $ z_i^{\{iter \}} \leftarrow  z_i^{\{iter -1\}}$
\EndIf
\EndFor
\If{ $\texttt{Procrustes Check} \bigl( \mathbf{z}^{\{old\}},\mathbf{z}^{\{iter\}} \bigr)== 0 $}
\State $\mathbf{z}^{\{iter \}} \leftarrow \mathbf{z}^{\{iter\}} $
\Else
\State $\mathbf{z}^{\{iter \}} \leftarrow \mathbf{z}^{\{iter -1\}} $
\EndIf
 
      \EndFor
\EndProcedure
\end{algorithmic}
\end{algorithm}
If edge-specific covariates are considered, the parameters $\lambda_f$ are updated after the latent positions, with a Metropolis-Hastings step, using their proposal distribution (\ref{app1}). The nuisance parameters $\mu_{\lambda_f}$ and $\sigma_{\lambda_f}^2$ are updated via their full conditional distribution, right after the update of the nuisance parameters for the prior distributions of $\alpha$ and $\beta$. In addition, the proposal distributions to use  for the coordinates, the coefficients and the intercepts are the ones described in appendix \ref{app1}.

\end{document}